\documentclass[12pt]{article}
\usepackage{a4wide}
\usepackage{latexsym}
\usepackage{amsmath}
\usepackage{amsfonts}
\usepackage{amscd}
\usepackage{cite}
\usepackage{graphicx}
\usepackage{axodraw}
\usepackage{float}

\usepackage{pslatex}
\usepackage[latin1]{inputenc}
\usepackage[OT2,T1]{fontenc}

\def\bq{\begin{eqnarray}}
\def\eq{\end{eqnarray}}
\def\l{\langle}
\def\r{\rangle} 
\def\eps{\varepsilon}

\def\mynosign{}
\def\mysign{-}

\DeclareSymbolFont{cyrletters}{OT2}{wncyr}{m}{n}
\DeclareMathSymbol{\Sha}{\mathalpha}{cyrletters}{"58}

\usepackage{color}

\begin{document}

\thispagestyle{empty}

\begin{flushright}
  MZ-TH/10-38
\end{flushright}

\vspace{1.5cm}

\begin{center}
  {\Large\bf Numerical NLO QCD calculations\\
  }
  \vspace{1cm}
  {\large Sebastian Becker, Christian Reuschle and Stefan Weinzierl\\
\vspace{2mm}
      {\small \em Institut f{\"u}r Physik, Universit{\"a}t Mainz,}\\
      {\small \em D - 55099 Mainz, Germany}\\
  } 
\end{center}

\vspace{2cm}

\begin{abstract}\noindent
  {
We present an algorithm for the numerical calculation of one-loop QCD amplitudes.
The algorithm consists of subtraction terms, approximating the soft, collinear and
ultraviolet divergences of one-loop amplitudes and a method to deform the integration contour
for the loop integration into the complex space.
The algorithm is formulated at the amplitude level and does not rely on Feynman graphs.
Therefore all required ingredients can be calculated efficiently using recurrence relations.
The algorithm applies to massless partons as well as to massive partons.
   }
\end{abstract}

\vspace*{\fill}

\newpage

\tableofcontents

\newpage

\section{Introduction}

Multi-jet final states play an important role for the experiments at the LHC.
An accurate description of jet physics is therefore desirable.
Although jet observables can rather easily be modelled at leading order (LO)
in perturbation theory,
this description suffers several drawbacks.
A leading order calculation depends strongly on the renormalisation scale
and can therefore give only an order-of-magnitude estimate on absolute rates.
Secondly, at leading order a jet is modelled by a single parton. This is a very crude
approximation and oversimplifies inter- and intra-jet correlations.
The situation is improved by including higher order corrections in perturbation theory.

At present, there are many next-to-leading order (NLO) calculations for $2 \rightarrow 2$ processes at
hadron colliders, but only a few for $3$ or more partons in the 
final state \cite{
Kilgore:1996sq,Nagy:2001fj,Nagy:2003tz,
Dittmaier:2007wz,Dittmaier:2008uj,Melnikov:2010iu,
Beenakker:2002nc,Dawson:2003zu,
Lazopoulos:2008de,
PengFei:2009ph,
Harris:2002md,Cao:2004ky,Cao:2004ap,Cao:2005pq,Campbell:2009ss,Heim:2009ku,
Campbell:2007ev,Dittmaier:2007th,Dittmaier:2009un,
Binoth:2009wk,
Melia:2010bm,
Bredenstein:2008zb,Bredenstein:2009aj,Bredenstein:2010rs,Bevilacqua:2009zn,Bevilacqua:2010ve,
Campbell:2000bg,Campbell:2002tg,Campbell:2003hd,Campbell:2005zv,Campbell:2006cu,
DelDuca:2001eu,DelDuca:2001fn,Campbell:2006xx,
Ellis:2009zw,KeithEllis:2009bu,Berger:2009zg,Berger:2009ep,Berger:2010vm,
Berger:2010zx,
Lazopoulos:2007ix,Binoth:2008kt,
Binoth:2009rv,
Jager:2006zc,Jager:2006cp,Bozzi:2007ur,Hankele:2007sb,Campanario:2008yg,Jager:2009xx
}.
It is desirable to have NLO calculations for $2 \rightarrow n$ processes in hadron-hadron
collisions with $n$ in the range of $n=3,...,7$.
However, the complexity of the calculation increases with the number of final state particles.
For any NLO calculation there are two parts to be calculated: the real and the virtual corrections.
Almost without exceptions all examples cited above use the 
dipole formalism \cite{Catani:1997vz,Dittmaier:1999mb,Phaf:2001gc,Catani:2002hc}
to subtract out the infrared divergences
from the real corrections. The subtracted real correction term is then integrable in four dimensions and can be
calculated numerically by Monte Carlo techniques.
By now there are several implementations for the automated construction of 
subtraction terms \cite{Weinzierl:2005dd,Gleisberg:2007md,Seymour:2008mu,Hasegawa:2009tx,Frederix:2008hu,Czakon:2009ss}.
The required Born amplitudes can be calculated efficiently with the help of 
recurrence relations\cite{Berends:1987me,Kosower:1989xy,Caravaglios:1995cd,Kanaki:2000ey,Moretti:2001zz,Dinsdale:2006sq,Duhr:2006iq}.
The calculation of the virtual corrections for QCD processes 
with many external legs has been considered to be a bottle neck for a long time.
The past years have witnessed significant progress in this direction.
The main lines of investigation focus on a perfection of the traditional 
Feynman graph approach \cite{Dittmaier:2003bc,Denner:2002ii,Denner:2005nn,Denner:2010tr,Giele:2004iy,vanHameren:2005ed,Ellis:2005zh,
delAguila:2004nf,Pittau:2004bc,Binoth:2002xh,Binoth:2005ff}
or are based on unitary methods \cite{Bern:1995cg,Berger:2008sj,Forde:2007mi,
Ossola:2006us,Ossola:2007ax,Ossola:2008xq,Mastrolia:2008jb,Draggiotis:2009yb,Garzelli:2009is,
Kilgore:2007qr,Anastasiou:2006jv,Anastasiou:2006gt,Ellis:2007br,Giele:2008ve,Ellis:2008ir}.

In this paper we would like to discuss a third and purely numerical approach. To this aim we extend the subtraction method
to the loop integration of the virtual corrections and we evaluate the subtracted virtual corrections numerically with
the help of a suitable chosen contour deformation.
The method is formulated in terms of amplitudes and does not rely on Feynman graphs. Therefore all ingredients can
be calculated efficiently using recurrence relations.
Purely numerical approaches have been discussed in the past
\cite{Soper:1998ye,Soper:1999xk,Soper:2001hu,Kramer:2002cd,Nagy:2003qn,Nagy:2006xy,Gong:2008ww,Passarino:2001wv,Ferroglia:2002mz,Anastasiou:2007qb}.
The literature focuses either on individual Feynman graphs and subtraction terms for individual graphs or on a contour deformation
for infrared and ultraviolet finite amplitudes, where no subtraction terms are needed.
Unfortunately the methods discussed in the literature for the subtraction terms on the one hand
and for the contour deformation on the other hand
are not compatible with each other and cannot be combined.
Furthermore it is not clear if the methods discussed so far in the literature are sufficiently efficient to be applied to 
multi-parton processes.
What is new in this paper is the development of compatible methods for the subtraction terms and the contour deformation and 
the combination of all relevant aspects into one formalism.
Inspired by recent work on the structure of infrared singularities of multi-loop amplitudes \cite{Becher:2009cu,Gardi:2009qi}
we found that the soft and collinear subtraction terms can be formulated at the level of amplitudes, without referring to individual
Feynman graphs \cite{Assadsolimani:2009cz}. 
This is a significant simplification and opens the door to an efficient implementation.
In addition we need subtraction terms for  the ultraviolet divergences. In this paper we present a set of ultraviolet subtraction terms
which have the expected form of local counterterms and which are particularly well suited for the numerical contour integration in the
sense that this set does not introduce additional singularities along the contour.
The second important ingredient of our method is an algorithm for the contour deformation. 
The subtraction terms eliminate only singularities, where the contour is pinched but leave singularities where a deformation into the complex
plane is possible. The contour deformation takes care of these remaining singularities.
It is highly non-trivial to find a general algorithm which avoids these singularities and which leads to stable Monte Carlo results.
We achieve this goal by first introducing Feynman parameters.
After Feynman parametrisation the contour deformation of the loop momentum is straightforward.
In order to avoid all singularities we have to deform the integration over the loop momentum and 
the integration over the Feynman parameters.
For an amplitude with a large number of external particles we take additional measures to improve
the efficiency of the numerical Monte Carlo integration.
Our method works for massless and massive particles.

This paper is organised as follows:
In section~\ref{sect:setup} we give an overview of the general ideas behind our approach.
In a shortened form these ideas have been presented in \cite{Assadsolimani:2010ka}.
In section~\ref{sect:subtraction} we provide a complete list of all subtraction terms.
The infrared subtraction terms have been given for the first time in \cite{Assadsolimani:2009cz}, 
we list them here again with a minor modification. The minor modification in the collinear subtraction terms
adapts the subtraction terms to the chosen method of contour deformation.
The ultraviolet subtraction terms are new.
In section~\ref{sect:contour_deformation} we discuss in detail the contour deformation.
Together with the subtraction terms this part constitutes the core of our method.
In section~\ref{sect:additional_remarks} we discuss a few points which might help to understand our method 
or which might help to avoid possible pitfalls.
Examples of a possible pitfall are diagrams like massless tadpoles, which give zero in an analytical calculation.
These diagrams have to be included in the numerical calculation in order not to spoil the local cancellation of
singularities.
Section~\ref{sect:examples} discusses checks and simple examples. 
NLO results on more complicated processes will be published in a separate publication.
Finally, section~\ref{sect:conclusions} contains a summary and our conclusions.
In an appendix we have documented certain technical aspects of our method.

\section{General setup}
\label{sect:setup}

In this section we give an overview of our method and define our notation.
In subsection~\ref{subsect:subtraction} we start from the subtraction method for real corrections and present the
extension to the virtual corrections.
The pole structure in the dimensional regularisation parameter 
of one-loop QCD amplitudes is very well understood and recalled in subsection~\ref{subsect:poles}.
Throughout this paper we work with colour ordered amplitudes. These are defined in subsection~\ref{subsect:colour_decomp}.
Subsection~\ref{subsect:kinematics} introduces the notation which we use for the kinematics.
Our method works at the level of amplitudes. These can be calculated efficiently with the help
of recurrence relations without relying on Feynman graphs. Recurrence relations are discussed 
in subsection~\ref{subsect:recurrence}.
For the construction of the subtraction terms we need to know from which integration region divergences arise.
This is reviewed in subsection~\ref{subsect:singular_regions}.

\subsection{The subtraction method}
\label{subsect:subtraction}

The starting point for the calculation of an infrared safe observable $O$ in
hadron-hadron collisions is the following formula:
\bq
\l O \r & = & \sum\limits_{a,b} \int dx_1 f_a(x_1) \int dx_2 f_b(x_2) 
             \frac{1}{2 K(\hat{s}) n_{\mathrm{spin}}(1) n_{\mathrm{spin}}(2) n_{\mathrm{colour}}(1) n_{\mathrm{colour}}(2)}
 \nonumber \\
 & &
             \sum\limits_n \int d\phi_{n}\left(p_1,p_2;p_3,...,p_{n+2}\right)
             O\left(p_1,...,p_{n+2}\right)
             \left| {\cal A}_{n+2} \right|^2.
\eq
In this equation we have written explicitly the sum over the flavours $a$ and $b$ of the two partons
in the initial state.
In addition there is a sum over the flavours of all final state particles, which is not shown 
explicitly.
The momenta of the two incoming particles are labelled $p_1$ and $p_2$, while $p_3$ to $p_{n+2}$ denote
the momenta of the final state particles.
$f_a(x)$ gives the probability of finding a parton $a$ with momentum fraction $x$ inside
the parent hadron $h$.
$2K(\hat{s})$ is the flux factor, for massless partons it is given by $2K(\hat{s})=2\hat{s}$.
The quantity $n_{\mathrm{spin}}(i)$ denotes the number of spin degrees of freedom of the parton $i$ and equals
two for quarks and gluons.
Correspondingly, $n_{\mathrm{colour}}(i)$ denotes the number of colour degrees of freedom of the parton $i$.
For quarks, this number equals three, while for gluons we have eight colour degrees of freedom.
The matrix element $| {\cal A}_{n+2} |^2$ is summed over all colours and spins.
Dividing by the appropriate number of degrees of freedom in the initial state
corresponds to an averaging.
$d\phi_n$ is the phase space measure for $n$ final state particles, including (if appropriate) the identical particle factors.
The matrix element $| {\cal A}_{n+2} |^2$ is calculated perturbatively.

The contributions at leading and next-to-leading order are written as
\bq
\l O \r^{LO} & = & 
 \int\limits_n O_n d\sigma^B,
 \nonumber \\ 
\l O \r^{NLO} & = & 
 \int\limits_{n+1} O_{n+1} d\sigma^R + \int\limits_n O_n d\sigma^V 
 + \int\limits_n O_n d\sigma^C.
\eq
Here a rather condensed notation is used. $d\sigma^B$ denotes the Born
contribution,
whose matrix elements are given by the square of the Born amplitudes with $(n+2)$ partons
$| {\cal A}^{(0)}_{n+2} |^2$.
Similar, $d\sigma^R$ denotes the real emission contribution,
whose matrix elements are given by the square of the Born amplitudes with $(n+3)$ partons
$| {\cal A}^{(0)}_{n+3} |^2$.
$d\sigma^V$ gives the virtual contribution, whose matrix elements are given by the interference term
of the one-loop amplitude ${\cal A}^{(1)}_{n+2}$, with $(n+2)$ partons, with the corresponding
Born amplitude ${\cal A}^{(0)}_{n+2}$.
$d\sigma^C$ denotes a collinear subtraction term, which subtracts the initial state collinear
singularities.
Taken separately, the individual contributions at next-to-leading order 
are divergent and only their sum is finite.
In order to render the individual contributions finite, such that the phase space integrations
can be performed by Monte Carlo methods, one adds and subtracts a suitable chosen piece
\cite{Catani:1997vz,Dittmaier:1999mb,Phaf:2001gc,Catani:2002hc}:
\bq
\l O \r^{NLO} & = & 
 \int\limits_{n+1} \left( O_{n+1} d\sigma^R - O_n d\sigma^A \right)
 + \int\limits_n \left( O_n d\sigma^V + O_n d\sigma^C + O_n \int\limits_1 d\sigma^A \right).
\eq
The term $(O_{n+1}  d\sigma^R - O_n d\sigma^A)$ in the first bracket is by construction integrable over the
$(n+1)$-particle phase space and can be evaluated numerically.
The subtraction term can be integrated analytically over the unresolved one-particle phase space.
Due to this integration all spin-correlations average out, but colour correlations still remain.
In a compact notation the result of this integration is often written as
\bq
 d\sigma^C + \int\limits_1 d\sigma^A  
 & = & {\bf I} \otimes d\sigma^B + {\bf K} \otimes d\sigma^B + {\bf P} \otimes d\sigma^B.
\eq
The notation $\otimes$ indicates that colour correlations due to the colour charge operators
${\bf T}_i$ still remain.
The action of a colour charge operator ${\bf T}_i$ for a quark, gluon and antiquark in the final state is given by
\bq
\label{colour_charge_operator_final}
\mbox{quark :} & & 
 {\cal A}^\ast\left(  ... q_i ... \right) \left( T_{ij}^a \right) {\cal A}\left(  ... q_j ... \right), \nonumber \\
\mbox{gluon :} & & 
 {\cal A}^\ast\left(  ... g^c ... \right) \left( i f^{cab} \right) {\cal A}\left(  ... g^b ... \right), \nonumber \\
\mbox{antiquark :} & & 
 {\cal A}^\ast\left(  ... \bar{q}_i ... \right) \left( - T_{ji}^a \right) {\cal A}\left(  ... \bar{q}_j ... \right).
\eq
The corresponding formulae for colour charge operators for a quark, gluon or antiquark in the initial state are
\bq
\label{colour_charge_operator_initial}
\mbox{quark :} & & 
 {\cal A}^\ast\left(  ... \bar{q}_i ... \right) \left( - T_{ji}^a \right) {\cal A}\left(  ... \bar{q}_j ... \right), \nonumber \\
\mbox{gluon :} & & 
 {\cal A}^\ast\left(  ... g^c ... \right) \left( i f^{cab} \right) {\cal A}\left(  ... g^b ... \right), \nonumber \\
\mbox{antiquark :} & & 
 {\cal A}^\ast\left(  ... q_i ... \right) \left( T_{ij}^a \right) {\cal A}\left(  ... q_j ... \right). 
\eq
In the amplitude an incoming quark is denoted as an outgoing antiquark and vice versa.
The terms with the insertion operators ${\bf K}$ and ${\bf P}$ do not have any poles in the dimensional
regularisation parameter and pose no problem for a numerical evaluation.
The term ${\bf I} \otimes d\sigma^B$ lives on the phase space of the $n$-parton configuration and has the appropriate
singularity structure to cancel the infrared divergences coming from the one-loop amplitude.
Therefore $d\sigma^V + {\bf I} \otimes d\sigma^B$ is infrared finite.
We emphasise that this cancellation occurs after the loop integration 
has been performed analytically in $D$ dimensions.
$d\sigma^V$ is given by
\bq
 d\sigma^V & = & 2 \;\mbox{Re}\; \left(\left.{\cal A}^{(0)}\right.^\ast {\cal A}^{(1)} \right){\cal O}_n d\phi_n.
\eq
${\cal A}^{(1)}$ denotes the renormalised one-loop amplitude. It is related to the bare amplitude by
\bq
\label{eq_one_loop}
 {\cal A}^{(1)} & = & {\cal A}^{(1)}_{\mathrm{bare}} + {\cal A}^{(1)}_{\mathrm{CT}}.
\eq
${\cal A}^{(1)}_{\mathrm{CT}}$ denotes the ultraviolet counterterm from renormalisation.
The bare one-loop amplitude involves the loop integration
\bq
\label{integrand_one_loop}
{\cal A}^{(1)}_{\mathrm{bare}} & = & \int \frac{d^Dk}{(2\pi)^D} {\cal G}^{(1)}_{\mathrm{bare}},
\eq
where ${\cal G}^{(1)}_{\mathrm{bare}}$ denotes the integrand of the bare one-loop amplitude.
In this paper we extend the subtraction method to the integration over the virtual particles circulating in the loop.
To this aim we rewrite eq.~(\ref{eq_one_loop}) as
\bq
\label{basic_subtraction_loop}
 {\cal A}_{\mathrm{bare}}^{(1)} + {\cal A}_{\mathrm{CT}}^{(1)} 
 & = & 
 \left( {\cal A}_{\mathrm{bare}}^{(1)} - {\cal A}_{\mathrm{soft}}^{(1)} - {\cal A}_{\mathrm{coll}}^{(1)} - {\cal A}_{\mathrm{UV}}^{(1)} \right)
 + \left( {\cal A}_{\mathrm{CT}}^{(1)}  
 + {\cal A}_{\mathrm{soft}}^{(1)} + {\cal A}_{\mathrm{coll}}^{(1)} + {\cal A}_{\mathrm{UV}}^{(1)} \right).
\eq
The subtraction terms ${\cal A}_{\mathrm{soft}}^{(1)}$, ${\cal A}_{\mathrm{coll}}^{(1)}$ and ${\cal A}_{\mathrm{UV}}^{(1)}$
are chosen such that they match locally the singular behaviour of the integrand of ${\cal A}_{\mathrm{bare}}^{(1)}$ in $D$ dimensions.
The first bracket in eq.~(\ref{basic_subtraction_loop})
can therefore be integrated numerically in four dimensions.
The term ${\cal A}_{\mathrm{soft}}^{(1)}$ approximates the soft singularities, 
${\cal A}_{\mathrm{coll}}^{(1)}$ approximates
the collinear singularities and 
the term ${\cal A}_{\mathrm{UV}}^{(1)}$ approximates the ultraviolet singularities.
These subtraction terms have a local form similar to eq.~(\ref{integrand_one_loop}):
\bq
{\cal A}^{(1)}_{\mathrm{soft}} = \int \frac{d^Dk}{(2\pi)^D} {\cal G}^{(1)}_{\mathrm{soft}},
\;\;\;\;\;\;
{\cal A}^{(1)}_{\mathrm{coll}} = \int \frac{d^Dk}{(2\pi)^D} {\cal G}^{(1)}_{\mathrm{coll}},
\;\;\;\;\;\;
{\cal A}^{(1)}_{\mathrm{UV}} = \int \frac{d^Dk}{(2\pi)^D} {\cal G}^{(1)}_{\mathrm{UV}}.
\eq
The contribution from the terms in the first bracket of eq.~(\ref{basic_subtraction_loop})
can be written as
\bq
\label{eq_monte_carlo_integration}
\lefteqn{
 \int
 2 \;\mbox{Re}\; \left[\left.{\cal A}^{(0)}\right.^\ast 
   \left( {\cal A}_{\mathrm{bare}}^{(1)} - {\cal A}_{\mathrm{soft}}^{(1)} - {\cal A}_{\mathrm{coll}}^{(1)} - {\cal A}_{\mathrm{UV}}^{(1)} \right) 
                \right]{\cal O}_n d\phi_n
 = } & &
 \nonumber \\
 & & 
 \int d\phi_n \int \frac{d^4k}{(2\pi)^4} 
 2 \;\mbox{Re}\; 
 \left[\left.{\cal A}^{(0)}\right.^\ast 
   \left( {\cal G}_{\mathrm{bare}}^{(1)} - {\cal G}_{\mathrm{soft}}^{(1)} 
        - {\cal G}_{\mathrm{coll}}^{(1)} - {\cal G}_{\mathrm{UV}}^{(1)} \right) 
                \right]
 {\cal O}_n
 + {\cal O}\left(\eps\right).
\eq
The integral on the right-hand side is finite.
It is one of the key ingredients of the method proposed here, that this rather complicated and
process-dependent integral can be performed numerically with Monte Carlo techniques.
We recall that the error of a Monte Carlo integration depends on the variance of the integrand and scales
with the number of integrand evaluations $N$ like $1/\sqrt{N}$.
It is important to note that the error does not depend on the dimension of the integration region.
Eq.~(\ref{eq_monte_carlo_integration}) gives for an observable ${\cal O}$ a contribution to the next-to-leading
order prediction.
The right-hand side corresponds to a $(3n)$-dimensional integral.
(The phase-space integral is $(3n-4)$-dimensional, the loop integral $4$-dimensional.)
In practise this $(3n)$-dimensional integral is done with a single Monte Carlo integration.
There is no need to evaluate for a given phase-space point the inner four-dimensional loop integral
by a separate Monte Carlo integration.
This is essential for the efficiency of the method.

The building blocks of the subtraction terms are process-independent. When adding them back, we integrate analytically
over the loop momentum $k$. The result can be written as
\bq
 2 \;\mbox{Re}\; \left[\left.{\cal A}^{(0)}\right.^\ast 
   \left( {\cal A}_{\mathrm{CT}}^{(1)} + {\cal A}_{\mathrm{soft}}^{(1)} + {\cal A}_{\mathrm{coll}}^{(1)} + {\cal A}_{\mathrm{UV}}^{(1)}\right) 
                \right]{\cal O}_n d\phi_n
 & = & 
 {\bf L} \otimes d\sigma^B.
\eq
The insertion operator ${\bf L}$ contains the explicit poles in the dimensional regularisation parameter related
to the infrared singularities of the one-loop amplitude.
These poles cancel when combined with the insertion operator ${\bf I}$:
\bq
 \left( {\bf I} + {\bf L} \right) \otimes d\sigma^B & = &
 \mbox{finite}.
\eq
The operator ${\bf L}$ contains, as does the operator ${\bf I}$, colour correlations due to soft gluons.

\subsection{Explicit poles in the dimensional regularisation parameter}
\label{subsect:poles}

The explicit poles in the dimensional regularisation parameter $\eps$ of the individual pieces are well known.
These poles are either of ultraviolet or infrared origin.
Let us consider an amplitude with $n_q$ external quarks, $n_{\bar q}$ external anti-quarks and $n_g$ external
gluons in massless QCD. We set $n=n_q+n_{\bar q}+n_g$. Obviously we have $n_q=n_{\bar q}$.
After an analytical integration the poles in the dimensional regularisation parameter $\eps$
of a bare massless one-loop QCD amplitude are given by
\bq
\label{poles_one_loop}
 {\cal A}^{(1)}_{\mathrm{bare}} & = &
 \frac{\alpha_s}{4\pi}
 \frac{e^{\eps \gamma_E}}{\Gamma(1-\eps)}
 \left[ 
 \frac{(n-2)}{2} \frac{\beta_0}{\eps}
 +
 \sum\limits_{i} \sum\limits_{j\neq i}
 {\bf T}_i {\bf T}_j 
 \left( \frac{1}{\eps^2} + \frac{\gamma_i}{{\bf T}_i^2} \frac{1}{\eps} \right)
 \left( \frac{-2p_ip_j}{\mu^2} \right)^{-\eps}
 \right]
 {\cal A}_{n}^{(0)}
 + {\cal O}(\eps^0).
 \nonumber \\
\eq
$\beta_0$ is the first coefficient of the QCD $\beta$-function and given by
\bq
\beta_0 = \frac{11}{3} C_A - \frac{4}{3} T_R N_f.
\eq
The constants $\gamma_i$ are given by
\bq
 \gamma_q = \gamma_{\bar q} = \frac{3}{2} C_F, 
 & &
 \gamma_g = \frac{1}{2} \beta_0.
\eq
The colour factors are as usual
\bq
C_A = N_c, \;\;\; C_F = \frac{N_c^2-1}{2N_c}, \;\;\; T_R = \frac{1}{2}.
\eq
The poles of eq.~(\ref{poles_one_loop}) are cancelled by ${\cal A}_{\mathrm{CT}}^{(1)}$
and the infrared poles obtained from integrating the real emission contribution over
the unresolved phase space.
The leading order QCD amplitude with $n$ partons is proportional to $\alpha_s^{(n-2)/2}$.
For a massless QCD amplitude with $n$ partons the ultraviolet counterterm is given by
\bq
 {\cal A}_{\mathrm{CT}}^{(1)} & = &
 - \frac{\alpha_s}{4\pi} 
 \frac{(n-2)}{2} \frac{\beta_0}{\eps} {\cal A}_{n}^{(0)},
\eq
The insertion operator ${\bf I}$ contains the infrared poles obtained from integrating the real emission
contribution over the unresolved phase space.
The insertion operator ${\bf I}$ is given in massless QCD by
\bq
 {\bf I} 
 & = &
 \frac{\alpha_s}{2\pi} 
 \frac{e^{\eps \gamma_E}}{\Gamma(1-\eps)}
 \sum\limits_{i} \sum\limits_{j \neq i}
 {\bf T}_i {\bf T}_j 
 \left( - \frac{1}{\eps^2} + \frac{\pi^2}{3} - \frac{\gamma_i}{{\bf T}_i^2} \frac{1}{\eps} 
 - \frac{\gamma_i}{{\bf T}_i^2} - \frac{K_i}{{\bf T}_i^2} \right)
 \left( \frac{\left|2p_ip_j\right|}{\mu^2} \right)^{-\eps}
 + {\cal O}(\eps).
\eq
with
\bq
 K_q = K_{\bar q} = \left( \frac{7}{2} - \frac{\pi^2}{6} \right) C_F,
 & &
 K_g = \left( \frac{67}{18} - \frac{\pi^2}{6} \right) C_A - \frac{10}{9} T_R N_f.
\eq

\subsection{Colour decomposition}
\label{subsect:colour_decomp}

Amplitudes in QCD may be decomposed into group-theoretical factors (carrying the colour structures)
multiplied by kinematic functions called partial amplitudes
\cite{Cvitanovic:1980bu,Berends:1987cv,Mangano:1987xk,Kosower:1987ic,Bern:1990ux}. 
These partial amplitudes do not contain any colour information and are gauge invariant objects. 
The colour decomposition is obtained by replacing the structure constants $f^{abc}$
by
\bq
 i f^{abc} & = & 2 \left[ \mbox{Tr}\left(T^a T^b T^c\right) - \mbox{Tr}\left(T^b T^a T^c\right) \right],
\eq
which follows from $ \left[ T^a, T^b \right] = i f^{abc} T^c$.
In this paper we use the normalisation
\bq
 \mbox{Tr}\;T^a T^b & = & \frac{1}{2} \delta^{a b}
\eq
for the colour matrices.
The resulting traces and strings of colour matrices can be further simplified with
the help of the Fierz identity :
\bq
 T^a_{ij} T^a_{kl} & = &  \frac{1}{2} \left( \delta_{il} \delta_{jk}
                         - \frac{1}{N} \delta_{ij} \delta_{kl} \right).
\eq
There are several possible choices for a basis in colour space. A convenient choice is the
colour-flow basis \cite{'tHooft:1973jz,Maltoni:2002mq,Weinzierl:2005dd}. 
This choice is obtained by attaching a factor
\bq
 \sqrt{2} T^a_{ij}
\eq
to each external gluon.
As an example we consider the colour decomposition of the pure gluon amplitude with $n$ external gluons.
The colour decomposition of the tree level amplitude may be written in 
the form
\bq
\label{colour_decomp_pure_gluon}
{\cal A}_{n}^{(0)}(g_1,g_2,...,g_n) & = & 
 \left(\frac{g}{\sqrt{2}}\right)^{n-2} \sum\limits_{\sigma \in S_{n}/Z_{n}} 
 \delta_{i_{\sigma_1} j_{\sigma_2}} \delta_{i_{\sigma_2} j_{\sigma_3}} 
 ... \delta_{i_{\sigma_n} j_{\sigma_1}}  
 A_{n}^{(0)}\left( g_{\sigma_1}, ..., g_{\sigma_n} \right),
\eq
where the sum is over all non-cyclic permutations of the external gluon legs.
The quantities $A_n(g_{\sigma_1},...,g_{\sigma_n})$, called the partial amplitudes, contain the 
kinematical information.
They are colour ordered, e.g. only diagrams with a particular cyclic ordering of the gluons contribute.
For the convenience of the reader we have listed the colour ordered Feynman rules
in appendix \ref{sect:feynman_rules}.
Similar decompositions exist for all other Born QCD amplitudes.
As a further example we give the colour decomposition 
for a tree amplitude with a pair of quarks:
\bq
\label{colour_decomp_qqbar_gluon}
{\cal A}_{n}^{(0)}(q,g_1,...,g_{n-2},\bar{q}) 
& = & \left(\frac{g}{\sqrt{2}}\right)^{n-2} 
 \sum\limits_{S_{n-2}} 
 \delta_{i_q j_{\sigma_1}} \delta_{i_{\sigma_1} j_{\sigma_2}} 
 ... \delta_{i_{\sigma_{n-2}} j_{\bar{q}}} 
A_{n}^{(0)}(q,g_{\sigma_1},...,g_{\sigma_{n-2}},\bar{q}),
\;\;\;
\eq
where the sum is over all permutations of the external gluon legs. 
In squaring these amplitudes a colour projector
\bq
 \delta_{\bar{i} i} \delta_{j \bar{j}} - \frac{1}{N} \delta_{\bar{i} \bar{j} } \delta_{j i}
\eq
has to be applied to each gluon.
At the one-loop level the colour decomposition is slightly more involved.
Let us start with an example: The colour decomposition of the one-loop gluon amplitude into
partial amplitudes reads \cite{Bern:1990ux}:
\bq
{\cal A}^{(1)}_{n}(g_1,g_2,...,g_n) 
 & = & 
 \left(\frac{g}{\sqrt{2}}\right)^n 
 \left\{
 \sum\limits_{\sigma \in S_{n}/Z_{n}} 
 N \; \delta_{i_{\sigma_1} j_{\sigma_2}} \delta_{i_{\sigma_2} j_{\sigma_3}} 
 ... \delta_{i_{\sigma_n} j_{\sigma_1}}
 A^{(1)}_{n,0}\left( g_{\sigma_1}, ..., g_{\sigma_n} \right)
 \right. 
 \nonumber \\
 & &
 \left.
 + 
 \sum\limits_{m} \sum\limits_{\sigma \in S_{n}/(Z_{m} \times Z_{n-m})} 
 \left( \delta_{i_{\sigma_1} j_{\sigma_2}} ... \delta_{i_{\sigma_m} j_{\sigma_1}} \right)
 \left( \delta_{i_{\sigma_{m+1}} j_{\sigma_{m+2}}} ... \delta_{i_{\sigma_n} j_{\sigma_{m+1}}} \right)
 \right. \nonumber \\
 & & \left.
 A^{(1)}_{n,m}\left( g_{\sigma_1}, ..., g_{\sigma_m}; g_{\sigma_{m+1}}, ..., g_{\sigma_n} \right)
 \right\}.
\eq
The one-loop partial amplitudes $A_{n,m}^{(1)}$ are again gauge invariant.
The subleading partial amplitudes are related to the leading ones by
\bq
 A^{(1)}_{n,m}\left( g_1, ..., g_m; g_{m+1}, ..., g_n \right)
 & = & (-1)^m
 \sum\limits_{\sigma \in \{m,...,1\} \Sha \{m+1,...,n\}}  
 A^{(1)}_{n,0}\left( g_{\sigma_1}, ..., g_{\sigma_n} \right),
\eq
where the sum is over all shuffles of the set $\{m,...,1\}$ with the set $\{m+1,...,n\}$.
It is therefore sufficient to focus on the leading partial amplitudes $A_{n,0}^{(1)}$.
The leading partial amplitudes can be decomposed further into smaller objects called
primitive amplitudes. The primitive amplitudes are separately gauge invariant. 
For the example of the one-loop gluon amplitude we have the decomposition of the leading partial
amplitude into two primitive amplitudes, which are characterised by the particle content 
circulating in the loop:
\bq
 A^{(1)}_{n,0} & = & A^{(1)}_{n,0,\mathrm{lc}} + \frac{N_f}{N} A^{(1)}_{n,0,\mathrm{nf}}.
\eq
For the primitive amplitude $A^{(1)}_{n,0,\mathrm{lc}}$ there is either a gluon or a ghost circulating in the
loop, while for the primitive amplitude $A^{(1)}_{n,0,\mathrm{nf}}$ there is a quark circulating in the loop.

Similar decompositions exist for all other one-loop QCD amplitudes.
If the external legs of a one-loop QCD amplitude involve a quark-antiquark pair
we can distinguish the two cases where the loop lies to right or to the left of the fermion line
if we follow the fermion line in the direction of the flow of the fermion number.
We call a fermion line ``left-moving'' if, following the
arrow of the fermion line, the loop is to the right.
Analogously, we call a fermion line ``right-moving'' if, following the
arrow of the fermion line, the loop is to the left.
\begin{figure}
\begin{center}
\includegraphics[bb= 140 610 420 715]{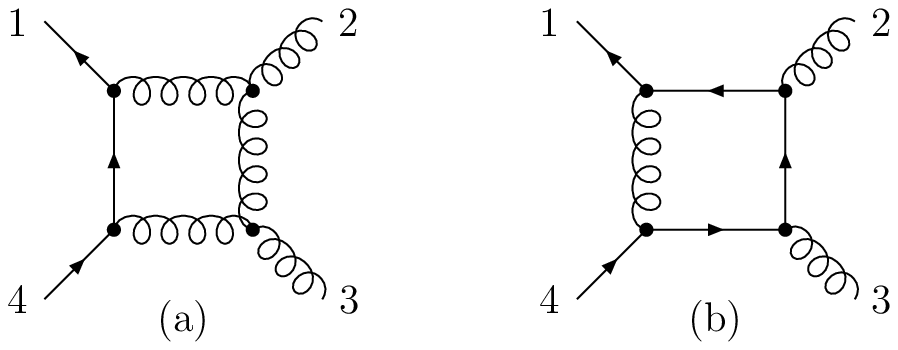}
\caption{\label{figure_example_qqgg}
Examples of diagrams regarding left-moving and right-moving primitive amplitudes: 
Diagrams (a) contributes to the left-moving primitive amplitude,
while diagram (b) contributes to the right-moving amplitude.
}
\end{center}
\end{figure}
Examples of diagrams regarding these types of primitive amplitudes are shown in fig.~(\ref{figure_example_qqgg}).
It turns out that in the decomposition into primitive amplitudes a specific quark line 
is, in all diagrams which contribute to a specific primitive amplitude, either always
left-moving or always right-moving \cite{Bern:1994fz}.
Therefore in the presence of external fermions primitive amplitudes are in addition characterised
by the routing of the fermion lines through the amplitude.
In summary we can always write 
a full one-loop QCD amplitude as a linear combination of primitive amplitudes:
\bq
 {\cal A}^{(1)}_n & = & \sum\limits_{j,k} C_j A^{(1)}_{n,j,k}.
\eq
The colour structures are denoted by $C_j$, while
the primitive amplitudes are denoted by $A^{(1)}_{n,j,k}$.
In the colour-flow basis the colour structures are linear combinations of monomials in 
Kronecker $\delta_{ij}$'s.
In order to construct a primitive one-loop amplitude one starts 
to draw all possible planar one-loop diagrams with
a fixed cyclic ordering of the external legs, subject to the constraint that each fermion line is either
only left-moving or only right-moving. 
This set of diagrams is then further divided into the subset of diagrams with a closed fermion
line and the ones without.
The set of diagrams with a closed fermion line form a separate primitive amplitude.

In the following we will work exclusively with primitive amplitudes.
In order to simplify the notation we will drop the subscripts and 
simply write
\bq
 A^{(1)}
\eq
for a primitive one-loop amplitude.
The full one-loop amplitude is just the sum of several primitive amplitudes multiplied 
by colour structures.
We stress a few important properties of primitive one-loop amplitudes:
\begin{enumerate}
\item Primitive amplitudes are gauge invariant.
\item Primitive amplitudes have a fixed cyclic ordering of the external legs 
and a definite routing of the external fermion lines.
\item The flavour of each propagator in the loop is unique: Either it is a quark propagator
or a gluon/ghost propagator.
\end{enumerate}
Our method exploits these facts.
The first property of gauge invariance is crucial for the proof of the method.
The second point ensures that there are at maximum $n$ different loop propagators in the problem, 
where $n$ is the number of external legs.

\subsection{Kinematics}
\label{subsect:kinematics}

We introduce some notation which will be used throughout this paper.
Let $A^{(1)}_{\mathrm{bare}}$ be a primitive amplitude with $n$ external legs.
Since the cyclic ordering of the external partons is fixed, there are only $n$ different propagators occurring
in the loop integral.
\begin{figure}
\begin{center}
\includegraphics[bb= 180 510 440 720]{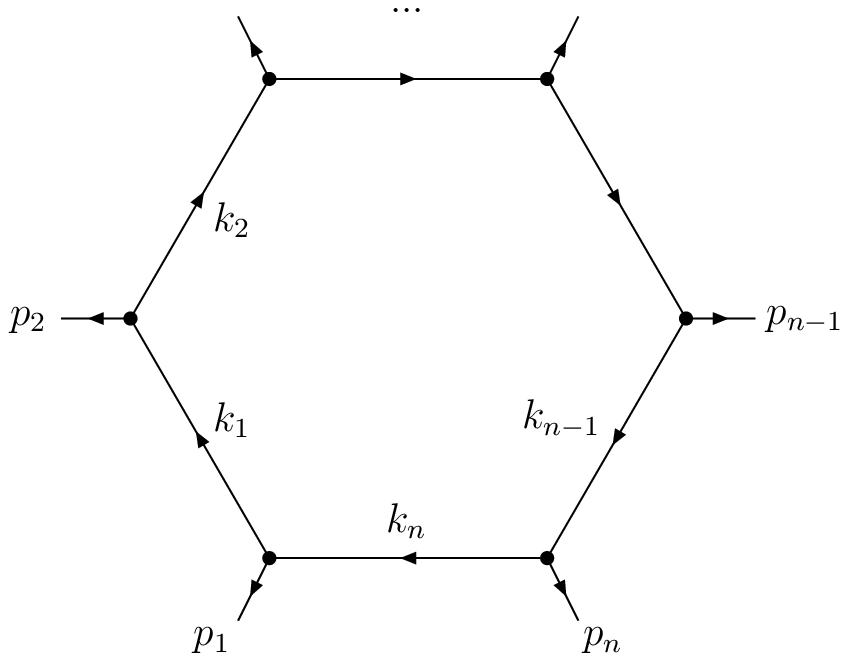}
\caption{\label{figure_momenta_one_loop}
The labelling of the momenta for a primitive one-loop amplitude. The arrows denote the momentum flow.}
\end{center}
\end{figure}
With the notation as in fig.~(\ref{figure_momenta_one_loop}) we define
\bq
 k_j & = & k - q_j,
 \;\;\; q_j = \sum\limits_{l=1}^j p_l.
\eq
We can write the bare primitive one-loop amplitude as
\bq
 A^{(1)}_{\mathrm{bare}} & = & \int \frac{d^Dk}{(2\pi)^D} G^{(1)}_{\mathrm{bare}},
 \;\;\;\;\;
 G^{(1)}_{\mathrm{bare}} = P(k) \prod\limits_{j=1}^n \frac{1}{k_j^2 - m_j^2 + i \delta}.
\eq
$P(k)$ is a polynomial in the loop momentum $k$.
The $+i \delta$-prescription in the propagators indicates into which direction the poles of
the propagators should be avoided.
We denote by $I_g$ the set of indices $j$, for which the 
propagator $j$ in the loop corresponds to a gluon.
If we take the subset of diagrams which have the gluon loop propagator $j$ and if we remove from each diagram of this subset 
the loop propagator $j$ we obtain a set of tree diagrams.
After removing multiple copies of identical diagrams this set 
forms a Born partial amplitude which we denote by $A^{(0)}_{j}$.

We introduce two matrices, which depend on the external momenta (and the internal masses), 
but not on the loop momentum $k$.
The kinematical matrix $S$ is a $n \times n$-matrix and is defined by
\bq
\label{def_S}
 S_{ij} & = & \left( q_i - q_j \right)^2 - m_i^2 - m_j^2.
\eq
The Gram matrix $G_{ij}^{(a)}$ is a $(n-1) \times (n-1)$-matrix and is defined by
\bq
\label{def_Gram}
 G_{ij}^{(a)} & = & 2 \left( q_i - q_a \right) \cdot \left( q_j - q_a \right),
\eq
where the indices $i$ and $j$ take the values $i,j \neq a$.

\subsection{Recurrence relations}
\label{subsect:recurrence}

Although we use in the proof of our method the fact that a primitive amplitude can be written
as a sum of Feynman diagrams, it is important to note that in the final formulae, which enter the
numerical Monte Carlo program, only amplitudes occur -- and not individual Feynman diagrams.
There are several possibilities how these amplitudes can be calculated.
A particular efficient method is based on recurrence relations.
We first review the recursive method for tree-level partial amplitudes and discuss afterwards the
necessary modifications for the computation of the integrand of a one-loop primitive amplitude.

We start with the computation of tree-level partial amplitudes.
Berends-Giele type recurrence relations \cite{Berends:1987me}
build tree-level partial amplitudes from smaller building blocks, usually
called colour ordered off-shell currents.
Off-shell currents are objects with $n$ on-shell legs and one additional leg off-shell.
Momentum conservation is satisfied. It should be noted that
off-shell currents are not gauge invariant objects.
Recurrence relations relate off-shell currents with $n$ legs 
to off-shell currents with fewer legs.
As an example we discuss the pure gluon current, which can be used to calculate the pure gluon amplitude.
The recursion starts with $n=1$:
\bq
J^\mu(p^{\lambda}) & = & \eps^\mu_{\lambda}(p).
\eq
$\eps^\mu_\lambda$ is the polarisation vector of the gluon corresponding to the polarisation $\lambda$.
The recursive relation states that in the pure gluon off-shell current
a gluon couples to other gluons only via the three- or four-gluon
vertices :
\bq
\label{Berends_Giele_recursion}
 J^\mu(p_1^{\lambda_1},...,p_n^{\lambda_n}) & = & 
 \frac{-i}{P^2_{1,n}} 
 \left[ 
        \sum\limits_{j=1}^{n-1} V_3^{\mu\nu\rho}(-P_{1,n},P_{1,j},P_{j+1,n})
                                J_\nu(p_1^{\lambda_1},...,p_j^{\lambda_j}) J_\rho(p_{j+1}^{\lambda_{j+1}},...,p_n^{\lambda_n}) 
 \right. \nonumber \\
 & & \left. 
        + \sum\limits_{j=1}^{n-2} \sum\limits_{l=j+1}^{n-1} V_4^{\mu\nu\rho\sigma} 
            J_\nu(p_1^{\lambda_1},...,p_j^{\lambda_j}) J_\rho(p_{j+1}^{\lambda_{j+1}},...,p_l^{\lambda_l}) J_\sigma(p_{l+1}^{\lambda_{l+1}},...,p_n^{\lambda_n}) 
     \right], 
\;\;\;\;\;
\eq
where
\bq
P_{i,j} & = & p_i + p_{i+1} + ... + p_j 
\eq
\begin{figure}
\begin{center}
\includegraphics[bb=115 525 590 660]{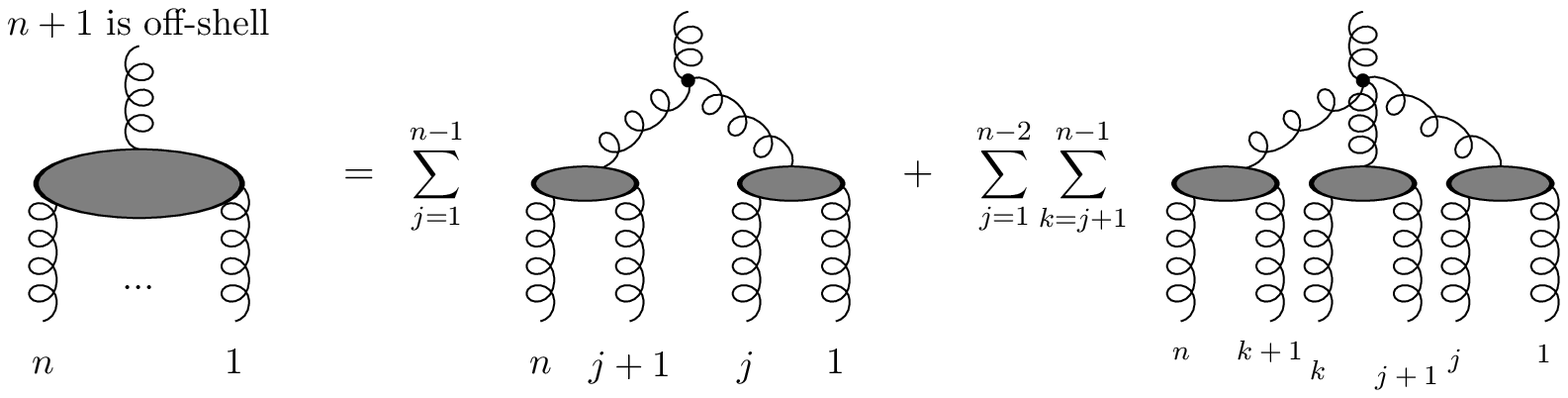}
\end{center}
\caption{\label{fig:recurrencerelation} The recurrence relation for the gluon current.
In an off-shell current particle $n+1$ is kept off-shell.
This allows to express an off-shell current with $n$ on-shell legs in terms of currents with fewer legs.}
\end{figure}
and $V_3$ and $V_4$ are the colour ordered three-gluon and four-gluon vertices
\bq
\label{Feynman_rules}
 V_3^{\mu\nu\rho}(p_1,p_2,p_3) 
 & = & 
 i \left[
          g^{\mu\nu} \left( p_2^\rho - p_1^\rho \right)
        + g^{\nu\rho} \left( p_3^\mu - p_2^\mu \right)
        + g^{\rho\mu} \left( p_1^\nu - p_3^\nu \right)
   \right],
 \nonumber \\
 V_4^{\mu\nu\rho\sigma} & = & i \left( 2 g^{\mu\rho} g^{\nu\sigma} - g^{\mu\nu} g^{\rho\sigma} -g^{\mu\sigma} g^{\nu\rho} \right).
\eq
The recurrence relation is shown pictorially in fig.~(\ref{fig:recurrencerelation}).
The gluon current $J_\mu$ is conserved:
\bq
\left( \sum\limits_{i=1}^n p_i^\mu \right) J_\mu & = & 0.
\eq
From an off-shell current one easily recovers the on-shell amplitude by 
removing the extra propagator,
taking the leg $(n+1)$ on-shell
and contracting with the appropriate polarisation vector.
Similar recurrence relations can be written down for the quark and antiquark currents, as well as the gluon
currents in full QCD.
The guiding principle is to follow the off-shell leg into the ``blob'', representing the sum of all diagrams,
and to sum on the r.h.s of the
recurrence relation over all vertices involving this off-shell leg and off-shell currents with less external
legs.

\begin{figure}
\begin{center}
\includegraphics[bb=125 410 545 645]{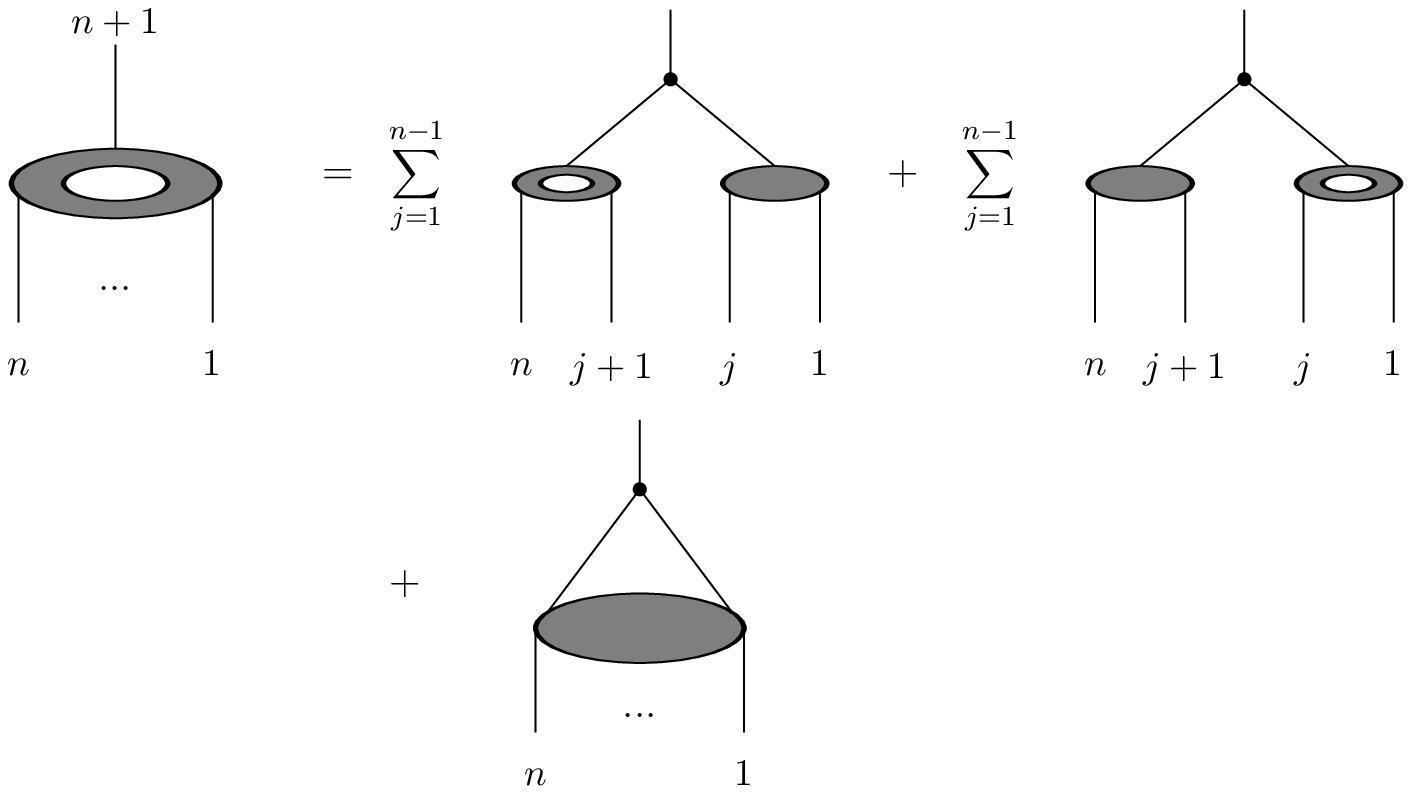}
\end{center}
\caption{\label{fig:onelooprecurrencerelation} The recurrence relation for the one-loop current
in a toy model with a single field and a single three-valent vertex.
The one loop currents are represented by an oval with a hole, tree-level currents are
represented by an oval without a hole.
}
\end{figure}

There are only a few modifications needed to compute 
the integrand of a one-loop primitive amplitude.
We can divide all vertices into two classes: either a vertex is directly connected to the loop or it is not.
Again we can follow the off-shell leg into the ``blob''. 
If the first vertex which we encounter belongs to the second class, then there is attached to this vertex
a one-loop off-shell current with fewer legs.
The other objects attached to this vertex are one or more tree-level off-shell currents.
If on the other hand the first vertex belongs to the first class, then two edges of the vertex are connected
to loop propagators. The object connected to these two edges is a tree-level off-shell current with two legs
off-shell.
It can be computed with methods similar to the computation for tree-level off-shell currents with one leg off-shell.
We illustrate this principle in fig.~(\ref{fig:onelooprecurrencerelation}) for the case of a toy theory with a single field and a single three-valent
vertex.
\begin{figure}
\begin{center}
\includegraphics[bb=115 410 540 645]{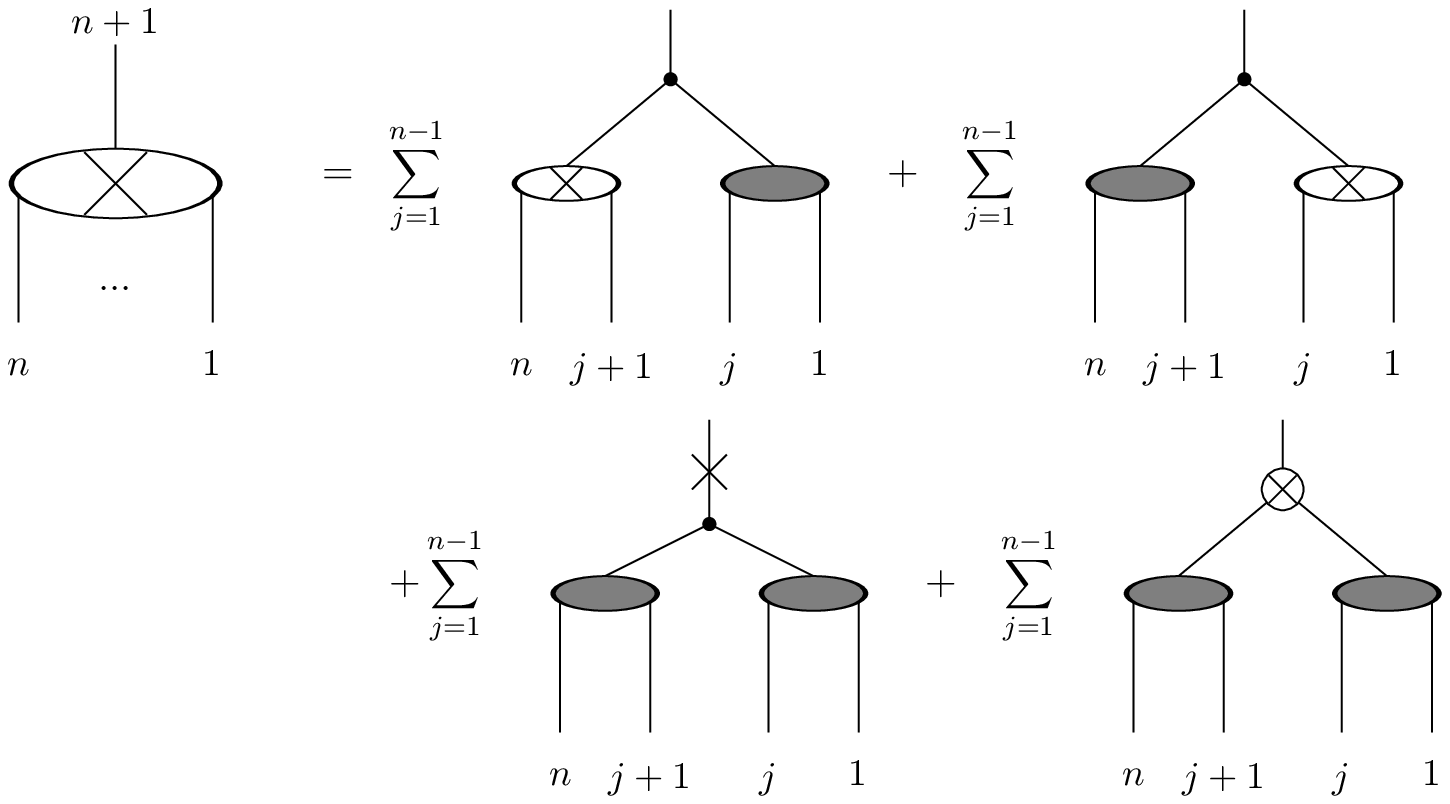}
\end{center}
\caption{\label{fig:UVrecurrencerelation} The recurrence relation for the ultraviolet subtraction terms
in a toy model with a single field and a single three-valent vertex.
Objects with a cross represent ultraviolet subtraction terms.
}
\end{figure}
Within our method we also need an ultraviolet subtraction term. The complete ultraviolet subtraction term
has a structure similar to ordinary ultraviolet counterterms: We can represent the ultraviolet subtraction term
as a sum over diagrams, where each diagram has a tree-structure with exactly one propagator or vertex replaced by a
basic ultraviolet subtraction term.
The complete ultraviolet subtraction term can again be calculated recursively.
As example we consider again the toy model with a single field and a single three-valent vertex.
The recursion relation for the ultraviolet subtraction term is shown in fig.~(\ref{fig:UVrecurrencerelation}).

\subsection{Singular regions of the integrand of one-loop amplitudes}
\label{subsect:singular_regions}

In order to construct the subtraction terms we have to know
in which regions of the integration domain for the loop momentum divergences arise.
There are three types of singularities in one-loop amplitudes, which have to be 
approximated by suitable local subtraction terms. 
The singularity types are soft, collinear and ultraviolet.
Let us briefly review under which conditions these singularities 
occur \cite{Kinoshita:1962ur,Nagy:2003qn,Dittmaier:2003bc}.
Soft singularities occur when a massless particle is exchanged between two on-shell particles.
In the amplitude
\bq
\label{example_amplitude}
 A^{(1)}_{\mathrm{bare}} & = & 
 \int \frac{d^Dk}{(2\pi)^D} P(k) \prod\limits_{j=1}^n \frac{1}{\left(k-q_j\right)^2 - m_j^2 + i \delta}.
\eq
this corresponds to the case
\bq
 m_j=0, \;\;\; p_j^2-m_{j-1}^2=0, \;\;\; p_{j+1}^2-m_{j+1}^2=0.
\eq
In that case the propagators $(j-1)$, $j$ and $(j+1)$ are on-shell. The singularity comes from the integration region
\bq
 k & \sim & q_j.
\eq
A collinear singularity occurs if a massless external on-shell particle is attached to two massless propagators.
In the amplitude of eq.~(\ref{example_amplitude}) this corresponds to
\bq
 p_j^2=0,
\;\;\;
 m_{j-1}=0,
\;\;\;
 m_{j}=0.
\eq
In that case the propagators $(j-1)$ and $j$ are on-shell. The singularity comes from the integration region
\bq
 k & \sim & q_j - x p_j,
\eq
where $x$ is a real variable.

Finally ultraviolet singularities arise when components of the loop momentum tend to infinity.
We will work in Feynman gauge throughout this paper. Therefore any loop integral with $n$ propagators
in the loop is maximally of rank $n$. Power counting arguments show immediately that all diagrams with
five or more propagators in the loop are ultraviolet finite.
Therefore all ultraviolet divergent diagrams have four or less propagators in the loop.
It can be shown that the ultraviolet divergent diagrams are only those which are propagator or vertex
corrections. Of course this has to be the case for a renormalisable theory.

\section{The subtraction terms}
\label{sect:subtraction}

In this section we present all subtraction terms for the numerical calculation of one-loop QCD amplitudes.
In subsection~\ref{subsect:ir_subtraction_massless} we give the infrared subtraction terms for massless QCD.
The generalisation to massive QCD is presented in subsection~\ref{subsect:ir_subtraction_massive}.
The ultraviolet subtraction terms for massless and massive particles can be found in subsection~\ref{subsect:uv_subtraction}.
All subtraction terms are added back in integrated form. The sum of all integrated subtraction terms
(plus the ultraviolet counterterms) defines the insertion operator ${\bf L}$. This is discussed in 
subsection~\ref{subsect:insertion_operator_L}.

\subsection{The infrared subtraction terms for massless QCD}
\label{subsect:ir_subtraction_massless}

In this section we present the infrared subtraction terms for massless QCD.
The extension of the infrared subtraction terms 
to massive particles is discussed in the next section.
We start with the soft subtraction term, which is given by
\bq
\label{soft_subtraction_term}
 G_{\mathrm{soft}}^{(1)} & = &
 \mynosign
 i
 \sum\limits_{j \in I_g}
 \frac{4 p_j \cdot p_{j+1}}{k_{j-1}^2 k_j^2 k_{j+1}^2}  A^{(0)}_j.
\eq
We recall that $I_g$ denotes the set of indices $j$, for which the 
propagator $j$ in the loop corresponds to a gluon.

The soft subtraction term is derived as follows:
In the case where gluon $i$ is soft, the corresponding propagator goes on-shell 
and we may replace in all Feynman diagrams which have the propagator $i$ the metric tensor $g_{\mu\nu}$ of this propagator by
a polarisation sum and gauge terms:
\bq
\label{gluon_prop_replacement}
 \frac{-ig^{\mu\nu}}{k_j^2} \rightarrow 
 \frac{i}{k_j^2} \left( d^{\mu\nu}(k_j^\flat,n) 
                      - 2 \frac{k_j^{\flat \mu} n^\nu + n^\mu k_j^{\flat \nu} }{2 k_j^\flat \cdot n} \right).
\eq
Here $k_j^\flat$ denotes the on-shell limit of $k_j$ and $d^{\mu\nu}$ denotes the sum over the physical polarisations:
\bq
 d^{\mu\nu}(k,n) & = &
 \sum\limits_{\lambda} \varepsilon^\mu_{\lambda}(k,n) \varepsilon^\nu_{-\lambda}(k,n)
 = 
 -g^{\mu\nu} + 2 \frac{k^\mu n^\nu + n^\mu k^\nu}{2 k \cdot n}.
\eq
$n^\mu$ is a light-like reference vector.
We note that self-energy diagrams are not singular in the soft limit, therefore adding them to the loop diagrams
will not change the soft limit.
With the inclusion of the self-energy diagrams and a corresponding replacement as in eq.~(\ref{gluon_prop_replacement})
the contribution from the polarisation sum in eq.~(\ref{gluon_prop_replacement})
makes up a tree-level partial amplitude, 
where two gluons with momenta $k_j^\flat$ and $-k_j^\flat$ have been inserted between the
external legs $j$ and $j+1$.
\begin{figure}
\begin{center}
\includegraphics[bb= 160 550 480 630]{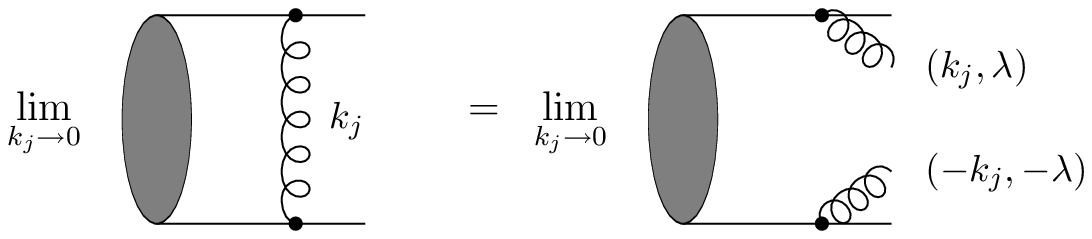}
\end{center}
\caption{The configuration for the soft limit.}
\label{fig2}
\end{figure}
This is illustrated in fig.~(\ref{fig2}).
In the soft limit this tree-level partial amplitude is given by two eikonal factors times the tree-level
partial amplitude without these two additional gluons:
\bq
\label{two_eikonal}
 \left( \frac{p_j^\mu}{p_j \cdot k_j^\flat} \right) 
 g_{\mu\nu} 
 \left( \frac{p_{j+1}^\nu}{p_{j+1}\cdot(-k_j^\flat)} \right) A^{(0)}_j.
\eq
In the soft limit we may replace $2 p_j \cdot k_j^\flat$ by $k_{j-1}^2$ and
$2 p_{i+j}\cdot(-k_j^\flat)$ by $k_{j+1}^2$.
Eq.~(\ref{two_eikonal}) then leads to eq.~(\ref{soft_subtraction_term}).
The terms with $k_j^{\flat \mu} n^\nu$ and $n^\mu k_j^{\flat \nu}$ in eq.~(\ref{gluon_prop_replacement})
vanish for the sum of all diagrams due to gauge invariance.
Integrating the soft subtraction term we obtain
\bq
\label{integrated_soft_subtraction_term}
 S_\eps^{-1} \mu^{2\eps} \int \frac{d^Dk}{(2\pi)^D} G_{\mathrm{soft}}^{(1)} 
 & = &
 \mysign 
 \frac{1}{(4\pi)^2} 
 \frac{e^{\eps \gamma_E}}{\Gamma(1-\eps)}
 \sum\limits_{j \in I_g}
 \frac{2}{\eps^2} 
 \left( \frac{-2p_j\cdot p_{j+1}}{\mu^2} \right)^{-\eps}
 A^{(0)}_j
 + {\cal O}(\eps).
\eq
We have multiplied the left-hand side by $S_\eps^{-1} \mu^{2\eps}$, where $S_\eps = (4\pi)^\eps e^{-\eps\gamma_E}$ is the 
typical volume factor of dimensional regularisation, $\gamma_E$ is Euler's constant  and $\mu$ is the renormalisation scale.

Let us now consider collinear singularities.
The collinear subtraction is given by
\bq
\label{coll_subtraction_term}
 G_{\mathrm{coll}}^{(1)} & = & 
 \mynosign
 i
 \sum\limits_{j\in I_g}
 (-2) 
 \left( 
 \frac{S_j g_{\mathrm{UV}}\left(k_{j-1}^2,k_j^2\right) }{k_{j-1}^2 k_j^2}
+
 \frac{S_{j+1} g_{\mathrm{UV}}\left(k_{j}^2,k_{j+1}^2\right) }{k_{j}^2 k_{j+1}^2}
 \right)
 A^{(0)}_j,
\eq
where the symmetry factors are given by
\bq
 S_q = S_{\bar q} = 1,
 \;\;\;
 S_g = \frac{1}{2}.
\eq
The function $g_{\mathrm{UV}}$ ensures that the integration over the loop momentum 
of the subtraction terms in eq.~(\ref{coll_subtraction_term}) is ultraviolet finite.
The function $g_{\mathrm{UV}}$ has the properties
\bq
 \lim\limits_{k_{j-1} || k_j} g_{\mathrm{UV}}\left(k_{j-1}^2,k_j^2\right) = 1,
 & & 
 \lim\limits_{k \rightarrow \infty} g_{\mathrm{UV}}\left(k_{j-1}^2,k_j^2\right) 
 = {\cal O}\left( \frac{1}{k} \right).
\eq
There are many possible choices for the function $g_{\mathrm{UV}}$. We use here a choice
which is compatible with the contour deformation discussed in sect.~(\ref{sect:contour_deformation}).
We take the function $g_{\mathrm{UV}}$ as
\bq
\label{def_g_UV}
 g_{\mathrm{UV}}\left(k_{j-1}^2,k_j^2\right)
 & = &
 1 - \frac{k_{j-1}^2 k_j^2}{\left[ (k-Q)^2-\mu_{\mathrm{UV}}^2 \right]^2}.
\eq
$Q$ is an arbitrary four-vector independent of the loop momentum $k$ and 
$\mu_{\mathrm{UV}}$ is an arbitrary scale. 
Since these two quantities are arbitrary, there are no restrictions on them, they even may have
complex values.
We will later choose $\mu_{\mathrm{UV}}^2$ purely imaginary with 
$\mbox{Im}\;\mu_{\mathrm{UV}}^2 < 0$.
This will ensure that the denominator of eq.~(\ref{def_g_UV}) does not introduce additional
singularities for the contour integration.

The collinear subtraction term is derived as follows:
We have to consider configurations where two adjacent propagators go on-shell with a massless external leg
in between.
\begin{figure}
\begin{center}
\includegraphics[bb= 250 550 330 630,width=0.2\textwidth]{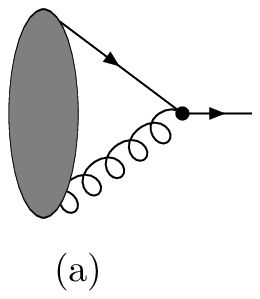}
\hspace*{5mm}
\includegraphics[bb= 250 550 330 630,width=0.2\textwidth]{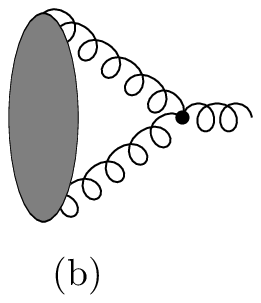}
\hspace*{5mm}
\includegraphics[bb= 250 550 330 630,width=0.2\textwidth]{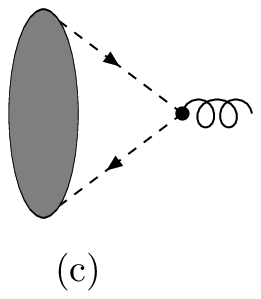}
\hspace*{5mm}
\includegraphics[bb= 250 550 330 630,width=0.2\textwidth]{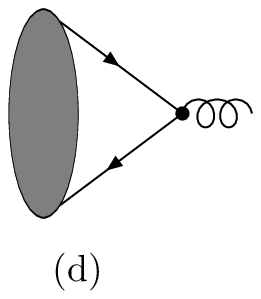}
\end{center}
\caption{Configurations for the collinear limit.
Only diagrams (a) and (b) lead to a divergence after integration.
Diagrams (c) and (d) are not singular enough to yield a divergence after integration.}
\label{fig3}
\end{figure}
These configurations are shown in fig.~(\ref{fig3}).
The diagrams (c) and (d) of fig.~(\ref{fig3}), where an external gluon splits into a 
ghost-antighost pair or into a quark-antiquark pair are in the collinear limit 
not singular enough to yield a divergence after integration.
Therefore we are left with diagrams (a) and (b).
Let us first consider the $q \rightarrow q g$ splitting as in diagram (a).
In Feynman gauge one can show that only the longitudinal polarisation of the gluon
contributes to the collinear limit.
The same holds true for the $g \rightarrow gg$ splitting of diagram (b).
In this case the collinear limit receives contributions when one of the two gluons in the loop
carries a longitudinal polarisation (but not both).
The external gluon has of course physical transverse polarisation.
It is well-known that the contraction of a longitudinal polarisation into a gauge invariant set of diagrams 
yields zero.
The shaded ``blobs'' of picture (a) and (b) of fig.~(\ref{fig3}) consist almost of a gauge invariant set of
diagrams. There is only one diagram missing, where the longitudinal polarised gluon couples directly
to the other parton connected to the shaded blob.
This is a self-energy insertion on an external line, which by definition is absent from
the amputated one-loop amplitude.
We can now turn the argument around and replace the sum of collinear singular diagrams by the negative
of the self-energy insertion on the external line.
\begin{figure}
\begin{center}
\includegraphics[bb= 250 550 330 630]{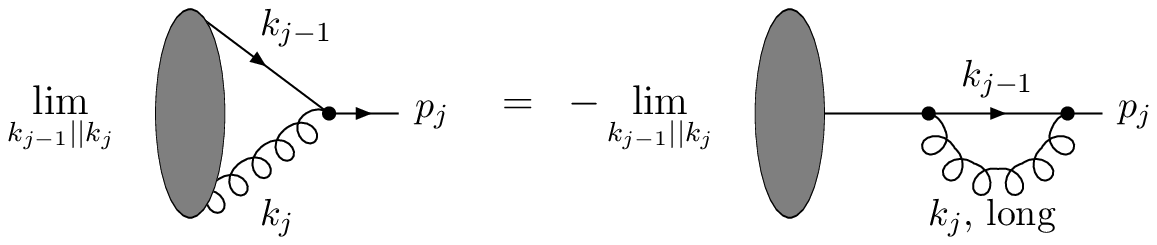}
\end{center}
\caption{In the collinear limit the sum of the singular diagrams in Feynman gauge of the amputated one-loop amplitude
is equal to the negative of the self-energy insertion on the external line where the gluon carries a longitudinal polarisation.}
\label{fig7}
\end{figure}
This is indicated in fig.~(\ref{fig7}) for the $q \rightarrow q g$ splitting and in fig~(\ref{fig8})
for the $g \rightarrow g g$ splitting.
\begin{figure}
\begin{center}
\includegraphics[bb= 50 550 560 630,width=0.9\textwidth]{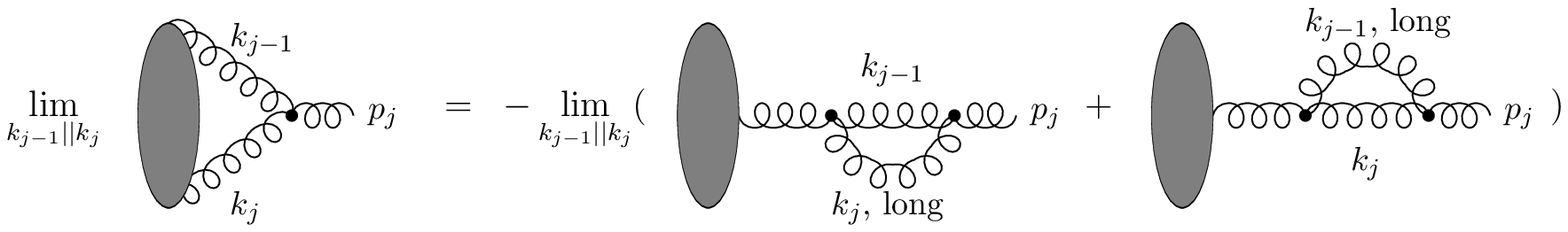}
\end{center}
\caption{In the collinear limit the sum of the singular diagrams in Feynman gauge of the amputated one-loop amplitude
is equal to the negative of the self-energy insertion on the external line where one of the gluons
carries a longitudinal polarisation.}
\label{fig8}
\end{figure}
The self-energy insertions on the external lines introduce a spurious $1/p_j^2$-singularity.
In order to calculate the singular part of the self-energies we regulate this spurious singularity
by putting $p_j^2$ slightly off-shell, but keeping $k_{j-1}$ and $k_j$ on-shell and imposing momentum conservation.
We can use the same parametrisation as in the real emission case:
\bq
 k_{j-1} & = & x p + k_\perp - \frac{k_\perp^2}{x} \frac{n}{(2p\cdot n)},
 \nonumber \\
 -k_{j} & = & (1-x) p - k_\perp - \frac{k_\perp^2}{(1-x)} \frac{n}{(2p\cdot n)},
\eq
with $p^2=n^2=0$ and $2 p\cdot k_\perp = 2 n \cdot k_\perp = 0$.
The singular part of the self-energies is proportional to 
\bq
 P_{q\rightarrow q g}^{long} & = & 
 -
 \frac{2}{2k_{j-1}\cdot k_j} \left( - \frac{2}{1-x} + 2 \right) p\!\!\!/,
 \nonumber \\
 P_{g\rightarrow g g}^{long} & = & 
 -
 \frac{2}{2k_{j-1}\cdot k_j} 
 \left( - \frac{2}{x} - \frac{2}{1-x} + 2 \right)
 \left( - g^{\mu\nu} + 2 \frac{p^\mu n^\nu + n^\mu p^\mu}{2 p \cdot n} \right).
\eq
The terms with $2/x$ and $2/(1-x)$ correspond to soft singularities and have already been subtracted out 
with the soft subtraction term $G_{\mathrm{soft}}^{(1)}$. 
In the collinear limit we therefore just have to subtract out the terms which are non-singular in the
soft limit. These terms are independent of $x$ and lead to eq.~(\ref{coll_subtraction_term}).
If we compare the soft and collinear subtraction terms for the integrand of a one-loop amplitude
with the subtraction terms for the real emission, we observe that there are no spin correlations
in the subtraction terms for the integrand of the one-loop amplitude. In the real emission case spin
correlations occur in the collinear limit.
This can be understood as follows:
From the proof of eq.~(\ref{coll_subtraction_term}) one can see that in the collinear limit always one of the collinear gluons
carries an unphysical longitudinal polarisation. Hence, there are no correlations between two transverse polarisations
of a gluon.

Integrating the collinear subtraction term we obtain
\bq
 S_\eps^{-1} \mu^{2\eps} \int \frac{d^Dk}{(2\pi)^D} G_{\mathrm{coll}}^{(1)} 
 & = & 
 \mysign 
 \frac{1}{(4\pi)^2} 
 \frac{e^{\eps \gamma_E}}{\Gamma(1-\eps)}
 \sum\limits_{j \in I_g}
 \left( S_j + S_{j+1} \right)
 \left( \frac{\mu_{\mathrm{UV}}^2}{\mu^2} \right)^{-\eps} \frac{2}{\eps}
 A^{(0)}_j
 + {\cal O}(\eps).
\eq

\subsection{The infrared subtraction terms for massive QCD}
\label{subsect:ir_subtraction_massive}

In this section we present the infrared subtraction terms for massive QCD.
This is in particular relevant to top quark physics. There are only a few modifications
necessary with respect to the massless case, which will be discussed in the following.
The modification for the unintegrated soft subtraction terms is straightforward:
\bq
 G_{\mathrm{soft}}^{(1)} & = &
 \mynosign
 i
 \sum\limits_{j \in I_g}
 \frac{4 p_j \cdot p_{j+1}}{\left( k_{j-1}^2 - m_{j-1}^2\right) k_j^2 \left( k_{j+1}^2 - m_{j+1}^2\right)}
 A^{(0)}_j.
\eq
As before, the sum runs over all particles in the loop which are gluons.
Integrating the soft subtraction term we have to distinguish whether the masses $m_{j-1}$ and $m_{j+1}$ 
are zero or not.
The result can be written as
\bq
 S_\eps^{-1} \mu^{2\eps}  \int \frac{d^Dk}{(2\pi)^D}
 G_{\mathrm{soft}}^{(1)} 
 = 
 \mysign 
 \frac{1}{(4\pi)^2} 
 \frac{e^{\eps \gamma_E}}{\Gamma(1-\eps)}
 \sum\limits_{j\in I_g}
 C\left((p_j+p_{j+1})^2,m_{j-1}^2,m_{j+1}^2,\mu^2\right)
 A^{(0)}_j
 + {\cal O}(\eps),
\eq
where the function $C(s,m_1^2,m_2^2)$ is given in the four different cases by \cite{Dittmaier:2003bc,Ellis:2007qk}
\bq
\lefteqn{
C\left(s,0,0,\mu^2\right) 
 =  
 \frac{2}{\eps^2} 
 \left( \frac{-s}{\mu^2} \right)^{-\eps},
 } & &
 \nonumber \\
\lefteqn{
C\left(s,m^2,0,\mu^2\right) 
 = C\left(s,0,m^2,\mu^2\right)
} & &
 \nonumber \\
 & = &
 \left( \frac{m^2}{\mu^2} \right)^{-\eps}
 \left[ \frac{1}{\eps^2} + \frac{2}{\eps} \ln\left(\frac{m^2}{m^2-s}\right) + \frac{\pi^2}{6}
 + \ln^2\left(\frac{m^2}{m^2-s}\right) - 2 \; \mbox{Li}_2\left(\frac{-s}{m^2-s}\right) \right],
 \nonumber \\
\lefteqn{
C\left(s,m_1^2,m_2^2,\mu^2\right) 
 =  
 \frac{2x(s-m_1^2-m_2^2)}{m_1m_2(1-x^2)} 
 \left\{ \ln(x) \left[ -\frac{1}{\eps} 
                       - \frac{1}{2} \ln(x) + 2 \ln(1-x^2) + \ln\left(\frac{m_1m_2}{\mu^2}\right) \right]
 \right. 
} & & \nonumber \\
 & & \left.
         - \frac{\pi^2}{6} + \mbox{Li}_2(x^2) + \frac{1}{2} \ln^2\left(\frac{m_1}{m_2}\right)
         + \mbox{Li}_2\left(1-x\frac{m_1}{m_2}\right) + \mbox{Li}_2\left(1-x\frac{m_2}{m_1}\right)
 \right\},
\eq
with
\bq
 x & = & - \frac{1-\chi}{1+\chi},
 \;\;\;\;\;
 \chi = \sqrt{1-\frac{4m_1m_2}{s-\left(m_1-m_2\right)^2} }.
\eq
The modification for the collinear subtraction term is even simpler:
There is no collinear singularity if an external quark or antiquark is massive.
It suffices therefore to define $S_Q=S_{\bar{Q}}=0$ for a massive quark or antiquark.

\subsection{The ultraviolet subtraction terms}
\label{subsect:uv_subtraction}

A primitive one-loop QCD amplitude contains, apart from infrared divergences, also ultraviolet divergences.
In an analytical calculation regulated by dimensional regularisation these divergences manifest 
themselves in a single pole in the dimensional regularisation parameter $\eps$.
If we would do the calculation within a cut-off regularisation we would find a logarithmic dependence on the cut-off.

Within a numerical approach we need subtraction terms which approximate the ultraviolet behaviour of the
integrand locally.
We first note that the one-loop amplitude has, in a fixed direction in loop momentum space, up to quadratic
UV-divergences.
These are reduced to a logarithmic UV-divergence only after angular integration.
For a local subtraction term we have to match the quadratic, linear and logarithmic divergence.
Since we work in Feynman gauge throughout this paper it follows that any loop integral with $n$ propagators
in the loop is maximally of rank $n$.
Power counting arguments show immediately that all diagrams with
fiver or more propagators in the loop are ultraviolet finite.
Therefore all ultraviolet divergent diagrams have four or less propagators in the loop.
It can be shown that the ultraviolet divergent diagrams are only those which are propagator or vertex
corrections. 
The ultraviolet subtraction terms have to subtract out correctly the exact pointwise ultraviolet behaviour 
of the one-loop
amplitude. There are several possibilities how the ultraviolet subtraction terms can be chosen.
We present here a list of ultraviolet subtraction terms which have two additional properties:
\begin{enumerate}
\item The unintegrated ultraviolet subtraction terms have all singularities localised on a single surface
\bq 
\label{singular_UV_surface}
 \left( k - Q \right)^2 - \mu_{\mathrm{UV}}^2 & = & 0.
\eq
$Q$ is an arbitrary four-vector independent of the loop momentum $k$.
By choosing $\mu_{\mathrm{UV}}^2$ purely imaginary with $\mbox{Im}\;\mu_{\mathrm{UV}}^2<0$ we can ensure
that the contour for the integration over the loop momentum $k$ 
never comes close to the singular surface defined by eq.~(\ref{singular_UV_surface}).
We remark that also the four-vector $Q$ is allowed to have complex entries.
We will choose $Q$ as a function of the (real) external momenta and the (complex) Feynman parameters.
\item We choose the ultraviolet subtraction terms such that the finite part of the 
integrated ultraviolet subtraction terms
is independent of $Q$ and proportional to the pole part, 
with the same constant of proportionality for all ultraviolet subtraction terms.
This ensures that the sum of all integrated UV subtraction terms is again proportional to a tree-level amplitude.
\end{enumerate}
To present the ultraviolet subtraction terms we follow the notation of fig.~(\ref{figure_momenta_one_loop}).
This implies that we take all external momenta as outgoing and that the direction of the flow of the loop
momentum is clock-wise. We label the external momenta clock-wise with $p_1$, $p_2$, etc..
With the conventions of fig.~(\ref{figure_momenta_one_loop}) the loop momentum $k$ is the 
momentum of the loop propagator preceding the external leg with momentum $p_1$.
We set $D=4-2\eps$ for the number of space-time dimensions and we define $\bar{k}$ by
\bq
 \bar{k} & = & k - Q.
\eq
It is sufficient to present the ultraviolet subtraction terms with these conventions.
The general case can always be obtained by an appropriate substitution.
Technically, the ultraviolet subtraction terms are derived by expanding the loop propagators around the 
propagator corresponding to eq.~(\ref{singular_UV_surface}). For a single propagator we have
\bq
 \frac{1}{\left(k-p\right)^2-m^2}
 & = &
 \frac{1}{\bar{k}^2-\mu_{\mathrm{UV}}^2}
 \left\{ 
       1 + \frac{2\bar{k}\cdot\left(p-Q\right)}{\bar{k}^2-\mu_{\mathrm{UV}}^2}
 - \frac{\left(p-Q\right)^2-m^2+\mu_{\mathrm{UV}}^2}{\bar{k}^2-\mu_{\mathrm{UV}}^2}
 + \frac{\left[ 2\bar{k}\cdot\left(p-Q\right)\right]^2}{\left[\bar{k}^2-\mu_{\mathrm{UV}}^2\right]^2}
 \right\}
 \nonumber \\
 & &
 + {\cal O}\left(\frac{1}{|\bar{k}|^5}\right).
\eq
\begin{figure}
\begin{center}
\includegraphics[bb= 115 613 490 694]{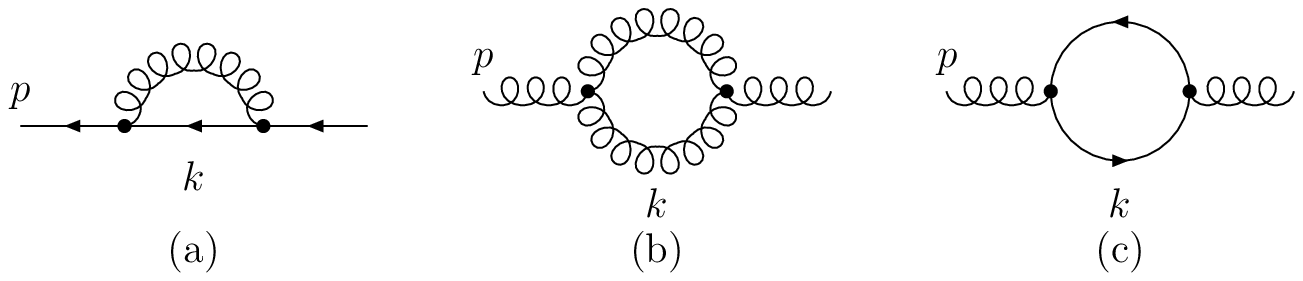}
\end{center}
\caption{
The master diagrams for the self-energies. 
Not shown are diagrams which are obtained from these by pinching some of the propagators.
Also not shown are diagrams with a closed ghost loop.
Diagram (a) gives the self-energy of the quark propagator,
diagram (b) corresponds to the leading colour self-energy of the gluon propagator while
diagram (c) gives the $N_f$-contribution to the self-energy of the gluon propagator.
}
\label{fig12}
\end{figure}
We start with the ultraviolet subtraction term for the quark propagator. 
The relevant diagram whose large $k$-behaviour has to be approximated is shown in fig.~(\ref{fig12} a).
For the quark propagator we take as UV-subtraction term
\bq
-i \Sigma^{(1)}_{\mathrm{UV}} & = &
 - i S_\eps^{-1} \mu^{2\eps} \int \frac{d^Dk}{(2\pi)^Di}
 \left[ 
        \frac{-2(1-\eps)\left(Q\!\!\!\!/+\bar{k}\!\!\!/\right)+4\left(1-\frac{1}{2}\eps\right)m}{\left(\bar{k}^2-\mu_{\mathrm{UV}}^2\right)^2}
        - 4 \left(1-\eps\right) \frac{\bar{k}\cdot \left( p - 2 Q \right) \; \bar{k}\!\!\!/}{\left(\bar{k}^2-\mu_{\mathrm{UV}}^2\right)^3} 
 \right. \nonumber \\
 & & \left.
        + \frac{2 \mu_{\mathrm{UV}}^2 \left( p\!\!\!/-2m \right)}{\left(\bar{k}^2-\mu_{\mathrm{UV}}^2\right)^3} 
 \right].
\eq
The term proportional to $\mu_{\mathrm{UV}}^2$ in the numerator is not divergent, but ensures that the 
finite part of the integrated expression is proportional to the pole part.
Integration yields
\bq
-i \Sigma^{(1)}_{\mathrm{UV}} & = &
 - i 
 \frac{1}{(4\pi)^2}
 \left( - p\!\!\!/ + 4 m \right) \left( \frac{1}{\eps} - \ln \frac{\mu_{\mathrm{UV}}^2}{\mu^2} \right)
 + {\cal O}(\eps).
\eq
We see that the finite part is given by $-\ln(\mu_{\mathrm{UV}}^2/\mu^2)$ times the pole part.
With the choice $\mu_{\mathrm{UV}}^2=\mu^2$ the finite part of the integrated ultraviolet subtraction terms
will be zero.
However this is not the way how we will use this formula. Our priority is to avoid additional
singularities for the numerical loop integration. Therefore we will choose $\mu_{\mathrm{UV}}^2$
along the negative imaginary axis. On the other hand $\mu^2$ is the usual renormalisation scale and 
conventionally taken as a positive real number.

Next we consider the gluon propagator. We divide the ultraviolet subtraction terms into a leading colour part
and a part proportional to $N_f$. 
An example of a diagram contributing to each part is shown in fig.~(\ref{fig12} b) and fig.~(\ref{fig12} c).
It is convenient to define
\bq
 R^\mu & = & \frac{1}{2} p^\mu - Q^\mu.
\eq
The leading colour UV-subtraction term is given by
\bq
\lefteqn{
 -i \Sigma^{(1) \; \mu\nu}_{\mathrm{UV}} = 
 - i S_\eps^{-1} \mu^{2\eps} \int \frac{d^Dk}{(2\pi)^Di} 
 \left\{
 \frac{4\left(1-\eps\right)g^{\mu\nu}}{\bar{k}^2-\mu_{\mathrm{UV}}^2}
 \right. } & &
 \\
 & & 
 +
 \frac{1}{\left(\bar{k}^2-\mu_{\mathrm{UV}}^2\right)^2}
 \left[ 
        - 4 g^{\mu\nu} p^2 
        + 4 p^\mu p^\nu 
        - 8 \left( 1 - \eps \right) 
            \left( \bar{k}^\mu \bar{k}^\nu - \bar{k}^\mu R^\nu - R^\mu \bar{k}^\nu + R^\mu R^\nu 
                   - g^{\mu\nu} \bar{k} \cdot R \right)
 \right]
 \nonumber \\
 & &
 -2 \left(1-\eps\right) g^{\mu\nu} \frac{Q^2+(p-Q)^2+2\mu_{\mathrm{UV}}^2}{\left(\bar{k}^2-\mu_{\mathrm{UV}}^2\right)^2}
-
 \frac{32 \left( 1 - \eps \right) \bar{k} \cdot R}{\left(\bar{k}^2-\mu_{\mathrm{UV}}^2\right)^3}
 \left[ 
        \bar{k}^\mu \bar{k}^\nu - \bar{k}^\mu R^\nu - R^\mu \bar{k}^\nu 
 \right]
 \nonumber \\
 & &
+ 8 \left( 1 - \eps \right) 
 \frac{Q^2+(p-Q)^2+2\mu_{\mathrm{UV}}^2}{\left(\bar{k}^2-\mu_{\mathrm{UV}}^2\right)^3}
        \bar{k}^\mu \bar{k}^\nu 
 + 8 \left( 1 - \eps \right) g^{\mu\nu}
 \frac{\left(\bar{k} \cdot Q \right)^2 + \left(\bar{k} \cdot \left(p-Q \right) \right)^2}{\left(\bar{k}^2-\mu_{\mathrm{UV}}^2\right)^3}
 \nonumber \\
 & &
 \left.
 + 
 \frac{4}{3} \frac{\left(-g^{\mu\nu} p^2 + p^\mu p^\nu\right)\mu_{\mathrm{UV}}^2}{\left(\bar{k}^2-\mu_{\mathrm{UV}}^2\right)^3}
- 32 \left( 1 - \eps \right) 
 \frac{\left(\bar{k} \cdot Q \right)^2 - \left(\bar{k} \cdot Q \right) \left(\bar{k} \cdot \left(p-Q \right) \right) + \left(\bar{k} \cdot \left(p-Q \right) \right)^2}{\left(\bar{k}^2-\mu_{\mathrm{UV}}^2\right)^4}
        \bar{k}^\mu \bar{k}^\nu 
 \right\}.
 \nonumber
\eq
For the fermionic contribution we have
\bq
\lefteqn{
 -i \Sigma^{(1) \; \mu\nu}_{\mathrm{UV},\mathrm{nf}} = 
 - i S_\eps^{-1} \mu^{2\eps} \int \frac{d^Dk}{(2\pi)^Di} 
 \left\{ 
 \frac{-4 g^{\mu\nu}}{\bar{k}^2-\mu_{\mathrm{UV}}^2}
 \right.
 } & & \\
 & &
 +
 \frac{1}{\left(\bar{k}^2-\mu_{\mathrm{UV}}^2\right)^2}
 \left[ 
          2 g^{\mu\nu} p^2 
        - 2 p^\mu p^\nu 
        + 8 \left( \bar{k}^\mu \bar{k}^\nu - \bar{k}^\mu R^\nu - R^\mu \bar{k}^\nu + R^\mu R^\nu 
                   - g^{\mu\nu} \bar{k} \cdot R\right)
 \right]
 \nonumber \\
 & &
 +2 g^{\mu\nu} \frac{Q^2+(p-Q)^2+2\mu_{\mathrm{UV}}^2}{\left(\bar{k}^2-\mu_{\mathrm{UV}}^2\right)^2}
 + 
 \frac{32 \bar{k} \cdot R}{\left(\bar{k}^2-\mu_{\mathrm{UV}}^2\right)^3}
 \left[ 
        \bar{k}^\mu \bar{k}^\nu
        - \bar{k}^\mu R^\nu - R^\mu \bar{k}^\nu 
 \right]
 \nonumber \\
 & &
 - 8
 \frac{Q^2+(p-Q)^2+2\mu_{\mathrm{UV}}^2}{\left(\bar{k}^2-\mu_{\mathrm{UV}}^2\right)^3}
        \bar{k}^\mu \bar{k}^\nu
 - 8 g^{\mu\nu}
 \frac{\left(\bar{k} \cdot Q \right)^2 + \left(\bar{k} \cdot \left(p-Q \right) \right)^2}{\left(\bar{k}^2-\mu_{\mathrm{UV}}^2\right)^3}
 \nonumber \\
 & &
 \left.
 + 32
 \frac{\left(\bar{k} \cdot Q \right)^2 - \left(\bar{k} \cdot Q \right) \left(\bar{k} \cdot \left(p-Q \right) \right) + \left(\bar{k} \cdot \left(p-Q \right) \right)^2}{\left(\bar{k}^2-\mu_{\mathrm{UV}}^2\right)^4}
        \bar{k}^\mu \bar{k}^\nu
 \right\}.
 \nonumber
\eq
Integration yields
\bq
 -i \Sigma^{(1) \; \mu\nu}_{\mathrm{UV}} & = &
 - i \frac{1}{(4\pi)^2} \left( -g ^{\mu\nu} p^2 + p^\mu p^\nu \right) 
     \left(\frac{10}{3} \right)
     \left( \frac{1}{\eps} - \ln \frac{\mu_{\mathrm{UV}}^2}{\mu^2} \right)
 + {\cal O}(\eps),
 \nonumber \\
 -i \Sigma^{(1) \; \mu\nu}_{\mathrm{UV},\mathrm{nf}} & = &
 -i \frac{1}{(4\pi)^2} \left( -g ^{\mu\nu} p^2 + p^\mu p^\nu \right)
     \left( - \frac{4}{3} \right)
     \left( \frac{1}{\eps} - \ln \frac{\mu_{\mathrm{UV}}^2}{\mu^2} \right)
 + {\cal O}(\eps).
\eq
\begin{figure}
\begin{center}
\includegraphics[bb= 150 590 460 725]{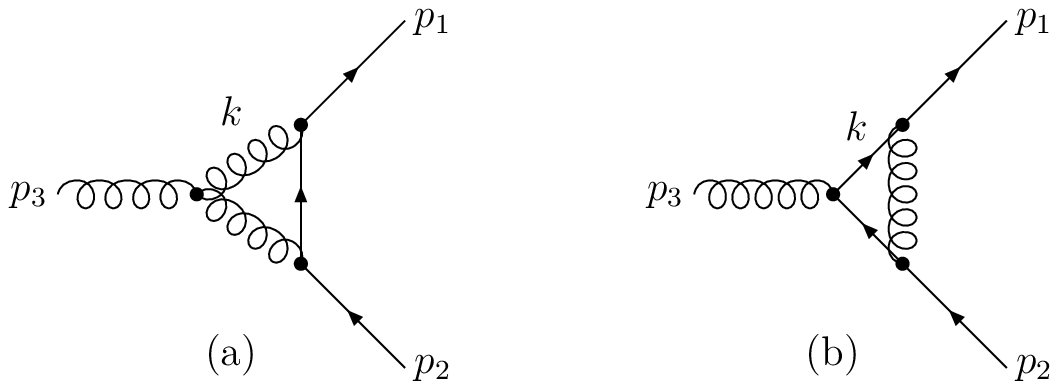}
\\
\includegraphics[bb= 150 590 460 725]{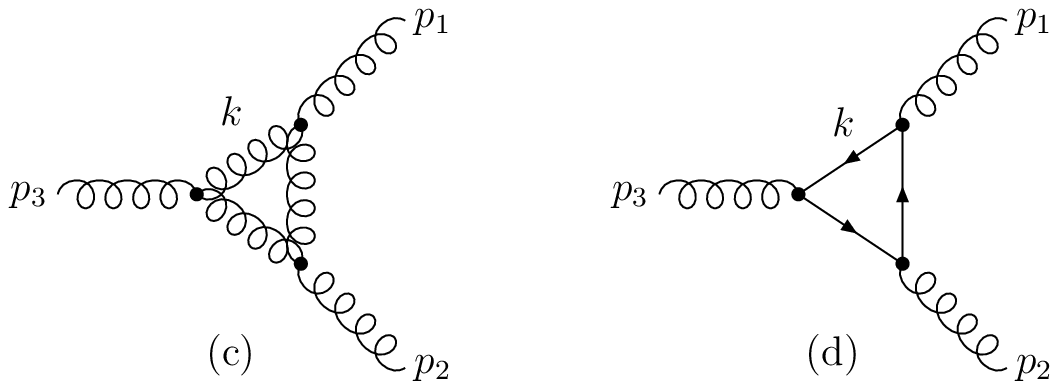}
\\
\includegraphics[bb= 150 610 460 725]{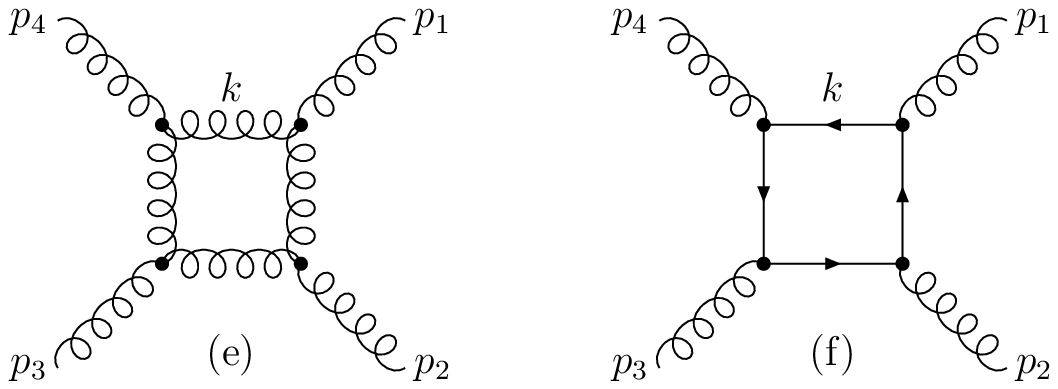}
\end{center}
\caption{
The master diagrams for the vertex corrections. 
Not shown are diagrams which are obtained from these by pinching some of the propagators.
Also not shown are diagrams with a closed ghost loop.
Diagram (a) gives the leading colour vertex correction to the quark-gluon vertex,
diagram (b) gives a subleading colour contribution.
Diagram (c) corresponds to the leading colour vertex correction to the three-gluon vertex,
diagram (d) gives the $N_f$-contribution.
Diagram (e) corresponds to the leading colour vertex correction to the four-gluon vertex,
diagram (f) gives the $N_f$-contribution.
The external momenta are always directed outwards, the loop momentum is always taken to flow clock-wise.
}
\label{fig13}
\end{figure}
For the quark-gluon vertex there is a leading colour contribution and a subleading colour contribution.
Representative diagrams are shown in fig.~(\ref{fig13} a) and fig.~(\ref{fig13} b).
The subtraction term for the leading colour contribution is given by
\bq
 V^{(1)}_{qqg, \mathrm{lc}} & = &   
 i S_\eps^{-1} \mu^{2\eps} \int \frac{d^Dk}{(2\pi)^Di} 
 \left[
 \frac{2 \gamma^\mu}{\left(\bar{k}^2-\mu_{\mathrm{UV}}^2\right)^2}
 +
 \frac{4 \left(1-\eps \right) \bar{k}\!\!\!/ \; \bar{k}^\mu - 2 \mu_{\mathrm{UV}}^2 \gamma^\mu}{\left(\bar{k}^2-\mu_{\mathrm{UV}}^2\right)^3}
 \right].
\eq
For the subleading colour contribution we have
\bq
\label{uv_subtr_quark_gluon_sc}
 V^{(1)}_{qqg, \mathrm{sc}} & = &   
 i S_\eps^{-1} \mu^{2\eps} \int \frac{d^Dk}{(2\pi)^Di} 
 \left[
 \frac{2 \left(1-\eps \right) \bar{k}\!\!\!/ \gamma^\mu \bar{k}\!\!\!/ + 4 \mu_{\mathrm{UV}}^2 \gamma^\mu}{\left(\bar{k}^2-\mu_{\mathrm{UV}}^2\right)^3}
 \right].
\eq
Integration leads to 
\bq
 V^{(1)}_{qqg, \mathrm{lc}} & = & 
 i \frac{1}{(4\pi)^2} \gamma^\mu 
 \left( 3 \right) \left( \frac{1}{\eps} - \ln \frac{\mu_{\mathrm{UV}}^2}{\mu^2} \right)
 + {\cal O}(\eps),
 \nonumber \\
 V^{(1)}_{qqg, \mathrm{sc}} & = & 
 i \frac{1}{(4\pi)^2} \gamma^\mu 
 \left( -1 \right) \left( \frac{1}{\eps} - \ln \frac{\mu_{\mathrm{UV}}^2}{\mu^2} \right)
 + {\cal O}(\eps).
\eq
For the three-gluon vertex we have a leading colour contribution and a contribution proportional to 
$N_f$.
Representative diagrams are shown in fig.~(\ref{fig13} c) and fig.~(\ref{fig13} d).
For the three-gluon vertex we define
\bq
 R_2^\mu & = & \frac{2}{3} p_1^\mu + \frac{1}{3} p_2^\mu - Q^\mu.
\eq
The subtraction terms are given by
\bq
\lefteqn{
 V^{(1)}_{ggg, \mathrm{lc}} =    
 i S_\eps^{-1} \mu^{2\eps} \int \frac{d^Dk}{(2\pi)^Di} 
 \left\{ 
 \frac{- 96 \left( 1 - \eps \right) \bar{k}^\mu \bar{k}^\nu \bar{k}^\rho \; \bar{k} \cdot R_2}{\left(\bar{k}^2-\mu_{\mathrm{UV}}^2\right)^4}
 +
 \frac{8\left( 1 - \eps \right) }{\left(\bar{k}^2-\mu_{\mathrm{UV}}^2\right)^3}
 \left[ 
        - 2 \bar{k}^\mu \bar{k}^\nu \bar{k}^\rho
        + p_1^\mu \bar{k}^\nu \bar{k}^\rho
 \right. \right.
 } & & 
 \nonumber \\
 & & \left.
        + 2 \bar{k}^\mu \left( p_1^\nu + \frac{1}{2} p_2^\nu \right) \bar{k}^\rho
        + \bar{k}^\mu \bar{k}^\nu \left( p_1^\rho + p_2^\rho \right)
        - 2 Q^\mu \bar{k}^\nu \bar{k}^\rho
        - 2 \bar{k}^\mu Q^\nu \bar{k}^\rho
        - 2 \bar{k}^\mu \bar{k}^\nu Q^\rho
 \right. \nonumber \\
 & & \left.
        + 2 g^{\mu\nu} \bar{k}^\rho \; \bar{k} \cdot \left( \frac{1}{2} p_1 + \frac{1}{2} p_2 - Q \right)
        + 2 g^{\nu\rho} \bar{k}^\mu \; \bar{k} \cdot \left( \frac{1}{2} p_1 - Q \right)
        + 2 g^{\rho\mu} \bar{k}^\nu \; \bar{k} \cdot \left( p_1 +\frac{1}{2} p_2 - Q \right)
 \right]
 \nonumber \\
 & &
 - \frac{4}{3}  
 \frac{\mu_{\mathrm{UV}}^2}{\left(\bar{k}^2-\mu_{\mathrm{UV}}^2\right)^3}
        \left( g^{\mu\nu} \left(p_2^\rho -p_1^\rho\right) 
                                      + g^{\nu\rho} \left(p_3^\mu-p_2^\mu \right) 
                                      + g^{\rho\mu} \left( p_1^\nu - p_3^\nu \right) \right) 
 \nonumber \\
 & &
 +
 \frac{4 \left( 1 - \eps \right) }{\left(\bar{k}^2-\mu_{\mathrm{UV}}^2\right)^2}
 \left[ 
        g^{\mu\nu} \left( \bar{k}^\rho + Q^\rho - p_2^\rho \right)
        + g^{\nu\rho} \left( \bar{k}^\mu + Q^\mu + p_2^\mu \right)
        + g^{\rho\mu} \left( \bar{k}^\nu + Q^\nu - 2 p_1^\nu - p_2^\nu \right)
 \right]
 \nonumber \\
 & &
 \left.
 -
 \frac{2\eps}{\left(\bar{k}^2-\mu_{\mathrm{UV}}^2\right)^2}
 \left[ 
        g^{\mu\nu} \left( p_2^\rho - p_1^\rho \right)
        + g^{\nu\rho} \left( p_3^\mu - p_2^\mu \right)
        + g^{\rho\mu} \left( p_1^\nu - p_3^\nu \right)
 \right] \right\},
 \\
\lefteqn{
 V^{(1)}_{ggg, \mathrm{nf}} =    
 i S_\eps^{-1} \mu^{2\eps} \int \frac{d^Dk}{(2\pi)^Di} 
 \left\{
 \frac{96 \bar{k}^\mu \bar{k}^\nu \bar{k}^\rho \; \bar{k} \cdot R_2}{\left(\bar{k}^2-\mu_{\mathrm{UV}}^2\right)^4}
 +
 \frac{8}{\left(\bar{k}^2-\mu_{\mathrm{UV}}^2\right)^3}
 \left[ 
        2 \bar{k}^\mu \bar{k}^\nu \bar{k}^\rho
        - p_1^\mu \bar{k}^\nu \bar{k}^\rho
 \right. \right.
 } & & \nonumber \\
 & & \left.
        - 2 \bar{k}^\mu \left( p_1^\nu + \frac{1}{2} p_2^\nu \right) \bar{k}^\rho
        - 1 \bar{k}^\mu \bar{k}^\nu \left( p_1^\rho + p_2^\rho \right)
        + 2 Q^\mu \bar{k}^\nu \bar{k}^\rho
        + 2 \bar{k}^\mu Q^\nu \bar{k}^\rho
        + 2 \bar{k}^\mu \bar{k}^\nu Q^\rho
 \right. \nonumber \\
 & & \left.
        - 2 g^{\mu\nu} \bar{k}^\rho \; \bar{k} \cdot \left( \frac{1}{2} p_1 + \frac{1}{2} p_2 - Q \right)
        - 2 g^{\nu\rho} \bar{k}^\mu \; \bar{k} \cdot \left( \frac{1}{2} p_1 - Q \right)
        - 2 g^{\rho\mu} \bar{k}^\nu \; \bar{k} \cdot \left( p_1 + \frac{1}{2} p_2 - Q \right)
 \right]
 \nonumber \\
 & & 
 \left.
 -
 \frac{4}{\left(\bar{k}^2-\mu_{\mathrm{UV}}^2\right)^2}
 \left[ 
        g^{\mu\nu} \left( \bar{k}^\rho + Q^\rho - p_2^\rho \right)
        + g^{\nu\rho} \left( \bar{k}^\mu + Q^\mu + p_2^\mu \right)
        + g^{\rho\mu} \left( \bar{k}^\nu + Q^\nu - 2 p_1^\nu - p_2^\nu \right)
 \right] \right\}.
 \nonumber
\eq
Integration leads to 
\bq
\lefteqn{
 V^{(1)}_{ggg, \mathrm{lc}} =  
 i \frac{1}{(4\pi)^2}
 \left[ g^{\mu\nu} \left( p_2^\rho - p_1^\rho \right) + g^{\nu\rho} \left( p_3^\mu - p_2^\mu \right)
        + g^{\rho\mu} \left( p_1^\nu - p_3^\nu \right) \right]
 \left( - \frac{4}{3} \right)
 \left( \frac{1}{\eps} - \ln \frac{\mu_{\mathrm{UV}}^2}{\mu^2} \right)
 + {\cal O}(\eps),
} & &
 \nonumber \\
\lefteqn{
 V^{(1)}_{ggg, \mathrm{nf}} =  
 i \frac{1}{(4\pi)^2}
 \left[ g^{\mu\nu} \left( p_2^\rho - p_1^\rho \right) + g^{\nu\rho} \left( p_3^\mu - p_2^\mu \right)
        + g^{\rho\mu} \left( p_1^\nu - p_3^\nu \right) \right]
 \left( \frac{4}{3} \right)
 \left( \frac{1}{\eps} - \ln \frac{\mu_{\mathrm{UV}}^2}{\mu^2} \right)
 + {\cal O}(\eps).
} & & 
 \hspace*{150mm}
 \nonumber \\
\eq
Finally we consider the four-gluon vertex.
Again we have a leading colour contribution and a contribution proportional to 
$N_f$.
Representative diagrams are shown in fig.~(\ref{fig13} e) and fig.~(\ref{fig13} f).
The ultraviolet subtraction terms are given by
\bq
\lefteqn{
 V^{(1)}_{gggg, \mathrm{lc}} =   
 i S_\eps^{-1} \mu^{2\eps} \int \frac{d^Dk}{(2\pi)^Di} 
 \left\{
 \frac{32 \left(1-\eps\right)\bar{k}^\mu \bar{k}^\nu \bar{k}^\rho \bar{k}^\sigma}{\left(\bar{k}^2-\mu_{\mathrm{UV}}^2\right)^4}
 \right.
 } & &
 \nonumber \\
 & &
 -
 \frac{8\left(1-\eps\right)}{\left(\bar{k}^2-\mu_{\mathrm{UV}}^2\right)^3}
 \left[ 
         g^{\mu\nu} \bar{k}^\rho \bar{k}^\sigma
       + g^{\nu\rho} \bar{k}^\sigma \bar{k}^\mu
       + g^{\rho\sigma} \bar{k}^\mu \bar{k}^\nu
       + g^{\sigma\mu} \bar{k}^\nu \bar{k}^\rho
 \right]
 \nonumber \\
 & &
 \left.
 - \frac{4}{3} 
 \frac{\mu_{\mathrm{UV}}^2}{\left(\bar{k}^2-\mu_{\mathrm{UV}}^2\right)^3}
 \left( 2 g^{\mu\rho} g^{\nu\sigma} - g^{\mu\nu} g^{\rho\sigma} - g^{\mu\sigma} g^{\nu\rho} \right)
 +
 \frac{2\left(1-\eps\right)}{\left(\bar{k}^2-\mu_{\mathrm{UV}}^2\right)^2}
 \left[ 
         g^{\mu\nu} g^{\rho\sigma} + g^{\mu\sigma} g^{\nu\rho}
 \right] \right\},
 \nonumber \\
\lefteqn{
 V^{(1)}_{gggg, \mathrm{nf}} =    
 i S_\eps^{-1} \mu^{2\eps} \int \frac{d^Dk}{(2\pi)^Di} 
 \left\{
 \frac{-32 \bar{k}^\mu \bar{k}^\nu \bar{k}^\rho \bar{k}^\sigma}{\left(\bar{k}^2-\mu_{\mathrm{UV}}^2\right)^4}
 \right.
 } & &
 \nonumber \\
 & &
 +
 \frac{8}{\left(\bar{k}^2-\mu_{\mathrm{UV}}^2\right)^3}
 \left[ 
         g^{\mu\nu} \bar{k}^\rho \bar{k}^\sigma
       + g^{\nu\rho} \bar{k}^\sigma \bar{k}^\mu
       + g^{\rho\sigma} \bar{k}^\mu \bar{k}^\nu
       + g^{\sigma\mu} \bar{k}^\nu \bar{k}^\rho
 \right]
 \nonumber \\
 & &
 \left.
 +
 \frac{4}{\left(\bar{k}^2-\mu_{\mathrm{UV}}^2\right)^2}
 \left[ 
         g^{\mu\rho} g^{\nu\sigma} - g^{\mu\nu} g^{\rho\sigma} - g^{\mu\sigma} g^{\nu\rho}
 \right] \right\}.
\eq
Integration leads to 
\bq
 V^{(1)}_{gggg, \mathrm{lc}} & = & 
 i \frac{1}{(4\pi)^2}
 \left( 2 g^{\mu\rho} g^{\nu\sigma} - g^{\mu\nu} g^{\rho\sigma} - g^{\mu\sigma} g^{\nu\rho} \right)
 \frac{2}{3} 
 \left( \frac{1}{\eps} - \ln \frac{\mu_{\mathrm{UV}}^2}{\mu^2} \right)
 + {\cal O}(\eps),
 \nonumber \\
 V^{(1)}_{gggg, \mathrm{nf}} & = & 
 i \frac{1}{(4\pi)^2}
 \left( 2 g^{\mu\rho} g^{\nu\sigma} - g^{\mu\nu} g^{\rho\sigma} - g^{\mu\sigma} g^{\nu\rho} \right)
 \frac{4}{3}
 \left( \frac{1}{\eps} - \ln \frac{\mu_{\mathrm{UV}}^2}{\mu^2} \right)
 + {\cal O}(\eps).
\eq

\subsection{The insertion operator ${\bf L}$}
\label{subsect:insertion_operator_L}

The integrated versions of the subtraction terms for the integrand of a one-loop
amplitude have to be added back.
The sum of all these subtraction terms is combined with 
the ultraviolet counterterm ${\cal A}^{(1)}_{\mathrm{CT}}$
and defines the insertion operator ${\bf L}$:
\bq
 \left.{\cal A}^{(0)}\right.^\ast {\bf L} \; {\cal A}^{(0)}
 & = & 
   2 \; \mbox{Re} \; \left.{\cal A}^{(0)}\right.^\ast
     \left( {\cal A}_{\mathrm{CT}}^{(1)} + {\cal A}_{\mathrm{soft}}^{(1)} + {\cal A}_{\mathrm{coll}}^{(1)} + {\cal A}_{\mathrm{UV}}^{(1)}\right).
\eq
In massless QCD this operator is rather simple and given by
\bq
 {\bf L} & = &
 \frac{\alpha_s}{2\pi}
 \frac{e^{\eps \gamma_E}}{\Gamma(1-\eps)}
 \; \mbox{Re}
 \left[
  \sum\limits_i \sum\limits_{j\neq i} {\bf T}_i {\bf T}_j \frac{1}{\eps^2} 
                \left( \frac{-2p_ip_j}{\mu^2} \right)^{-\eps}
 - \sum\limits_i \frac{\gamma_i}{\eps} \left( \frac{\mu_{\mathrm{UV}}^2}{\mu^2} \right)^{-\eps}
 -  \frac{(n-2)}{2} \beta_0 \ln \frac{\mu_{\mathrm{UV}}^2}{\mu^2}
 \right]
\nonumber \\ 
& &
 + {\cal O}(\eps).
\eq
Combining the insertion operator ${\bf L}$ with the insertion operator ${\bf I}$ we obtain in the massless case
\bq
 {\bf I} + {\bf L}
 & = &
 \frac{\alpha_s}{2\pi}
 \; \mbox{Re}
 \left[
  \sum\limits_i \sum\limits_{j\neq i} {\bf T}_i {\bf T}_j 
      \left( 
              \frac{\gamma_i}{{\bf T}_i^2} \ln \frac{\left|2p_ip_j\right|}{\mu_{\mathrm{UV}}^2}
            - \frac{\pi^2}{2} \theta(2p_ip_j)
      \right)
 \right. \nonumber \\
 & & \left.
 + \sum\limits_i \left( \gamma_i + K_i - \frac{\pi^2}{3} {\bf T}_i^2 \right)
 -  \frac{(n-2)}{2} \beta_0 \ln \frac{\mu_{\mathrm{UV}}^2}{\mu^2}
 \right]
 + {\cal O}(\eps).
\eq
We see that the sum of the insertion operators ${\bf I}$ and ${\bf L}$
is free of poles in the dimensional regularisation parameter $\eps$.
 
Let us now consider the massive case.
It is sufficient to discuss the case of QCD amplitudes with one heavy flavour, the generalisation to
several heavy flavours is straightforward.
There are a few modifications.
We have to take into account the heavy quark field renormalisation constant, which is given by
\bq
 Z_{2,Q} & = & 1 + \frac{\alpha_s}{4\pi} C_F \left( -\frac{3}{\eps} - 4 + 3 \ln \frac{m^2}{\mu^2} \right)
 + {\cal O}\left(\alpha_s^2\right).
\eq
Secondly, the mass of the heavy quark is renormalised.
For the heavy quark mass we have to choose a renormalisation scheme.
In the on-shell scheme the mass renormalisation constant is given by
\bq
 Z_{m,\mathrm{on-shell}} & = & 1 + \frac{\alpha_s}{4\pi} C_F \left( -\frac{3}{\eps} - 4 + 3 \ln \frac{m^2}{\mu^2} \right)
 + {\cal O}\left(\alpha_s^2\right).
\eq
In the $\overline{MS}$-scheme the mass renormalisation constant is simply given by
\bq
 Z_{m,\overline{MS}} & = & 1 + \frac{\alpha_s}{4\pi} C_F \left( -\frac{3}{\eps} \right)
 + {\cal O}\left(\alpha_s^2\right).
\eq
The renormalisation of the mass and of the heavy quark field will modify the ultraviolet counterterm
and the insertion operator ${\bf L}$.
In order to present the counterterm related to the mass renormalistion it is convenient to define
the quantity ${\cal B}^{(0)}(p_1,...,p_n,g,m)$ through
\bq
 {\cal A}^{(0)}\left(p_1,...,p_n,g,m+\delta m\right)
 =  
 {\cal A}^{(0)}\left(p_1,...,p_n,g,m\right)
 + \frac{\delta m}{m}
 {\cal B}^{(0)}\left(p_1,...,p_n,g,m \right)
 + 
 {\cal O}\left( (\delta m)^2 \right).
\eq
We consider a QCD amplitude with $n$ external partons, out of which $n_g$ are gluons, $n_q$ are massless quarks,
$n_{\bar{q}}$ are massless anti-quarks, $n_Q$ are massive quarks with mass $m$ and $n_{\bar{Q}}$ are massive 
anti-quarks with mass $m$.
Obviously we have $n_q=n_{\bar{q}}$ and $n_Q=n_{\bar{Q}}$
We find for the operator ${\bf L}$
\bq
\lefteqn{
 \left.{\cal A}^{(0)}\right.^\ast {\bf L} \; {\cal A}^{(0)} 
 =} & & \nonumber \\
 & &
 \frac{\alpha_s}{2\pi}
 \frac{e^{\eps \gamma_E}}{\Gamma(1-\eps)}
 \; \mbox{Re}
 \left\{ \left[
  \sum\limits_i \sum\limits_{j\neq i} \frac{1}{2} {\bf T}_i {\bf T}_j  
                C\left( \left(p_i+p_j\right)^2, p_i^2, p_j^2, \mu^2 \right)
 - \sum\limits_i \frac{\gamma_i}{\eps} \left( \frac{\mu_{\mathrm{UV}}^2}{\mu^2} \right)^{-\eps}
 \right. \right. \nonumber \\
 & & \left. \left.
 -  \frac{(n-2)}{2} \beta_0 \ln \frac{\mu_{\mathrm{UV}}^2}{\mu^2}
 - \left( n_Q + n_{\bar{Q}} \right) C_F \left( 2 + \frac{3}{2} \ln \frac{\mu_{\mathrm{UV}}^2}{m^2} \right)
 \right] \left| {\cal A}^{(0)} \right|^2
 - C_M \left.{\cal A}^{(0)}\right.^\ast {\cal B}^{(0)} \right\}
\nonumber \\ 
& &
 + {\cal O}(\eps).
\eq
Here we defined for heavy quarks
\bq
 \gamma_Q = \gamma_{\bar{Q}} = 0.
\eq
The coefficient $C_M$ depends on the renormalisation scheme for the mass and is given by
\bq
 C_M & = & 
 \left\{
 \begin{array}{ll}
 \left( 4 + 3 \ln \frac{\mu_{\mathrm{UV}}^2}{m^2} \right) C_F & \mbox{on-shell scheme,}\\
 3 C_F \ln \frac{\mu_{\mathrm{UV}}^2}{\mu^2} & \mbox{$\overline{MS}$-scheme.}\\
\end{array} \right.
\eq

\section{Contour deformation}
\label{sect:contour_deformation}

In this section we present the algorithm for the contour deformation.
The guiding principles of the method are outlined in subsection~\ref{subsect:contour_overview}.
The individual parts are subsequently discussed in detail:
Subsection~\ref{subsect:contour_feynman} describes the details of the deformation.
In order to resolve the direction by which we approach singular points we introduce 
additional Feynman parameters. 
Issues related to the numerical stability of the Monte Carlo integration are discussed in
subsection~\ref{subsect:stability}.

\subsection{Overview of the contour deformation}
\label{subsect:contour_overview}

Having a complete list of ultraviolet and infrared subtraction terms at hand, we can ensure that the integration
over the loop momentum gives a finite result and can therefore be performed in four dimensions.
However, this does not yet imply that we can simply or safely integrate each of the four components of the loop momentum $k^\mu$
from minus infinity to plus infinity along the real axis.
There is still the possibility that some of the loop propagators go on-shell for real values of the loop momentum.
If the contour is not pinched this is harmless, as we may escape into the complex plane in a direction indicated by
Feynman's $+i\delta$-prescription.
However, it implies that the integration should be done over a region of real dimension four in the complex space
${\mathbb C}^4$.
There are many choices for the contour deformation which are formally correct, in the sense that they avoid the poles
of the propagators where a deformation is possible and deform the contour in the right direction.
However, not all choices are suitable for a numerical Monte Carlo integration. Most choices will lead to large
cancellations between different integration regions and therefore to large Monte Carlo integration errors.
Finding an algorithm which determines a correct deformation of the integration contour and which leads to a numerical
stable integration is a challenging problem.
We use several steps and techniques to achieve this goal. The critical regions in the integration domain are the regions
where one or more propagators go on-shell.
\begin{enumerate}
\item In a first step we introduce Feynman parameters as additional integration variables. 
This transforms the singular variety defined by  the vanishing of some of the $n$ propagators 
into a simple variety consisting of a single cone.
For a single cone the required deformation for the loop momentum $k$ can easily be stated.
The deformation vanishes at the origin of the cone.
\item In a second step we deform the Feynman parameters into the complex plane. 
This is necessary because the deformation of the loop momentum vanishes at the origin of the cone.
The deformation of the Feynman parameters vanishes whenever Landau's equations are satisfied.
\item The total deformation vanishes therefore if the loop momentum $k$ is at the origin
and if the Feynman parameters satisfy Landau's equations.
Let us now consider an integrable singularity corresponding to the vanishing of $n_{\mathrm{sing}}$
propagators in an amplitude with $n$ external particles.
The Feynman parametrisation has the disadvantage that it raises the denominator to the power $n$.
This leads to numerical instabilities in the Monte Carlo integration for large values of $n$.
In order to improve the stability of the Monte Carlo integration we expand the true
propagators around propagators where we have added a small mass $\mu^2_{\mathrm{IR}}$.
We will take $\mu^2_{\mathrm{IR}}$ purely imaginary with $\mbox{Im}\;\mu^2_{\mathrm{IR}}<0$.
(Comparing $\mu^2_{\mathrm{IR}}$ with $\mu^2_{\mathrm{UV}}$ we first note that we take them both
purely imaginary. In practise we choose $\mbox{Im}\;\mu^2_{\mathrm{UV}} < \mbox{Im}\;\mu^2_{\mathrm{IR}} <0$.)
\end{enumerate}
We remark that the singular region defined by the vanishing of the propagator of the $\mathrm{UV}$-counterterm
does not introduce additional problems: By choosing $\mu_{\mathrm{UV}}^2$ along the negative imaginary axis we can 
ensure that this singular region is moved away from the integration contour.

Let us now discuss how the contour deformation is done in detail.
We would like to evaluate numerically the expression
\bq
 A^{(1)}_{\mathrm{subtr}}
 & = & 
 A^{(1)}_{\mathrm{bare}} - A^{(1)}_{\mathrm{soft}} - A^{(1)}_{\mathrm{coll}} - A^{(1)}_{\mathrm{UV}}
 = 
 \int \frac{d^Dk}{(2\pi)^D} 
  \left( G^{(1)}_{\mathrm{bare}} - G^{(1)}_{\mathrm{soft}} - G^{(1)}_{\mathrm{coll}} - G^{(1)}_{\mathrm{UV}} \right).
\eq
The subtraction terms $G^{(1)}_{\mathrm{soft}}$, $G^{(1)}_{\mathrm{coll}}$ and $G^{(1)}_{\mathrm{UV}}$
are local counterterms. Therefore the final result is finite and the integral can be performed in $4$ dimensions.
We can write the integrand in the form
\bq
\label{integrand_form}
G^{(1)}_{\mathrm{bare}} - G^{(1)}_{\mathrm{soft}} - G^{(1)}_{\mathrm{coll}} - G^{(1)}_{\mathrm{UV}} 
 & = &
 \frac{P(k)}{\prod\limits_{j=1}^n \left( k_j^2 - m_j^2 \right)}
 -
 \frac{P_{\mathrm{UV}}(k)}{\left( \bar{k}^2 - \mu_{\mathrm{UV}}^2 \right)^{n_{\mathrm{UV}}}}.
\eq
$P(k)$ and $P_{\mathrm{UV}}(k)$ are polynomials in the loop momentum $k$.
$G^{(1)}_{\mathrm{bare}}$ and $G^{(1)}_{\mathrm{soft}}$ contribute only to the first term of the
r.h.s of eq.~(\ref{integrand_form}), while $G^{(1)}_{\mathrm{UV}}$ contributes only to the second 
term on the r.h.s of eq.~(\ref{integrand_form}).
On the other hand, $G^{(1)}_{\mathrm{coll}}$ contributes to both terms on the r.h.s of eq.~(\ref{integrand_form}).
The number $n_{\mathrm{UV}}$ is a positive integer, which by construction of the ultraviolet
counterterms is not larger than four.
The difference on the right-hand side of eq.~(\ref{integrand_form}) falls off at least like
$1/|k|^5$ for large $|k|$, which corresponds to the fact that the difference is UV-finite.
The two individual terms
\bq
 \frac{P(k)}{\prod\limits_{j=1}^n \left( k_j^2 - m_j^2 \right)},
 & &
 \frac{P_{\mathrm{UV}}(k)}{\left( \bar{k}^2 - \mu_{\mathrm{UV}}^2 \right)^{n_{\mathrm{UV}}}}
\eq
fall off at least like $1/|k|^2$ for large $|k|$.
This corresponds to the fact that all contributions are maximally quadratic divergent.

Eq.~(\ref{integrand_form}) shows clearly all possible singularities of the integrand.
The integrand has singularities when one of the propagators $1/(k_j^2-m_j^2)$ goes on-shell as well as when
the propagator of the ultraviolet counterterm $1/(\bar{k}^2-\mu_{\mathrm{UV}}^2)$ goes on-shell.
The singularities of the ultraviolet counterterm are all localised 
on a single cone $\bar{k}^2-\mu_{\mathrm{UV}}^2=0$ and rather easy to handle.
With the definition of $P(k)$ and $P_{\mathrm{UV}}(k)$
we can write the integral as
\bq
 A^{(1)}_{\mathrm{subtr}}
 & = & 
 \int \frac{d^4k}{(2\pi)^4}
 \left[
 \frac{P(k)}{\prod\limits_{j=1}^n \left(k_j^2 - m_j^2\right)}
  -
 \frac{P_{\mathrm{UV}}(k)}{\left( \bar{k}^2 - \mu_{\mathrm{UV}}^2 \right)^{n_{\mathrm{UV}}}}
 \right].
\eq
In the next steps we define an integration contour which avoids the poles at
$k_j^2-m_j^2=0$ and $\bar{k}^2-\mu_{\mathrm{UV}}^2=0$.

\subsection{Feynman parametrisation}
\label{subsect:contour_feynman}

Let us now consider an integral of the form
\bq
 I & = & 
 \int \frac{d^4k}{(2\pi)^4} 
 \left[
 \frac{P(k)}{\prod\limits_{j=1}^n \left(k_j^2 - m_j^2 + i \delta \right)}
 -
 \frac{P_{\mathrm{UV}}(k)}{\left( \bar{k}^2 - \mu_{\mathrm{UV}}^2 + i \delta \right)^{n_{\mathrm{UV}}}}
 \right]
 =
 \int \frac{d^4k}{(2\pi)^4} 
 \frac{R(k)}{\prod\limits_{j=1}^n \left(k_j^2 - m_j^2 + i \delta \right)},
 \nonumber \\
 & & R(k) =  
 P(k) 
 - \frac{P_{\mathrm{UV}}(k)}{\left( \bar{k}^2 - \mu_{\mathrm{UV}}^2 + i \delta \right)^{n_{\mathrm{UV}}}}
   \prod\limits_{j=1}^n \left(k_j^2 - m_j^2 + i \delta \right),
\eq
where $P(k)$ and $P_{\mathrm{UV}}(k)$ are polynomials in the loop momentum $k$, $R(k)$ is 
a rational function in the loop momentum.
The integration is over a complex contour in order to avoid -- whenever possible --
the poles of the propagators. The direction of the deformation is indicated by the $+i\delta$-prescription.
We now proceed as follows: We first introduce Schwinger parameters. We then deform the loop momentum 
and the Schwinger parameters into the complex plane. Each deformation will introduce a Jacobian.
We can then integrate out one Schwinger parameter and we arrive at a formula equivalent to the Feynman
parametrisation.
We could have started directly with the Feynman parametrisation. The detour through the Schwinger
parameters has the advantage that we obtain the correct Jacobian in a simple way.

Assume that $A$ is a complex number with a positive imaginary part. Then we have
\bq
 \frac{1}{A} & = & -i \int\limits_0^\infty dt e^{i A t}.
\eq
We use this identity to rewrite the propagators with the help of the Schwinger parametrisation:
\bq
 \prod\limits_{j=1}^n \frac{1}{k_j^2 - m_j^2}
 & = &
 \left( -i \right)^n
 \int\limits_0^\infty dt_1 ... \int\limits_0^\infty dt_n
 \;
 \exp\left( i \sum\limits_{j=1}^n t_j \left( k_j^2 - m_j^2 \right) \right).
\eq
For the argument of the exponential function we have
\bq
\sum\limits_{j=1}^n t_j \left[ k_j^2 - m_j^2 \right]
 & = &
 t \left( k - \frac{1}{t} \sum\limits_{j=1}^n t_j q_j \right)^2
 +
 \frac{1}{2t} \sum\limits_{i=1}^n \sum\limits_{j=1}^n t_i S_{ij} t_j,
 \;\;\;\;\;\;
 t = \sum\limits_{l=1}^n t_l.
\eq
We now set
\bq
\label{redef_k}
 {k}^\mu & = & \tilde{k}^\mu + i g_{\mu\nu} \tilde{k}^\nu + \frac{1}{t} \sum\limits_{i=1}^n t_i q_i^\mu.
\eq
We note that $k$ as a function of the variables $t_j$ is homogeneous of degree $0$.
The Jacobian is
\bq
 \left| \frac{\partial {k}^\mu}{\partial \tilde{k}^\nu} \right|
 & = & - 4 i.
\eq
Then we have
\bq
\label{res_imag_part_k_deform}
\sum\limits_{j=1}^n t_j \left( k_j^2 - m_j^2 \right)
 & = &
 2 i t \tilde{k} \circ \tilde{k} 
 +
 \frac{1}{2 t} \sum\limits_{i=1}^n \sum\limits_{j=1}^n t_i S_{ij} t_j,
\eq
where $\tilde{k} \circ \tilde{k}$ denotes the Euclidean scalar product.
The imaginary part of eq.~(\ref{res_imag_part_k_deform}) vanishes for
$\tilde{k} = 0$. 
The resulting singularities can still be avoided by deforming 
the Schwinger parameters into the complex plane. Therefor we set
\bq
\label{deformation_t}
 t_j & = & \tilde{t}_j + i \lambda \tilde{t}_j \beta_j(\tilde{t}_1,...,\tilde{t}_n),
 \;\;\;\;\;\;
 \beta_j(\tilde{t}_1,...,\tilde{t}_n) = \frac{\sum\limits_{k=1}^n S_{jk} \tilde{t}_k}{\sqrt{\sum\limits_{a=1}^n \left( \sum\limits_{b=1}^n S_{ab} \tilde{t}_b \right)^2}}, 
\eq
such that all $\tilde{t}_j$ are real and positive.
We note that $t_j$ as a function of the variables $\tilde{t}_1$, ..., $\tilde{t}_n$ is homogeneous of degree $1$.
We denote by $J$ the Jacobian of this transformation
\bq
 J(\tilde{t}_1,...,\tilde{t}_n) & = & \left| \frac{\partial t_i}{\partial \tilde{t}_j} \right|.
\eq
Note that the Jacobian is a homogeneous function of degree $0$ in the variables $\tilde{t}_1$, ..., $\tilde{t}_n$.
Since $R(k)$ and $J(\tilde{t}_1,...,\tilde{t}_n)$ are both homogeneous of degree $0$ in the variables $\tilde{t}_1$, ..., $\tilde{t}_n$
we can integrate out one variable.
We make use of the identity
\bq
 \int\limits_{\tilde{t}_j \ge 0} d^n\tilde{t} \; f(\tilde{t}) 
 & = & 
 \int\limits_{\tilde{t}_j \ge 0} d^n\tilde{t} \; \int\limits_{\hat{t} \ge 0} d\hat{t} \; \delta\left(\hat{t} - \tilde{t}\right) \; f(\tilde{t})
 =
 \int\limits_{\tilde{x}_j \ge 0} d^n\tilde{x} \; \delta\left(1-\sum\limits_{j=1}^n \tilde{x}_j \right) \; 
 \int\limits_{\hat{t} \ge 0} d\hat{t} \; \hat{t}^{n-1} \; f(\tilde{t}),
 \nonumber 
\eq
with $\tilde{x}_j = \tilde{t}_j/\hat{t}$. We also write
\bq
\label{deformation_x}
 x_j = \tilde{x}_j + i \lambda \tilde{x}_j \beta_j\left(\tilde{x}_1,...,\tilde{x}_n\right),
 & &
 x = \sum\limits_{k=1}^n x_k.
\eq
We arrive at
\bq
\label{final_feynman_integral}
 I & = &
 4 \Gamma(n) \int \frac{d^4\tilde{k}}{(2\pi)^4 i} 
 \int d^n\tilde{x} \; \delta\left(1-\sum\limits_{j=1}^n \tilde{x}_j \right) 
 f\left(\tilde{x}_1,...,\tilde{x}_n\right),
\eq
with
\bq
\label{def_f_func}
 f\left(\tilde{x}_1,...,\tilde{x}_n\right) 
 & = &
 J(\tilde{x}_1,...,\tilde{x}_n) 
 \left[ 2 i x \tilde{k} \circ \tilde{k} 
 + \frac{1}{2 x} \sum\limits_{a=1}^n \sum\limits_{b=1}^n x_a S_{ab} x_b \right]^{-n}
 R(k).
\eq
This is the standard formula for the Feynman parametrisation supplemented with the correct 
Jacobian corresponding to the deformation given in eq.~(\ref{deformation_t}) 
and eq.~(\ref{deformation_x}).
The deformation for the parameters $\tilde{x}_j$ vanishes for $\tilde{x}_j=0$. The deformation has the additional
property that it does not necessarily vanish in regions where two parameters are equal.

We further note that 
$f$ is a function homogeneous of degree $(-n)$ in the variables $\tilde{x}_1$, ..., $\tilde{x}_n$.
We may use this property to replace the integration over the $(n-1)$-dimensional simplex
by an integration over the $n$-dimensional hyper-cube.
This is discussed in appendix \ref{sect:generating_feynman_parameters}.
The integration over the loop momentum $\tilde{k}^\mu$ is over four real variables from 
minus infinity to plus infinity. It is convenient for a numerical Monte Carlo integration to map
this region onto the finite region $[0,1]^4$. 
This is discussed in appendix \ref{sect:generating_loop_momentum}.

It remains to discuss the ultraviolet subtraction terms.
For the arbitrary four-vector $Q^\mu$ occurring in the ultraviolet subtraction terms
we can make the choice
\bq
\label{choice_1_for_Q_mu}
 Q^\mu & = & \frac{1}{x} \sum\limits_{i=1}^n x_i q_i^\mu.
\eq
Since the Feynman parameters $x_i$ are deformed into the complex space it follows that
$Q^\mu$ is in general a four-vector with complex entries.
With this choice of $Q^\mu$ the relation between $\bar{k}^\mu$ and $k^\mu$ is given by
\bq
 \bar{k}^\mu & = & k^\mu - Q^\mu
 = k^\mu - \frac{1}{x} \sum\limits_{i=1}^n x_i q_i^\mu.
\eq
For the propagators of the ultraviolet subtraction terms we have then
\bq
 \bar{k}^2 - \mu^2_{\mathrm{UV}} & = & 2 i \tilde{k} \circ \tilde{k} - \mu^2_{\mathrm{UV}}.
\eq
The Euclidean norm $\tilde{k} \circ \tilde{k}$ is always non-negative.
If we set $\mu^2_{\mathrm{UV}}$ purely imaginary with $\mbox{Im}\;\mu^2_{\mathrm{UV}}<0$ we ensure that this
quantity is never zero.

\subsection{Improving the numerical stability}
\label{subsect:stability}

In the integrand of eq.~(\ref{final_feynman_integral}) we have the function
\bq
 L(\tilde{k};\tilde{x}_1,...,\tilde{x}_n) & = &
 2 i x \tilde{k} \circ \tilde{k} 
 + \frac{1}{2x} \sum\limits_{a=1}^n \sum\limits_{b=1}^n x_a S_{ab} x_b.
\eq
Let us discuss the conditions under which this function vanishes.
To this aim we focus on the imaginary part.
In order to simplify the following discussion we multiply this function by $x$.
The factor $x$ is never zero and therefore uncritical.
We find that to order $\lambda$ the imaginary part of $(x L)$ is given by
\bq
 \mbox{Im}\;\left( x L\right) & = &
 2 \tilde{x}^2 \tilde{k} \circ \tilde{k}
 + \lambda  
\frac{\sum\limits_{j=1}^n \tilde{x}_j \left( \sum\limits_{l=1}^n S_{jl} \tilde{x}_l \right)^2}
     {\sqrt{\sum\limits_{a=1}^n \left( \sum\limits_{b=1}^n S_{ab} \tilde{x}_b \right)^2}}
 + {\cal O}(\lambda^2). 
\eq
The expression on the right-hand side is always non-negative.
The expression is zero if $\tilde{k}=0$ and if for all $j\in\{1,...,n\}$ we have either
\bq
\label{landau_1}
 \tilde{x}_j & = & 0
\eq
or
\bq
\label{landau_2}
 \sum\limits_{l=1}^n S_{jl} \tilde{x}_l & = & 0.
\eq
Eq.~(\ref{landau_1}) and eq.~(\ref{landau_2}) are the well-known equations of Landau.
These two equations together with the condition $\tilde{k}=0$ give therefore 
necessary conditions for the vanishing of the function $L$.
Let us now consider a subset $S \subset \{1,...,n\}$ of $n_{\mathrm{sing}}$ elements, such that
eq.~(\ref{landau_1}) is fulfilled for all $j \in \{1,...,n\}\backslash S$ and eq.~(\ref{landau_2}) is fulfilled
for all $j \in S$.
For $n_{\mathrm{sing}}<n$ this corresponds to a non-leading Landau singularity.
In the final formula for the contour deformation in eq.~(\ref{final_feynman_integral})
the function $L$ occurs to the power $(-n)$, but only
a power of $(-n_{\mathrm{sing}})$ can be attributed to the underlying physical configuration.
The remaining $(n-n_{\mathrm{sing}})$ powers are artificially introduced by the Feynman parametrisation.
These powers are compensated by the integration over the Feynman parameters in the
directions corresponding to $\{1,...,n\}\backslash S$.
However this integration is done numerically with Monte Carlo methods.
The net effects are numerical instabilities for large $n$ from the regions where the function
$L$ is close to zero.

In order to improve the efficiency of the Monte Carlo integration we have either to avoid
that the function $L$ gets close to zero 
or to reduce the power $n$ to which the function $L$ is raised.
We present here an efficient method which is based on the first strategy.
In appendix \ref{sect:reduce_power} we describe a method based on the second strategy.

In order to avoid that the function $L$ gets close to zero we start from the identity
\bq
\label{hypergeometric_1F0}
 L^{-n} & = &
 \left( L - x \mu^2_{\mathrm{IR}} \right)^{-n}
 \frac{1}{\Gamma(n)}
 \sum\limits_{n_{\mathrm{IR}}=0}^\infty
 \frac{\Gamma\left(n_{\mathrm{IR}}+n\right)}{\Gamma\left(n_{\mathrm{IR}}+1\right)}
 \left( \frac{-x \mu^2_{\mathrm{IR}}}{L - x \mu^2_{\mathrm{IR}}} \right)^{n_{\mathrm{IR}}}.
\eq
Here we introduced an additional parameter $\mu^2_{\mathrm{IR}}$, which we will take purely imaginary
with $\mbox{Im}\;\mu^2_{\mathrm{IR}} < 0$.
This ensures that
\bq
 \mbox{Im} \; \left( L - x \mu^2_{\mathrm{IR}} \right) & > & 0.
\eq
It follows that the expression $(L-x \mu^2_{\mathrm{IR}})$ is never zero.

The sum on the right-hand side of eq.~(\ref{hypergeometric_1F0}) converges for 
\bq
 \left| \frac{-x \mu^2_{\mathrm{IR}}}{L - x \mu^2_{\mathrm{IR}}} \right| & < & 1.
\eq
To leading order in $\lambda$ we can show that we always have
\bq
 \left| \frac{-x \mu^2_{\mathrm{IR}}}{L - x \mu^2_{\mathrm{IR}}} \right| & \le & 1,
\eq
and that the equal sign implies $L=0$.
For $L=0$ the sum on the right-hand side of eq.~(\ref{hypergeometric_1F0}) is divergent, 
but in this case at the same time 
$L^{-n}$ on the left-hand side is an ill-defined expression.

We then truncate the series at order $N_{\mathrm{IR}}$. This amounts to the replacement
\bq
\label{replacement}
 L^{-n} & \rightarrow &
 \frac{1}{\Gamma(n)}
 \sum\limits_{n_{\mathrm{IR}}=0}^{N_{\mathrm{IR}}}
 \frac{\Gamma\left(n_{\mathrm{IR}}+n\right)}{\Gamma\left(n_{\mathrm{IR}}+1\right)}
 \frac{\left(-x \mu^2_{\mathrm{IR}}\right)^{n_{\mathrm{IR}}}}
      {\left(L - x \mu^2_{\mathrm{IR}}\right)^{n+n_{\mathrm{IR}}}}.
\eq
The net effect is that the critical expression $L^{-n}$ is replaced by
$(L-x\mu^2_{\mathrm{IR}})^{-n-n_{\mathrm{IR}}}$, where $(L-x\mu^2_{\mathrm{IR}})$ is always non-zero.
The replacement depends on two parameters $\mu^2_{\mathrm{IR}}$ and $N_{\mathrm{IR}}$, which we can use
to control the quality of the approximation.
We note that the replacement in eq.~(\ref{replacement}) is exact in the limit $\mu^2_{\mathrm{IR}} \rightarrow 0$.

\section{Additional remarks}
\label{sect:additional_remarks}

In this section we offer a few additional remarks, which might help to avoid possible pitfalls.
Subsection~\ref{subsect:locality} is devoted to diagrams like massless tadpoles, which give zero in an analytical calculation.
For this reason they are often discarded in an analytical calculation from the very beginning.
Discarding these diagrams in a numerical calculation will spoil the local cancellation of the singularities,
so these diagrams have to be kept.
Subsection~\ref{subsect:scheme} discusses the scheme-independence of the infrared subtraction terms.
Subsection~\ref{subsect:subleading} is devoted to subleading tree-level partial amplitudes, 
which can occur in the infrared subtraction terms.

\subsection{Locality of the subtraction terms and vanishing diagrams}
\label{subsect:locality}

A one-loop amplitude can be represented by Feynman diagrams.
The Feynman diagrams contributing to an amplitude are the ones contributing to the corresponding
amputated Green's function.
For a one-loop amplitude all Feynman diagrams which are not self-energy insertions on external
lines contribute.
\begin{figure}
\begin{center}
\includegraphics[bb= 125 610 555 715]{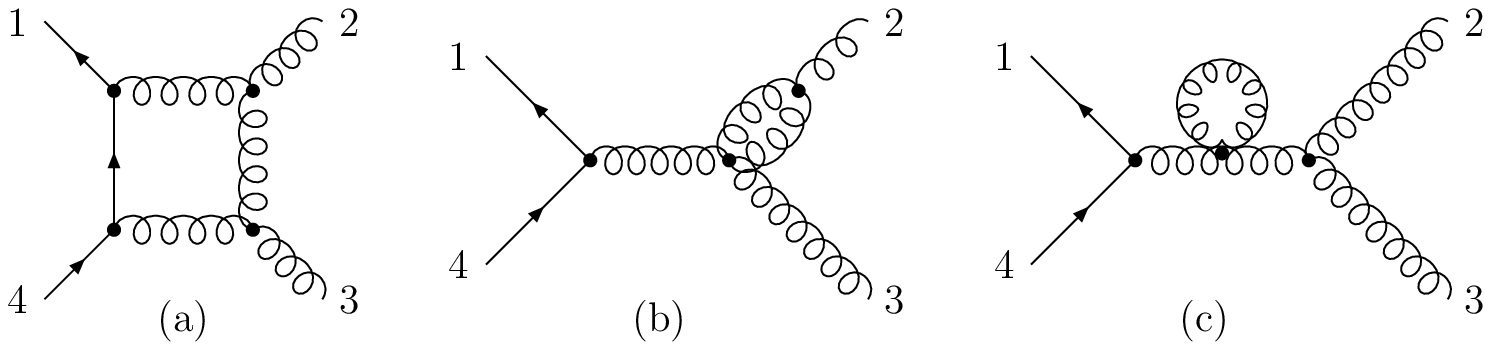}
\caption{\label{figure_zero_diagram}
Diagram (a) shows the top-level diagram for the left-moving primitive $qgg\bar{q}$ amplitude.
Diagrams (b) and (c) show lower-point diagrams contributing to the same amplitude.
In an analytical calculation within dimensional regularisation these diagrams yield zero after integration.
In a numerical calculation these diagrams have to be included in order not spoil the locality
of the subtraction terms.}
\end{center}
\end{figure}
Fig.~(\ref{figure_zero_diagram}) shows some diagrams contributing 
to the left-moving primitive $qgg\bar{q}$ amplitude.
We focus here on the diagrams (b) and (c) of fig.~(\ref{figure_zero_diagram}).
We first note that diagram (b) is not a self-energy insertion on the external line with momentum
$p_2$.
This diagram belongs to the amputated Green's function.
In an analytical calculation using dimensional regularisation the diagrams (b) and (c) yield
zero after integration.
For this reason they are often discarded in an analytical calculation from the very beginning.
This cannot be done in a numerical calculation.
Diagram (b) yields zero because of an exact cancellation between an ultraviolet divergence and an infrared
divergence. Neglecting this diagram in a numerical calculation would spoil the local
cancellations of the singularities.

Diagram (c) yields zero because the massless tadpole corresponds to a scaleless integral.
The corresponding integral is set to zero within dimensional regularisation.
Diagram (c) has a quadratically divergent ultraviolet singularity. In a numerical calculation this 
singularity is subtracted out with the ultraviolet subtraction terms for the gluon propagator.

\subsection{Scheme independence}
\label{subsect:scheme}

In this subsection we investigate the scheme independence of the infrared subtraction terms.
Contrary to the subtraction terms for the real emission, there is no dependence on the variant
of dimensional regularisation (conventional dimensional regularisation, 't Hooft-Veltman scheme,
four-dimensional scheme) in the integrated result for the infrared subtraction terms for the loop
integrand.
At first sight this is puzzling: 
In the real emission case the variant of dimensional regularisation introduces a scheme-dependent finite term.
On the other hand unitarity requires that this scheme-dependent finite term cancels in the final result.
It is instructive to investigate how this cancellation occurs.
For simplicity we discuss the massless case.
The solution comes from the LSZ reduction formula:
The renormalised one-loop amplitude with $n_g$ gluons, $n_q$ quarks and $n_{\bar{q}}$ antiquarks
is related to the bare amplitude by
\bq
\label{LSZ}
 {\cal A}_{\mathrm{ren}}(p_1,...,p_n,\alpha_s)
 & = &
 \left(Z_2^{1/2} \right)^{n_q+n_{\bar q}} \left( Z_3^{1/2} \right)^{n_g}
 {\cal A}_{\mathrm{bare}}\left(p_1,...,p_n,Z_g^2 S_\eps^{-1}\mu^{2\eps} \alpha_s\right).
\eq
$Z_g$ is the renormalisation constant for the strong coupling, given by 
\bq
 Z_g & = & 1 + \frac{\alpha_s}{4\pi} \left( - \frac{\beta_0}{2} \right)  \frac{1}{\eps} + {\cal O}(\alpha_s^2).
\eq
$Z_2$ is the quark field renormalisation constant and $Z_3$ is the gluon field renormalisation constant.
The LSZ reduction formula instructs us to take as field renormalisation constants the residue
of the propagators at the pole.
In dimensional regularisation this residue is $1$ for massless particles and therefore the field renormalisation
constants are often omitted from eq.~(\ref{LSZ}).
However $Z_2=Z_3=1$ is due to a cancellation between ultraviolet and infrared divergences \cite{Harris:2002md}.
In Feynman gauge we have
\bq
 Z_{2} & = & 1 + \frac{\alpha_s}{4\pi} C_F \left( \frac{1}{\eps_{IR}} - \frac{1}{\eps_{UV}} \right) + {\cal O}(\alpha_s^2),
 \nonumber \\
 Z_{3} & = & 1 + \frac{\alpha_s}{4\pi} \left( 2 C_A -\beta_0 \right) \left( \frac{1}{\eps_{IR}} - \frac{1}{\eps_{UV}} \right) 
 + {\cal O}(\alpha_s^2).
\eq
Here we indicated explicitly the origin of the $1/\eps$-poles.
These poles introduce scheme-dependent finite terms of ultraviolet and infrared origin with opposite sign.
Now the cancellation of the scheme-dependent finite term of infrared origin is as follows:
The scheme-dependent finite term from the real emission contribution cancels with the scheme-dependent finite term
of infrared origin from the renormalisation constants.
This leaves the scheme-dependent finite terms of ultraviolet origin in the renormalisation constants and in the bare
one-loop amplitude.
These remain and give the result of the calculation in the chosen scheme.
One can convert from one scheme to another by a finite renormalisation.
We remark that
the scheme dependence of the bare one-loop amplitude is entirely of ultraviolet origin.

\subsection{Subleading tree-level partial amplitudes}
\label{subsect:subleading}

In this subsection we discuss the tree-level partial amplitudes, which occur in the infrared subtraction terms.
We would like to point out, that a tree-level partial amplitude  occurring in the infrared subtraction terms need
not be identical to the ones occurring in the colour decomposition of the leading order calculation.
This is best explained through an example.
\begin{figure}
\begin{center}
\includegraphics[bb= 110 610 495 722]{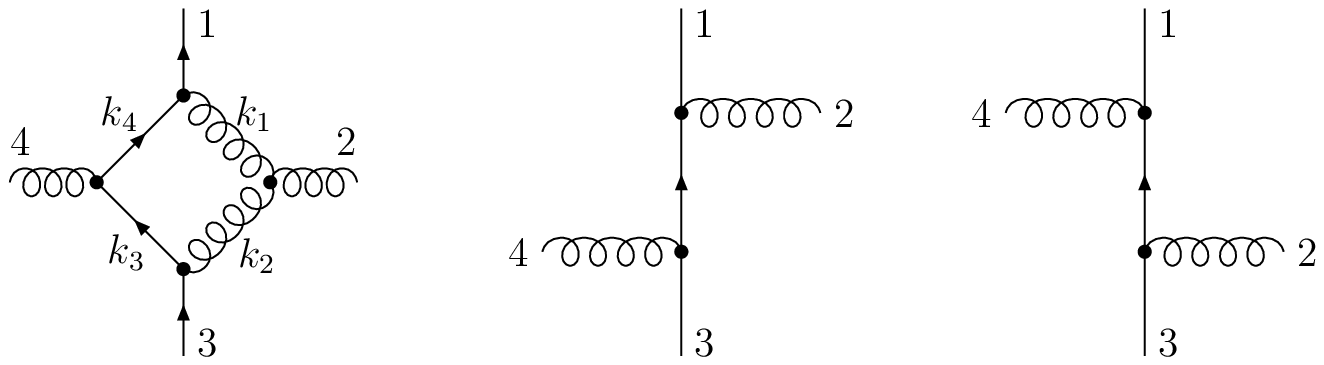}
\caption{\label{figure_subleading_tree}
The left diagram shows the top-level diagram for the one-loop primitive $q g \bar{q} g$ amplitude.
The tree-amplitude entering the infrared subtraction terms is given by the sum of the two diagrams
on the right.
}
\end{center}
\end{figure}
Let us consider the left-moving one-loop primitive amplitude $A^{(1)}(q_1,g_2,\bar{q}_3,g_4)$ with the cyclic order
$q_1,g_2,\bar{q}_3,g_4$. The top-level one-loop diagram for this amplitude is shown in the left diagram of 
fig.~(\ref{figure_subleading_tree}).
This primitive one-loop amplitude has soft singularities, when either the gluon in the loop with momentum
$k_1$ or the one with momentum $k_2$ becomes soft.
The amplitude has collinear singularities when either $k_4||k_1$, $k_1||k_2$ or $k_2||k_3$.
In all these cases the infrared subtraction terms are proportional to the tree-level partial amplitude
with the cyclic order $q_1,g_2,\bar{q}_3,g_4$.
The two tree diagrams contributing to this tree-level partial amplitude are shown in the middle diagram
and in the right diagram of fig.~(\ref{figure_subleading_tree}).
Note that these diagrams only involve the quark-gluon vertex, but no three-gluon vertex.
On the other hand the colour-decomposition of the leading order tree-level amplitude is given in 
eq.~(\ref{colour_decomp_qqbar_gluon}) and involves the partial amplitudes with the cyclic order
\bq
\label{partial_amplitudes_qggqbar}
 A^{(0)}\left( q_1, g_2, g_4, \bar{q}_3 \right),
 & &
 A^{(0)}\left( q_1, g_4, g_2, \bar{q}_3 \right).
\eq
In these two partial amplitudes the quark and the anti-quark are always adjacent in the cyclic order.
\begin{figure}
\begin{center}
\includegraphics[bb= 190 610 460 722]{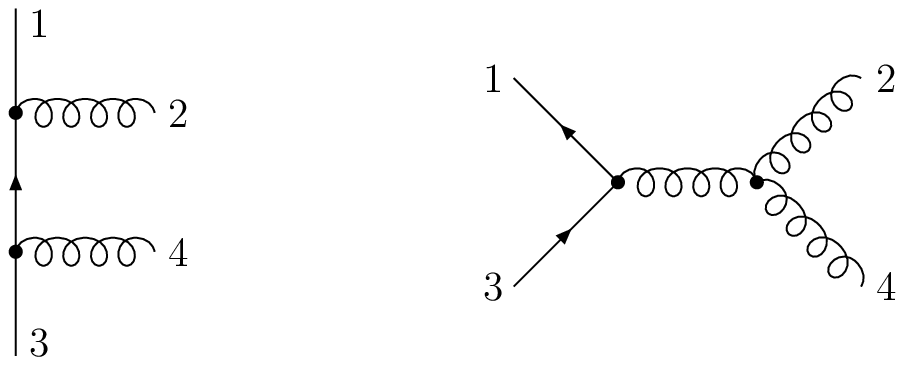}
\caption{\label{figure_tree_qggqbar}
The diagrams contributing to the partial amplitude $A^{(0)}( q_1, g_2, g_4, \bar{q}_3 )$.
}
\end{center}
\end{figure}
Each of these partial amplitudes consists of two Feynman diagram. For the amplitude 
$A^{(0)}( q_1, g_2, g_4, \bar{q}_3 )$ the diagrams are shown in fig.~(\ref{figure_tree_qggqbar}).
The corresponding diagrams for the amplitude $A^{(0)}( q_1, g_4, g_2, \bar{q}_3 )$
are obtained by exchanging $g_2 \leftrightarrow g_4$.

The tree-level partial amplitude $A^{(0)}(q_1,g_2,\bar{q}_3,g_4)$ can be calculated (as any other tree-level
partial amplitude) with the recursive techniques of subsection~\ref{subsect:recurrence}.
However, if the partial amplitudes in eq.~(\ref{partial_amplitudes_qggqbar}) are already known,
the partial amplitude $A^{(0)}(q_1,g_2,\bar{q}_3,g_4)$ is simply given as
\bq
 A^{(0)}\left(q_1,g_2,\bar{q}_3,g_4\right)
 & = &
 A^{(0)}\left( q_1, g_2, g_4, \bar{q}_3 \right)
 +
 A^{(0)}\left( q_1, g_4, g_2, \bar{q}_3 \right).
\eq
In the sum on the right-hand side the diagrams with the non-abelian three-gluon vertex drop out, due to the anti-symmetry
of the colour ordered three-gluon vertex.

This is easily generalised to amplitudes with more gluons: For an amplitude 
with the cyclic order $q,g_1,...,g_m,\bar{q},g_{m+1},...,g_n$ we have
\bq
\label{subleading_qqbar_n_gluon}
 A^{(0)}(q,g_1,...,g_m,\bar{q},g_{m+1},...,g_n)
 & = & 
 \sum\limits_{\sigma \in \{1,...,m\} \Sha \{n,...,m+1\}}
 A^{(0)}\left(q,g_{\sigma_1},...,g_{\sigma_n},\bar{q}\right).
\eq
This expresses the amplitude with the cyclic order $q,g_1,...,g_m,\bar{q},g_{m+1},...,g_n$
in terms of amplitudes with the cyclic order $q,g_{\sigma_1},...,g_{\sigma_n},\bar{q}$, in which the quark and
the anti-quark are adjacent.
The sum is over all shuffles of the set $\{1,...,m\}$ with the set $\{n,...,m+1\}$. A shuffle is a permutation of the $n$
elements, which preserves the relative order of the elements in the first set, as well as the relative order of the 
elements in the second set.
In order to prove the relation in eq.~(\ref{subleading_qqbar_n_gluon}) one observes that in the sum on the right-hand side
all non-abelian vertices involving gluons from both sets drop out.
Let $P_a$ and $P_b$ denote momenta, which are the sums of momenta of a subset from $\{p_1,...,p_m\}$.
Similar, we denote by $P_i$ a momentum, which is the sum of momenta of a subset from $\{p_{m+1},...,p_n\}$.
In the sum over shuffles each three-point vertex connected to lines with the (ordered) momenta $P_a,P_i$
will also occur in the order $P_i,P_a$.
Due to the antisymmetric nature of the colour ordered three-point vertex this will add up to zero.
For the four-point vertices we have a similar situation.
Four-point vertices with mixed momenta occur always in the combination $P_a,P_b,P_i$, in the combination $P_a,P_i,P_b$ 
as well as in the combination $P_i,P_a,P_b$.
Again, the three combinations will add up to zero.

\section{Checks and examples}
\label{sect:examples}

In this section we discuss a few checks and examples.
One of the simplest processes where our method can be applied are the NLO corrections to
the process $\gamma^\ast \rightarrow \bar{q} q$. This tests the locality of the subtraction terms
and is discussed in subsection~\ref{subsect:gamma_qbarq}.
In subsection~\ref{subsect:triangle} we describe a strong check for the contour deformation.
We study the accuracy by which we can evaluate numerically the one-loop three point function with massless internal propagators
and massive external lines. This integral is finite and we compute this integral for a contour based on a process
with $n$ external legs. We can think of this integral as occurring in an amplitude with $n$ external legs
where $(n-3)$ propagators are pinched.

\subsection{The process $\gamma^\ast \rightarrow q \bar{q}$}
\label{subsect:gamma_qbarq}

The NLO corrections to the process $\gamma^\ast \rightarrow q \bar{q}$
is a simple example where our method can be applied.
We discuss this process in detail. This example illustrates nicely how the subtraction terms
cancel the divergences locally.

The Born amplitude squared for $\gamma^\ast \rightarrow q(p_1) + \bar{q}(p_2)$ is given
in $D=4-2\eps$ dimensions by
\bq
\left| {\cal A}^{(0)}_3 \right|^2 
 & = & 
 4 N_c Q_q^2 e^2 \left( 1 - \eps \right) s,
\eq
with $s=s_{12}$. Here we use the notation $s_{ij} = (p_i+p_j)^2$.
The elementary electric charge is denoted by $e$ and $Q_q$ gives the electric charge of the quarks in units of
the elementary electric charge $e$.

The virtual correction is given with $k_1=k-p_1$, $k_2=k-p_1-p_2$ and $k_3=k$ by
\bq
2 \mbox{Re}\; {{\cal A}^{(0)}_3}^\ast {\cal A}^{(1)}_3
 & = & 
 2 \mbox{Re}\; C_F N_c Q_q^2 e^2 g^2 S_\eps^{-1} \mu^{2\eps}
 \int \frac{d^Dk}{(2\pi)^Di} 
\frac{-\mbox{Tr}\; p\!\!\!/_2 \gamma_\mu p\!\!\!/_1 \gamma_\nu k\!\!\!/_3 \gamma^\mu k\!\!\!/_2 \gamma^\nu}{k_1^2 k_2^2 k_3^2}.
\eq
Using momentum conservation and the Dirac algebra in $D$ dimensions this is equivalent to
\bq
\lefteqn{
2 \mbox{Re}\; {{\cal A}^{(0)}_3}^\ast {\cal A}^{(1)}_3
 = } & &
 \nonumber \\
 & & 
\left| {\cal A}^{(0)}_3 \right|^2 
 C_F g^2
 2 \mbox{Re}\; S_\eps^{-1} \mu^{2\eps}
 \int \frac{d^Dk}{(2\pi)^Di} 
 \frac{1}{k_1^2 k_2^2 k_3^2}
 \left\{
  -2s_{12} - 3 k_1^2 + 2 k_2^2 + 2 k_3^2 + (D-4) k_1^2
\right. \nonumber \\
 & & \left.
 - \frac{1}{s_{12}} \left[ 2p_1\cdot k_3 \; k_2^2 - 2 p_2 \cdot k_2 \; k_3^2 \right]
 - \frac{1}{2s_{12}} \left[ 2(p_2-p_1)\cdot k_2 + 2(p_2-p_1)\cdot k_3 \right] k_1^2 
 \right\}.
\eq
In an analytical calculation all terms in the last line will yield zero after integration.
The infrared subtraction terms are given by
\bq
\lefteqn{
2 \mbox{Re}\; {{\cal A}^{(0)}_3}^\ast \left( {\cal A}^{(1)}_{\mathrm{soft}} + {\cal A}^{(1)}_{\mathrm{coll}} \right)
 = } & &
 \\
 & & 
\left| {\cal A}^{(0)}_3 \right|^2 
 C_F g^2
 2 \mbox{Re}\; S_\eps^{-1} \mu^{2\eps}
 \int \frac{d^Dk}{(2\pi)^Di} 
 \left[
 -\frac{2s_{12}}{k_1^2 k_2^2 k_3^2}
 +\frac{2}{k_1^2 k_3^2}
 +\frac{2}{k_1^2 k_2^2}
 -\frac{4}{\left(\bar{k}^2-\mu_{\mathrm{UV}}^2\right)^2}
 \right].
\nonumber
\eq
As before we have set $\bar{k}=k-Q$.
As ultraviolet subtraction term we need only a subtraction term for the quark-photon vertex. The form of this subtraction term
is identical to the subtraction term for the subleading colour contribution for the quark-gluon vertex.
The latter has been given in eq.~(\ref{uv_subtr_quark_gluon_sc}).
We have
\bq
\lefteqn{
2 \mbox{Re}\; {{\cal A}^{(0)}_3}^\ast {\cal A}^{(1)}_{\mathrm{UV}}
 = } & &
 \\
 & & 
\left| {\cal A}^{(0)}_3 \right|^2 
 C_F g^2
 2 \mbox{Re}\; S_\eps^{-1} \mu^{2\eps}
 \int \frac{d^Dk}{(2\pi)^Di} 
 \frac{1}{\left(\bar{k}^2-\mu_{\mathrm{UV}}^2\right)^3}
 \left[
 \frac{2}{s_{12}} \left( 2p_1\cdot \bar{k} \right) \left( 2 p_2 \cdot \bar{k} \right) - 2 \eps \bar{k}^2 - 4 \mu_{\mathrm{UV}}^2 
 \right].
 \nonumber
\eq
The subtracted one-loop amplitude is finite and we may take the limit $D \rightarrow 4$. We obtain
\bq
\label{example_numerical_integral_virt}
\lefteqn{
2 \mbox{Re}\; {{\cal A}^{(0)}_3}^\ast \left( {\cal A}^{(1)}_3 - {\cal A}^{(1)}_{\mathrm{soft}} - {\cal A}^{(1)}_{\mathrm{coll}} - {\cal A}^{(1)}_{\mathrm{UV}}\right)
 = 
\left| {\cal A}^{(0)}_3 \right|^2 
 C_F g^2
} & &
 \nonumber \\
 & & 
 2 \mbox{Re}\; 
 \int \frac{d^4k}{(2\pi)^4i} 
 \left\{
 -\frac{1}{k_2^2 k_3^2} \left[ 3 + \frac{2\left(p_2-p_1\right)\cdot\left(k_2+k_3\right)}{2 s_{12}} \right]
 -\frac{2p_1\cdot k_3}{s_{12} k_1^2 k_3^2}
 +\frac{2p_2\cdot k_2}{s_{12} k_1^2 k_2^2}
 \right. \nonumber \\
 & & 
 \left.
 + \frac{2}{\left(\bar{k}^2-\mu_{\mathrm{UV}}^2\right)^3}
   \left[ 2 \bar{k}^2 - \frac{\left( 2 p_1 \cdot \bar{k} \right) \left( 2 p_2 \cdot \bar{k} \right) }{s_{12}} \right]
 \right\}
 + {\cal O}\left(\eps\right).
\eq
This integral is finite and within the numerical approach it will be evaluated with 
Monte Carlo techniques. For the example discussed here, the integral is rather simple and
can also easily be evaluated analytically.
The analytical result for this integral is given by
\bq
2 \mbox{Re}\; {{\cal A}^{(0)}_3}^\ast \left( {\cal A}^{(1)}_3 - {\cal A}^{(1)}_{\mathrm{soft}} - {\cal A}^{(1)}_{\mathrm{coll}} - {\cal A}^{(1)}_{\mathrm{UV}}\right)
 = 
 \frac{\alpha_s}{2\pi} C_F \left| {\cal A}^{(0)}_3 \right|^2 
 \left( -8 + 3 \;\mbox{Re} \; \ln \frac{-s}{\mu_{\mathrm{UV}}^2} \right)
 + {\cal O}\left(\eps\right).
\eq
We can therefore compare the numerical result with the exact analytical result.
With the methods of section~\ref{sect:contour_deformation} we easily achieve with a numerical Monte Carlo integration
a precision at the per mille level for the integral in eq.~(\ref{example_numerical_integral_virt}).

The subtraction terms for the integrand of the one-loop amplitude are added back in integrated form. 
For the process $\gamma^\ast \rightarrow q \bar{q}$ we have in addition
that there are no ultraviolet counterterms from renormalisation, i.e.
\bq
{\cal A}^{(1)}_{\mathrm{CT}} & = & 0.
\eq
We therefore have
\bq
\lefteqn{
2 \mbox{Re}\; {{\cal A}^{(0)}_3}^\ast 
 \left( {\cal A}^{(1)}_{\mathrm{CT}} + {\cal A}^{(1)}_{\mathrm{soft}} + {\cal A}^{(1)}_{\mathrm{coll}} + {\cal A}^{(1)}_{\mathrm{UV}}\right)
 = } & &
 \\
 & & 
 \frac{\alpha_s}{2\pi} C_F
 \left| {\cal A}^{(0)}_3 \right|^2 
 \mbox{Re}\; 
 \left[
         - \frac{2}{\eps^2} + \frac{1}{\eps}\left( -3 + 2 \ln \frac{-s}{\mu^2} \right)
         - \ln^2 \frac{-s}{\mu^2} + \frac{\pi^2}{6} 
         + 3 \ln \frac{\mu_{\mathrm{UV}}^2}{\mu^2}
 \right]
 + {\cal O}\left(\eps\right).
 \nonumber
\eq
This defines the insertion operator ${\bf L}$ for the process $\gamma^\ast \rightarrow q \bar{q}$ as
\bq
 {\bf L} = 
 \frac{\alpha_s}{2\pi} C_F
 \mbox{Re}\; 
 \left[
         - \frac{2}{\eps^2} + \frac{1}{\eps}\left( -3 + 2 \ln \frac{-s}{\mu^2} \right)
         - \ln^2 \frac{-s}{\mu^2} + \frac{\pi^2}{6}
         + 3 \ln \frac{\mu_{\mathrm{UV}}^2}{\mu^2}
 \right]
 + {\cal O}\left(\eps\right).
\eq
Let us also consider the real correction.
The Born amplitude squared for $\gamma^\ast \rightarrow q(p_1) + g(p_2) + \bar{q}(p_3)$ is
\bq
\left| {\cal A}^{(0)}_4 \right|^2 
 & = & 
 C_F N_c Q_q^2 e^2 g^2 
 8 ( 1 - \eps) 
 \nonumber \\
 & &
 \left[
         2 \frac{s_{123}^2}{s_{12} s_{23}} - 2 \frac{s_{123}}{s_{12}}
                                                    - 2 \frac{s_{123}}{s_{23}} 
         +(1-\eps) \frac{s_{12}}{s_{23}} + (1-\eps) \frac{s_{23}}{s_{12}}
         - 2 \eps 
        \right],
\eq
where we used the notation $s_{ijk}=(p_i+p_j+p_k)^2$.
The dipole subtraction terms read
\bq
 {\cal D}_{qg,\bar{q}} + {\cal D}_{\bar{q}g,q}
 & = &
 C_F N_c Q_q^2 e^2 g^2 
 8 ( 1 - \eps) \left\{
 \left[ 2 \frac{s_{123}^2}{s_{12} \left( s_{12} + s_{23} \right)} - 2 \frac{s_{123}}{s_{12}} 
        + \left(1-\eps\right) \frac{s_{23} s_{123}}{s_{12} \left(s_{13}+s_{23}\right)} \right]
 \right. \nonumber \\
 & & \left.
 +
 \left[ 2 \frac{s_{123}^2}{s_{23} \left( s_{12} + s_{23} \right)} - 2 \frac{s_{123}}{s_{23}} 
        + \left(1-\eps\right) \frac{s_{12} s_{123}}{s_{23} \left(s_{12}+s_{13}\right)} \right]
 \right\}.
\eq
For the real corrections one evaluates numerically the finite integral
\bq
\label{example_numerical_integral_real}
\lefteqn{
 \int d\phi_{\mathrm{unres}} \left( \left| {\cal A}^{(0)}_4 \right|^2 - {\cal D}_{qg,\bar{q}} - {\cal D}_{\bar{q}g,q} \right)
 =  } & & \\
 & &
 8 C_F N_c Q_q^2 e^2 g^2 
 \int d\phi_{\mathrm{unres}} 
  \left(
  - \frac{s_{12}}{s_{12}+s_{13}} - \frac{s_{23}}{s_{13}+s_{23}}
        \right)
 + {\cal O}\left(\eps\right)
 \nonumber
\eq
in four dimensions.
The unresolved phase space measure is given in $D$ dimensions by
\bq
 d\phi_{\mathrm{unres}} & = &
\frac{(4\pi)^{\eps-2}}{\Gamma\left(1-\eps\right)}  
\left( s_{123} \right)^{1-\eps} 
\int d^3 x \delta\left( 1 - \sum\limits_{i=1}^3 x_i \right) 
 x_1^{-\eps} x_2^{-\eps} x_3^{-\eps},
\eq
where
\bq
x_1 = \frac{s_{12}}{s_{123}},
\;\;\;
x_2 = \frac{s_{23}}{s_{123}},
\;\;\;
x_3 = \frac{s_{13}}{s_{123}}.
\eq
For the simple example considered here the integral in eq.~(\ref{example_numerical_integral_real}) can also be done analytically. 
The result reads with $s=s_{123}$
\bq
 \int d\phi_{\mathrm{unres}} \left( \left| {\cal A}^{(0)}_4 \right|^2 - {\cal D}_{qg,\bar{q}} - {\cal D}_{\bar{q}g,q} \right)
 & = &
 \frac{\alpha_s}{2\pi} C_F
 \left| {\cal A}^{(0)}_3 \right|^2 
 \cdot \left( - \frac{1}{2} \right)
 + {\cal O}\left(\eps\right).
\eq
Integrating the dipole subtraction terms analytically in $D$ dimensions one finds
\bq
\lefteqn{
 S_\eps^{-1} \mu^{2\eps} \int d\phi_{\mathrm{unres}} \left( {\cal D}_{qg,\bar{q}} + {\cal D}_{\bar{q}g,q} \right)
 = } & & 
 \\
 & &
 \frac{\alpha_s}{2\pi} C_F
 \left| {\cal A}^{(0)}_3 \right|^2 
 \left[
         \frac{2}{\eps^2} + \frac{1}{\eps} \left( 3 - 2 \ln \frac{s}{\mu^2} \right)
         + \ln^2 \frac{s}{\mu^2} - 3 \ln \frac{s}{\mu^2}
         + 10 - \frac{7}{6} \pi^2
\right]
 + {\cal O}\left(\eps\right).
 \nonumber
\eq
We may now sum up the integrated subtraction terms from the virtual part and the real part.
This yields the expression
\bq
\label{example_I_L}
\lefteqn{
2 \mbox{Re}\; {{\cal A}^{(0)}_3}^\ast 
 \left( {\cal A}^{(1)}_{\mathrm{CT}} + {\cal A}^{(1)}_{\mathrm{soft}} + {\cal A}^{(1)}_{\mathrm{coll}} + {\cal A}^{(1)}_{\mathrm{UV}}\right)
+
 S_\eps^{-1} \mu^{2\eps} \int d\phi_{\mathrm{unres}} \left( {\cal D}_{qg,\bar{q}} + {\cal D}_{\bar{q}g,q} \right)
 = } & &
 \nonumber \\
 & &
\hspace*{60mm}
 \frac{\alpha_s}{2\pi} C_F
 \left| {\cal A}^{(0)}_3 \right|^2 
 \left(
         10 - 3 \; \mbox{Re}\; \ln \frac{s}{\mu_{\mathrm{UV}}^2}
\right)
 + {\cal O}\left(\eps\right),
\eq
which corresponds to the sum of the insertion operators ${\bf I}+{\bf L}$. 
We thus have for the process $\gamma^\ast \rightarrow q \bar{q}$
\bq
 {\bf I}+{\bf L}
 & = &
 \frac{\alpha_s}{2\pi} C_F
 \left(
         10 - 3 \; \mbox{Re}\; \ln \frac{s}{\mu_{\mathrm{UV}}^2}
\right)
 + {\cal O}\left(\eps\right).
\eq
Note that this expression is finite.

In summary the complete NLO corrections to the process $\gamma^\ast \rightarrow q \bar{q}$
is the sum of three contributions, which are given in eq.~(\ref{example_numerical_integral_virt}), 
eq.~(\ref{example_numerical_integral_real}) and
eq.~(\ref{example_I_L}), respectively.
Each contribution is individually finite.
Summing them up one obtains the well-known result
\bq
2 \mbox{Re}\; {{\cal A}^{(0)}_3}^\ast {\cal A}^{(1)}_3
+
 S_\eps^{-1} \mu^{2\eps} \int d\phi_{\mathrm{unres}} \left| {\cal A}^{(0)}_4 \right|^2
 & = &
 \frac{3}{2} \cdot \frac{\alpha_s}{2\pi} C_F
 \left| {\cal A}^{(0)}_3 \right|^2
 + {\cal O}\left(\eps\right).
\eq

\subsection{Precision and accuracy of the contour integration}
\label{subsect:triangle}

The previous example shows nicely the local cancellation of all divergent parts.
However, the virtual correction of the previous example has only three external legs
and the resulting numerical integration in eq.~(\ref{example_numerical_integral_virt}) 
is rather trivial.
Complications due to a large number of external particles do not enter the previous example.
In order to test our method in situations with a large number of external particles 
we start by considering a configuration with
$n$ external particles. 
The top-level one-loop diagram associated with this configuration is
a diagram with $n$ internal loop propagators.
If we now pinch $(n-3)$ of the $n$ internal loop propagators, we obtain a
one-loop three-point function.
\begin{figure}
\begin{center}
\includegraphics[bb= 135 610 485 722]{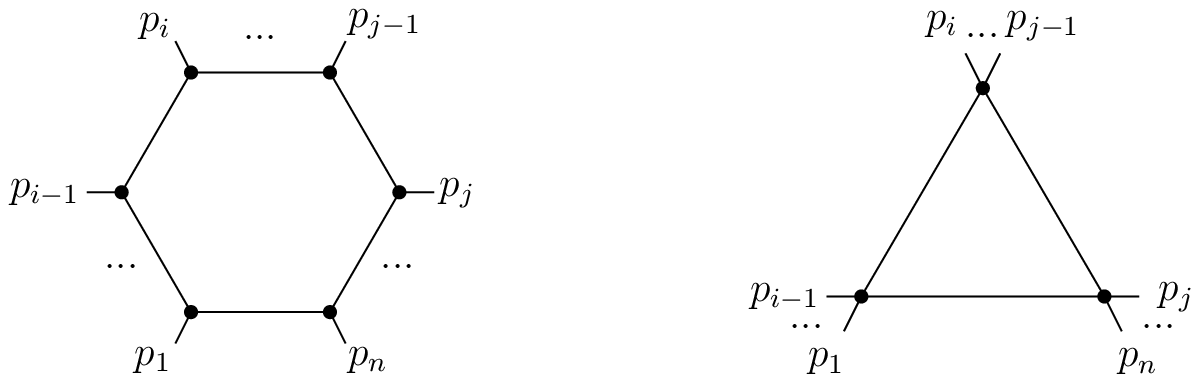}
\caption{\label{figure_triangle}
The diagram on the left shows the top-level diagram for a process with $n$ external legs.
Pinching $(n-3)$ loop propagators results in a three-point function, shown on the right.
}
\end{center}
\end{figure}
This is illustrated in fig.~(\ref{figure_triangle}).
The three-point function has the external momenta
\bq
 P_1 = p_1 + ... + p_{i-1},
 \;\;\;\;\;\;
 P_2 = p_i + ... + p_{j-1},
 \;\;\;\;\;\;
 P_3 = p_j + ... + p_n.
\eq
We now consider the massless one-loop scalar three-point function
\bq
 I & = &
 16 \pi^2 
 \int \frac{d^4k}{(2\pi)^4i} \frac{1}{k^2\left(k-P_1\right)^2\left(k-P_1-P_2\right)^2}
\eq
in the case $P_1^2 \neq 0$, $P_2^2 \neq 0$ and $P_3^2 \neq 0$.
This integral is finite and the analytical result is well known \cite{Ussyukina:1993jd,Lu:1992ny,Bern:1997ka}.
With the notation
\bq
 & &
 \delta_1 = P_1^2 - P_2^2 - P_3^2,
 \;\;\;
 \delta_2 = P_2^2 - P_3^2 - P_1^2,
 \;\;\;
 \delta_3 = P_3^2 - P_1^2 - P_2^2,
 \nonumber \\
 & &
 \Delta_3 = \left(P_1^2\right)^2 + \left(P_2^2\right)^2 + \left(P_3^2\right)^2 
               - 2 P_1^2 P_2^2 - 2 P_2^2 P_3^2 - 2 P_3^2 P_1^2,
\eq
the three-mass triangle $I$ is given in the region
$P_1^2,P_2^2,P_3^2 < 0$ and $\Delta_3 < 0$  by 
\bq
\lefteqn{
 I = 
   -\frac{2}{\sqrt{-\Delta_3}} 
} & & \\
 & & 
   \times \left[
   \mbox{Cl}_2\left( 2 \arctan \left( \frac{\sqrt{-\Delta_3}}{\delta_1}\right)\right)
   +\mbox{Cl}_2\left( 2 \arctan \left( \frac{\sqrt{-\Delta_3}}{\delta_2}\right)\right) 
   +\mbox{Cl}_2\left( 2 \arctan \left( \frac{\sqrt{-\Delta_3}}{\delta_3}\right)\right)
   \right].
 \nonumber 
\eq
$\mbox{Cl}_2(x)$ denotes the Clausen function.
In  the region $P_1^2,P_2^2,P_3^2 < 0$ and $\Delta_3 > 0$ as well as in the region
$P_1^2, P_3^2 < 0$, $P_2^2 > 0$ (for which $\Delta_3$ is always positive) the integral
$I$ is given by 
\bq
 I & = & \frac{1}{\sqrt{\Delta_3}} \mbox{Re} \left[
       2 \left( \mbox{Li}_2(-\rho x)+\mbox{Li}_2(-\rho y) \right) + \ln(\rho x) \ln(\rho y) 
       + \ln \left( \frac{y}{x} \right) \ln \left( \frac{1+\rho x}{1+\rho y} \right) + \frac{\pi^2}{3} \right] 
 \nonumber \\
 & & 
       + \frac{i \pi \theta(P_2^2)}{\sqrt{\Delta_3}} \ln \left( 
         \frac{\left(\delta_1+\sqrt{\Delta_3}\right) \left(\delta_3+\sqrt{\Delta_3}\right)}
              {\left(\delta_1-\sqrt{\Delta_3}\right) \left(\delta_3-\sqrt{\Delta_3}\right)} \right),
\eq
where
\bq
 x = \frac{P_1^2}{P_3^2}, \;\;\;\; y = \frac{P_2^2}{P_3^2}, \;\;\;\;
 \rho = \frac{2 P_3^2}{\delta_3+\sqrt{\Delta_3}}.
\eq
The step function $\theta(x)$ is defined as $\theta(x)=1$ for $x>0$ and $\theta(x)=0$ otherwise.
Equipped with the analytical result we may now test the 
precision and accuracy of the numerical integration.
We can think of the three-point function as a finite lower-point function occurring in the 
calculation of an $n$-point amplitude.
Therefore we have to check if the numerical integration is precise and accurate with respect
to a contour which avoids all the singular surfaces defined by the $n$ propagators
\bq
 \left( k - q_i \right)^2 = 0,
 & & i=1,..,n,
\eq
and not just the three propagators
\bq
 k^2=0,
 \;\;\;\;\;\;
 \left( k - P_1 \right)^2=0,
 \;\;\;\;\;\;
 \left( k - P_1 - P_2 \right)^2=0.
\eq
The former is more challenging than the latter.
In general, the three-point integral $I$ evaluates to a complex number.
We test the real part and the imaginary part separately.

We start with a configuration corresponding to just $n=3$ external particles and then increase $n$
step by step.
All numerical integrations are done with the help of the 
Vegas algorithm \cite{Lepage:1978sw,Lepage:1980dq}. 
After a warm-up phase with five iterations of $10^5$ evaluations each, we obtain
the numerical result from 20 iterations of $10^6$ evaluations each.
Up to $n=5$ we find no problems in obtaining the correct answer from the numerical Monte Carlo integration with
a precision better than one per cent by using just the methods of section~\ref{subsect:contour_feynman} alone.
However, starting from $n=6$ we find numerical instabilities (and as consequence large statistical errors) by using just
the methods of section~\ref{subsect:contour_feynman}.
The situation is significantly improved with the help of the method of section~\ref{subsect:stability}.
This method introduces two parameters $\mu^2_{\mathrm{IR}}$ and $N_{\mathrm{IR}}$.
We recall that we choose $\mu^2_{\mathrm{IR}}$ as a purely imaginary number with $\mbox{Im}\;\mu^2_{\mathrm{IR}}<0$.
It is convenient to parametrise $\mu^2_{\mathrm{IR}}$ by a dimensionless number $\eta_{\mathrm{IR}}$ 
through
\bq
 \mu^2_{\mathrm{IR}} & = & - i \eta_{\mathrm{IR}}^2 Q^2,
\eq
where $Q$ is the centre-of-mass energy of the process under consideration.
\begin{figure}
\begin{center}
\includegraphics[bb= 135 460 485 710]{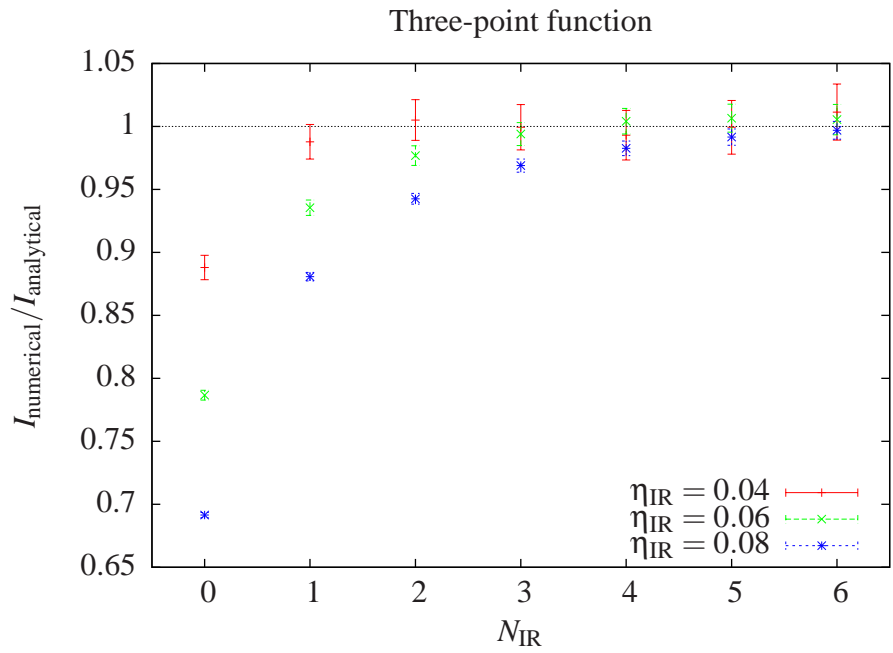}
\caption{\label{figure_res_triangle}
Results for the numerical integration of the three-point function 
normalised to the analytical result for a contour based on six
external particles for various values of $\eta_{\mathrm{IR}}$ and $N_{\mathrm{IR}}$.
The errorbars indicate the statistical error of the Monte Carlo integration.
}
\end{center}
\end{figure}
Fig.~(\ref{figure_res_triangle}) shows the result of the numerical integration 
normalised to the analytical result for $n=6$ external momenta for various values
of $\eta_{\mathrm{IR}}$ and $N_{\mathrm{IR}}$.
We have plotted the results for $\eta_{\mathrm{IR}}=0.04$, $\eta_{\mathrm{IR}}=0.06$ and $\eta_{\mathrm{IR}}=0.08$
as a function of $N_{\mathrm{IR}}$.
In all cases the correct value is approached as we increase $N_{\mathrm{IR}}$.
As expected, we observe that the series converges faster for smaller values of $\eta_{\mathrm{IR}}$.
However, as can be inferred from the errorbars in fig.~(\ref{figure_res_triangle})
smaller values of $\eta_{\mathrm{IR}}$ imply larger statistical errors.
By a suitable choice of $\eta_{\mathrm{IR}}$ and $N_{\mathrm{IR}}$ one obtains again a precision better than one per cent.
We have tested the numerical accuracy and precision of the three-point function with a contour 
up to $n=9$ propagators in the loop.

\section{Conclusions}
\label{sect:conclusions}

In this paper we have presented a complete algorithm for the numerical calculation of one-loop 
QCD amplitudes.
The algorithm consists of two main ingredients. First a set of subtraction terms, which approximate
locally the soft, collinear and ultraviolet divergences of one-loop amplitudes.
The integrals over these subtraction terms are simple and calculated analytically 
in $D$ dimensions once and for all.
The difference between the loop amplitude and the subtraction terms is finite and can be evaluated numerically
in four dimensions.

Secondly, the algorithm consists of a method to deform the integration contour into the complex space.
Here we add for a process with $n$ external particles to the four dimensions of momentum space
$n$ additional dimensions corresponding to the Feynman parameters.
This leads to a simple and general prescription for the contour deformation.
In order to improve the stability of the numerical Monte Carlo integration we expand all
propagators in the loop around propagators where a small fictitious width has been added.

The algorithm is formulated at the amplitude level and does not rely on Feynman graphs.
Therefore all required ingredients can be calculated efficiently using recurrence relations.
The algorithm applies to massless partons as well as to massive partons.

Although we have presented here the algorithm for QCD amplitudes, the extension to the 
electroweak sector and the Higgs sector, as well as to physics beyond the standard model seems to be
straightforward.

\subsection*{Acknowledgements}

Sebastian Becker acknowledges the support of the DFG through 
the ``Graduiertenkolleg Symmetriebrechung in fundamentalen Wechselwirkungen''.
Christian Reuschle is grateful for the support from the
``Schwerpunkt Rechnergest\"utzte Forschungsmethoden in den Naturwissenschaften''.

\begin{appendix}

\section{Colour ordered Feynman rules}
\label{sect:feynman_rules}

In this appendix we give a list of the colour ordered Feynman rules. 
They are obtained from the standard Feynman rules by extracting from each formula the coupling
constant and the colour part.
The propagators for quark, gluon and ghost particles are given by
\bq
\begin{picture}(85,20)(0,5)
 \ArrowLine(70,10)(20,10)
\end{picture} 
 & = &
 i\frac{k\!\!\!/+m}{k^2-m^2},
 \nonumber \\
\begin{picture}(85,20)(0,5)
 \Gluon(20,10)(70,10){-5}{5}
\end{picture} 
& = &
 \frac{-ig^{\mu\nu}}{k^2},
 \nonumber \\
\begin{picture}(85,20)(0,5)
 \DashArrowLine(70,10)(20,10){3}
\end{picture} 
 & = &
 \frac{i}{k^2}.
\eq
The colour ordered Feynman rules for the vertices are
\bq
\begin{picture}(100,35)(0,55)
\Vertex(50,50){2}
\Gluon(50,50)(50,80){3}{4}
\ArrowLine(80,50)(50,50)
\ArrowLine(50,50)(20,50)
\Text(50,82)[b]{\small $\mu$}
\end{picture}
 & = &
-i \gamma^\mu,
 \\ \nonumber
 \\ \nonumber
\begin{picture}(100,35)(0,55)
\Vertex(50,50){2}
\Gluon(50,50)(50,80){3}{4}
\Gluon(50,50)(76,35){3}{4}
\Gluon(50,50)(24,35){3}{4}
\LongArrow(56,70)(56,80)
\LongArrow(67,47)(76,42)
\LongArrow(33,47)(24,42)
\Text(60,80)[lt]{$k_1^\mu$}
\Text(78,35)[lc]{$k_2^\nu$}
\Text(22,35)[rc]{$k_3^\lambda$}
\end{picture}
 & = &
 i \left[ g^{\mu\nu} \left( k_2^\lambda - k_1^\lambda \right)
         +g^{\nu\lambda} \left( k_3^\mu - k_2^\mu \right)
         +g^{\lambda\mu} \left( k_1^\nu - k_3^\nu \right)
   \right],
 \nonumber \\
 \nonumber \\
 \nonumber \\
\begin{picture}(100,35)(0,55)
\Vertex(50,50){2}
\Gluon(50,50)(71,71){3}{4}
\Gluon(50,50)(71,29){3}{4}
\Gluon(50,50)(29,29){3}{4}
\Gluon(50,50)(29,71){3}{4}
\Text(72,72)[lb]{\small $\mu$}
\Text(72,28)[lt]{\small $\nu$}
\Text(28,28)[rt]{\small $\lambda$}
\Text(28,72)[rb]{\small $\rho$}
\end{picture}
 & = &
  i \left[
        2 g^{\mu\lambda} g^{\nu\rho} - g^{\mu\nu} g^{\lambda\rho} 
                                     - g^{\mu\rho} g^{\nu\lambda}
 \right],
 \nonumber \\
 \nonumber \\
 \nonumber \\
\begin{picture}(100,35)(0,55)
\Vertex(50,50){2}
\Gluon(50,50)(80,50){3}{4}
\DashArrowLine(50,50)(29,71){3}
\DashArrowLine(29,29)(50,50){3}
\LongArrow(36,59)(29,66)
\Text(82,50)[lc]{$\mu$}
\Text(28,71)[rb]{$k$}
\end{picture}
 & = &
 -i k_{\mu}.
 \nonumber \\
 \nonumber \\
\eq

\section{Generating the random points}
\label{sect:generating_random_points}

The numerical integration discussed in section~\ref{sect:contour_deformation} is over
the four-dimensional loop momentum space and an $(n-1)$-dimensional Feynman parameter space.
After the contour integration we integrate over real variables.
The integration over the four-dimensional loop momentum space extends for each dimension
from minus infinity to plus infinity.
The integration region over the Feynman parameters is an $(n-1)$-dimensional simplex embedded
into an $n$-dimensional space.
In order to be able to use standard methods for the Monte Carlo integration we map these domains
to the unit hyper-cube.

\subsection{Generating the loop momentum}
\label{sect:generating_loop_momentum}

In this appendix we show how to rewrite the integral
\bq
 \int d^4k f(k),
\eq
where the integration is over four real variables $k_0$, $k_1$, $k_2$ and $k_3$
from minus infinity to plus infinity,
as an integral over the finite domain $[0,1]^4$.
We use four uniformly in $[0,1]$ distributed random numbers $u_1$, ..., $u_4$ and
define the four quantities $k_E$, $\xi$, $\theta$ and $\phi$ by the equations
\bq
 k_E & = & \mu_1 \sqrt{ \tan \frac{\pi}{2} u_1 },
 \nonumber \\
 \frac{1}{\pi} \left( \xi - \sin \xi \cos \xi \right) - u_2 & = & 0,
 \nonumber \\
 \theta & = & \arccos\left(1-2u_3\right),
 \nonumber \\
 \phi & = & 2 \pi u_4.
\eq
$\mu_1$ is an arbitrary scale. 
The second equation is solved numerically for $\xi$.
We then set
\bq
 k_0 = k_E \cos \xi,
 \;\;\;
 k_r = k_E \sin \xi,
 \;\;\;
 \vec{k} = k_r \left( \begin{array}{c}
 \sin \theta \sin \phi \\
 \sin \theta \cos \phi \\
 \cos \theta
 \end{array} \right).
\eq
The (inverse) Jacobian of this transformation
is
\bq
 p(k) & = & \left| \frac{\partial u}{\partial k} \right| 
 = \frac{2}{\pi^3} 
   \frac{\mu_1^2}{\left[\left(k_0^2+k_r^2\right)^2+\mu_1^4\right]}
   \frac{1}{k_0^2+k_r^2}.
\eq
We then have
\bq
 \int d^4k f(k)
 & = & 
 \int\limits_{[0,1]^4} d^4u
 \frac{f(k)}{p(k)}.
\eq

\subsection{Generating the Feynman parameters}
\label{sect:generating_feynman_parameters}

In this appendix we consider an integral of the form
\bq
 I & = &
 \int d^nx \; \delta\left(1-\sum\limits_{j=1}^n x_j \right) 
 f\left(x_1,...,x_n\right)
\eq
over the standard $(n-1)$-dimensional simplex.
We assume that the function $f(x_1,...,x_n)$ is homogeneous of degree $(-n)$ in the variables
$x_j$.
We may then replace the integration 
over the simplex by an integration over the $n$ faces of the $n$-dimensional hyper-cube which are not
coordinate subspaces. These $n$ faces are defined by
\bq
 x_l = 1, & & 0 \le x_i \le 1, \;\;\; i \neq l.
\eq
We can rewrite the integral as
\bq
 I & = & 
 \sum\limits_{l=1}^n \int\limits_0^1 d^nx \; \delta\left(x_l-1\right) f\left(x_1,...,x_n\right)
 = 
 \sum\limits_{l=1}^n \int\limits_0^1 d^nx \; \delta\left(x_l-1\right) f\left(x_1,...,x_n\right)
 \int\limits_0^1 d\lambda \; n \lambda^{n-1}.
 \nonumber
\eq
Setting
\bq
 \left( \begin{array}{c}
 u_1 \\ ... \\ u_l \\ ... \\ u_n \\
 \end{array} \right)
 & = & 
 \lambda
 \left( \begin{array}{c}
 x_1 \\ ... \\ 1 \\ ... \\ x_n \\
 \end{array} \right)
\eq
we arrive at
\bq
 I & = & 
 n 
 \int\limits_0^1 d^nu \; f\left(x_1,...,x_n\right),
\eq
where 
\bq
 x_i & = & \frac{u_i}{u_{j_{\mathrm{max}}}}
\eq
and $j_{\mathrm{max}}$ is the index such that $u_{j_{\mathrm{max}}} \ge u_i$ for all $i$.

\section{Reducing the power in the denominator}
\label{sect:reduce_power}

In this appendix we show how to reduce the number of propagators, and in consequence 
also the power to which the denominator function $L$ occurs in eq.~(\ref{final_feynman_integral}).
We split the integration into different channels. 
In each channel the number of propagators is reduced.
In different channels different propagators are removed.
Let us consider a process with $n_{\mathrm{process}}$ external partons.
We denote the number of loop propagators in a specific channel by $n_{\mathrm{channel}}$. 
For each channel we have again the kinematical $n_{\mathrm{channel}} \times n_{\mathrm{channel}}$-matrix $S_{ij}$ and 
the $(n_{\mathrm{channel}}-1)\times(n_{\mathrm{channel}}-1)$-dimensional Gram matrix $G^{(a)}_{ij}$.
These kinematical quantities are obtained 
from the kinematics for the process with $n_{\mathrm{process}}$ propagators by
pinching $(n_{\mathrm{process}}-n_{\mathrm{channel}})$ propagators.

The reduction technique exploits the fact that in four dimensions we can have maximally four linear independently vectors and is 
well known from analytical calculations. Here we use it in a purely numerical context.
The technique makes use of a special property of the kinematical matrix $S$ defined in eq.~(\ref{def_S}),
namely that for each channel we can always find numbers $b_1$, $b_2$, ..., $b_{n_{\mathrm{channel}}}$ such that
\bq
 \sum\limits_{j=1}^{n_{\mathrm{channel}}} S_{ij} b_j & = & 1
 \;\;\;\;\;\; \mbox{for all} \; 1 \le i \le n_{\mathrm{channel}}.
\eq
As a standalone method the reduction technique is not as efficient as the method presented in section~\ref{subsect:stability},
mainly because the coefficients $b_i$ are in general large and alternating in sign.
However the reduction technique can be combined with the method of section~\ref{subsect:stability}
in a way we will indicate below. This improves the efficiency of the method of section~\ref{subsect:stability} even
further.

There is a minor modification which we have to make to the method presented in section~\ref{subsect:contour_feynman}:
The reduction technique introduces additional ultraviolet subtraction terms. These additional ultraviolet subtraction
terms cancel when summed over all channels.
However the cancellation requires that the four-vector $Q^\mu$ is chosen to be the same in all channels.
The choice in eq.~(\ref{choice_1_for_Q_mu}) does not meet this requirement, since this choice
depends on the Feynman parameters and therefore on the channel.
An alternative choice is
\bq
\label{choice_2_for_Q_mu}
 Q^\mu & = & \frac{1}{n_{\mathrm{process}}} \sum\limits_{j=1}^{n_{\mathrm{process}}} q_{\mathrm{process},j}^\mu.
\eq
We use the choice in eq.~(\ref{choice_2_for_Q_mu}) in connection with the reduction technique.
Again we can check that the ultraviolet propagators never go on-shell with a suitable choice of $\mu^2_{\mathrm{UV}}$.
For the propagators of the ultraviolet subtraction terms we have now
\bq
 \bar{k}^2 - \mu_{\mathrm{UV}}^2
 & = & 
 2 i \left( \tilde{k} + \frac{1}{2} V \right) \circ \left( \tilde{k} + \frac{1}{2} V \right)
 - \frac{i}{2} V \circ V - \mu_{\mathrm{UV}}^2
 + 2 \tilde{k} \cdot V + V^2,
\eq
where
\bq
 V^\mu & = & 
 \frac{1}{x} \sum\limits_{i=1}^{n_{\mathrm{channel}}} x_i q_{\mathrm{channel},i}^\mu
 - 
 \frac{1}{n_{\mathrm{process}}} \sum\limits_{j=1}^{n_{\mathrm{process}}} q_{\mathrm{process},j}^\mu.
\eq
Since the Euclidean norm is always non-negative we can ensure for infinitesimal $\lambda$
that the imaginary part of this expression is always positive by setting 
$\mu_{\mathrm{UV}}^2$ purely imaginary with 
\bq
 \mbox{Im}\;\mu_{\mathrm{UV}}^2 & < & - \frac{1}{2} \; \max \; V \circ V.
\eq
Doing so, we ensure that the propagators of the ultraviolet subtraction terms never go on-shell.

In practise we use the reduction technique in combination with the techniques presented in section~\ref{sect:contour_deformation}.
We rewrite an integral of the form
\bq
 I & = & 
 \int \frac{d^4k}{(2\pi)^4} 
 \frac{R(k)}{\prod\limits_{j=1}^{n_{\mathrm{process}}} \left(k_j^2 - m_j^2 \right)}
\eq
as
\bq
 I & = & 
 \int \frac{d^4k}{(2\pi)^4} 
 \frac{R(k)}{\prod\limits_{j=1}^{n_{\mathrm{process}}} \left(k_j^2 - m_j^2 - \mu_{\mathrm{IR}}^2 \right)}
 +
 \int \frac{d^4k}{(2\pi)^4} 
 \left[
       \frac{R(k)}{\prod\limits_{j=1}^{n_{\mathrm{process}}} \left(k_j^2 - m_j^2 \right)}
       -
       \frac{R(k)}{\prod\limits_{j=1}^{n_{\mathrm{process}}} \left(k_j^2 - m_j^2 - \mu_{\mathrm{IR}}^2 \right)}
 \right].
 \nonumber \\
\eq
The first term is calculated directly with the help of the Feynman parametrisation.
To the second term we apply first the reduction identities 
and then the methods of section~\ref{subsect:contour_feynman} and \ref{subsect:stability}.

\subsection{Reduction identities}
\label{sect:reduction_identities}

In this appendix we show how to reduce in a channel with $n$ propagators the number of propagators 
from $n$ to $(n-1)$.
In the reduction identities we distinguish the cases $n\ge 6$ and $n<6$.
In the former case there is a complete reduction to $(n-1)$, 
while in the latter case an integral with $n$ propagators remains.
It may seem that in the case $n<6$ not much is gained, but it can be shown that the remaining integral
with $n$ propagators is better behaved in the infrared.
We start with the case $n \ge 6$, the case $n < 6$ will be discussed later.

We will make use of the property that we can always find numbers $b_1$, $b_2$, ..., $b_n$ such that
\bq
 \sum\limits_{j=1}^n S_{ij} b_j & = & 1
 \;\;\;\;\;\; \mbox{for all} \; 1 \le i \le n.
\eq
The kinematical $n \times n$-matrix $S$ was defined in eq.~(\ref{def_S}).
The kinematical matrix $S$ and the numbers $b_j$ are independent of the loop momentum $k$.
We have collected useful information for the computation of the numbers $b_j$ in the appendix \ref{sect:coeff_b_i}.
For $n\ge 6$ the numbers $b_1$, ..., $b_n$ have the additional properties that
\bq
 \sum\limits_{j=1}^n b_j = 0,
 & &
 \sum\limits_{j=1}^n b_j q_j = 0.
\eq
This leads to the identity
\bq
\label{def_reduction}
 1
 & = &
 \sum\limits_{i=1}^n 
 b_i
 \left\{
 \left( k_i^2 - m_i^2 \right)
 - \left[ \bar{k}^2 - \mu_{\mathrm{UV}}^2 +  \sum\limits_{r=1,r\neq i}^n 2 \bar{k} \cdot \left(q_r-Q\right) \right]
   \frac{\prod\limits_{j=1}^n \left( k_j^2 - m_j^2 \right)}
        {\left( \bar{k}^2 - \mu_{\mathrm{UV}}^2 \right)^{n}}     
 \right\}.
\eq
The second term inside the curly bracket is an ultraviolet subtraction term and compensates the quadratic and linear growth
of $k_i^2-m_i^2$ with large $|k|$.
The complete expression inside the curly bracket goes to a constant for large $|k|$.
The sum of all the extra ultraviolet subtraction terms vanishes.
If we start from an integral of the form
\bq
 I & = & 
 \int \frac{d^4k}{(2\pi)^4} 
 \frac{R(k)}{\prod\limits_{j=1}^n \left(k_j^2 - m_j^2 \right)},
\eq
where $R(k)$ is a rational function in the loop momentum with poles only at
\bq
 \bar{k}^2 - \mu_{\mathrm{UV}}^2 & =& 0,
\eq
we obtain the following reduction:
\bq
\label{reduction_to_5}
 I & = &
 \sum\limits_{i=1}^n 
 \int \frac{d^4k}{(2\pi)^4} 
 \frac{R_i(k)}{\prod\limits_{j=1,j\neq i}^n \left(k_j^2 - m_j^2 \right)},
\eq
with
\bq
\label{def_R_i}
 R_i(k) & = &
 b_i R(k)
 \left\{
 1
 - \left[ \bar{k}^2 - \mu_{\mathrm{UV}}^2 +  \sum\limits_{r=1,r\neq i}^n 2 \bar{k} \cdot \left(q_r-Q\right) \right]
   \frac{\prod\limits_{j=1, j \neq i}^n \left( k_j^2 - m_j^2 \right)}
        {\left( \bar{k}^2 - \mu_{\mathrm{UV}}^2 \right)^{n}}     
 \right\}.
\eq
Eq.~(\ref{reduction_to_5}) splits an integral with $n$ propagators into $n$ integrals with $(n-1)$ propagators each.
This reduces the number of propagators which need to be taken into account for the contour deformation.
The rational functions $R_i(k)$ appearing in eq.~(\ref{reduction_to_5}) have again only poles at the location
of the ultraviolet subtraction propagator.
We can iterate this procedure until we arrive at channels with five propagators.

Let us now consider the case $n \le 5$.
Again we can find numbers $b_1$, ..., $b_n$, such that
\bq
 \sum\limits_{j=1}^n S_{ij} b_j & = & 1
 \;\;\;\;\;\; \mbox{for all} \; 1 \le i \le n.
\eq
But in contrast to the previous case we now have in general
\bq
 \sum\limits_{j=1}^n b_j \neq 0,
 & &
 \sum\limits_{j=1}^n b_j q_j \neq 0.
\eq
The reduction identity reads now
\bq
\label{reduction_5_to_4}
 I & = &
 \int \frac{d^4k}{(2\pi)^4} 
 \frac{\tilde{R}(k)}{\prod\limits_{j=1}^n \left(k_j^2 - m_j^2 \right)}
 +
 \sum\limits_{i=1}^n 
 \int \frac{d^4k}{(2\pi)^4} 
 \frac{R_i(k)}{\prod\limits_{j=1,j\neq i}^n \left(k_j^2 - m_j^2 \right)},
\eq
where $R_i(k)$ is defined as in eq.~(\ref{def_R_i}).
The rational function $\tilde{R}(k)$ is given by
\bq
 \tilde{R}(k) & = &
 R(k) \left[ C + C_{\mathrm{UV}} \frac{\prod\limits_{j=1}^n \left(k_j^2 - m_j^2\right) }
                                      {\left(\bar{k}^2 - \mu_{\mathrm{UV}}^2 \right)^{n}} \right]
\eq
$C$ and $C_{\mathrm{UV}}$ are given by
\bq
 C & = & 
    - B k^2 
    + \sum\limits_{i=1}^n 2 k \cdot \left( b_i q_i \right)  
    + \sum\limits_{i=1}^n \sum\limits_{j=1}^n b_i \left( q_j^2 - 2 q_i \cdot q_j - m_j^2 \right) \frac{x_j}{x},
 \nonumber \\ 
 C_{\mathrm{UV}} & = &
    B \left( \bar{k}^2 - \mu_{\mathrm{UV}}^2 \right)
    + 2 \bar{k} \cdot \left(-(n-1) B Q+ \sum\limits_{r=1}^n (B-b_r) q_r\right).
\eq
In the formula for the coefficient $C$ the variables $x_j$ ($j=1,...,n$) can be arbitrary.
We can take them equal to the Feynman parameters.
The quantities $B$ and $x$ are given by
\bq
 B = \sum\limits_{i=1}^n b_i, 
 & &
 x = \sum\limits_{i=1}^n x_i.
\eq
It is advantageous to slightly rearrange the formula for the coefficient $C$.
We first write the loop momentum $k^\mu$ as
\bq
\label{def_K_mu}
 k^\mu & = & K^\mu + \frac{1}{x} \sum\limits_{i=1}^n x_i q_i^\mu,
\eq
where the $x_i$'s are the Feynman parameters. 
In terms of the shifted loop momentum $K^\mu$ we have for the coefficient $C$
\bq
 C & = & 
 B \left[
          -K^2  
          + \frac{1}{2x^2} \sum\limits_{i=1}^n \sum\limits_{j=1}^n x_i S_{ij} x_j
   \right]
 - 2 K \cdot \left[ \sum\limits_{i=1}^n \left( B \frac{x_i }{x} - b_i \right) q_i 
             \right].
\eq
Eq.~(\ref{reduction_5_to_4}) does not eliminate the $n$-point integrals.
Instead this reduction identity provides an additional factor $C$ in the numerator
for the $n$-point integrals. The factor $C$ vanishes, whenever the denominator after 
Feynman parametrisation and contour deformation vanishes.
This improves the numerical stability.

\subsection{The coefficients $b_i$}
\label{sect:coeff_b_i}

In this appendix we collect useful information related to the fact that the vector $(1,1,...,1)$
is always in the range of the kinematical matrix $S$ \cite{Binoth:1999sp,Fleischer:1999hq,Denner:2002ii,Duplancic:2003tv,Giele:2004iy,vanHameren:2005ed}. 
We recall that the kinematical matrix $S$
was defined in eq.~(\ref{def_S}) as a $n \times n$-matrix by
\bq
 S_{ij} & = & \left( q_i - q_j \right)^2 - m_i^2 - m_j^2.
\eq
$S$ depends only on the external momenta and the internal masses, but not on the loop momentum.
It is a highly non-trivial statement, that we can always
determine $n$ numbers $b_1$, ..., $b_n$ such that
\bq
\label{def_eqn_b_coeff}
 \sum\limits_{j=1}^n S_{ij} b_j & = & 1
 \;\;\;\;\;\; \mbox{for all} \; 1 \le i \le n.
\eq
For generic external momenta the matrix $S$ is invertible for $n \le 6$. In this case $b_i$ is given by
\bq
 b_i & = & \sum\limits_{j=1}^n \left( S^{-1} \right)_{ij}
\eq
For $n \ge 7$ we always have $\det S = 0$ and therefore $S$ is not invertible.
However the vector $(1,...,1)$ is still in the range of the linear map defined by $S$.
In this case the solution for the numbers $b_i$ is not unique.
A possible set of coefficients $b_i$ can be
obtained from the singular value decomposition of the $(n-1) \times (n-1)$ Gram matrix 
$G^{(a)}_{ij} = 2 (q_i - q_a) (q_j- q_a)$ where the indices take the values $i,j \neq a$.
The starting point is to express the kinematical matrix $S_{ij}$ in terms of the
Gram matrix:
\bq
 S_{ij} & = & - G_{ij}^{(a)} + V_i^{(a)} + V_j^{(a)},
 \;\;\;\;\;\;
 V_i^{(a)} = \left( q_i - q_a \right)^2 - m_i^2.
\eq
The condition in eq.~(\ref{def_eqn_b_coeff}) translates into the two equations
\bq
 \sum\limits_{j=1, j \neq a}^n G_{ij}^{(a)} b_j & = & B \left( V_i^{(a)} - V_a^{(a)} \right),
 \nonumber \\
 \sum\limits_{j=1}^n \left( V_j^{(a)} - V_a^{(a)} \right) b_j & = & 1 - 2 B V_a^{(a)},
\eq
where we set $B=\sum\limits_{j=1}^n b_j$.
An equation of the form $G x = y$ has a solution if and only if $y=G H y$, where
$H$ is the pseudo-inverse of $G$. 
The pseudo-inverse $H$ of a symmetric matrix $G$ is uniquely defined by the properties
\bq
 H G H = H,
 \;\;\;\;\;\;
 G H G = G,
 \;\;\;\;\;\;
 G H = H G.
\eq
In the case where $y=G H y$ holds the solution to the equation $G x = y$ is given by
\bq
 x & = & H y + \left( 1 - H G \right) u,
\eq
where $u$ is an arbitrary vector. If $G$ does not have maximal rank one can show
that $y = G H y$ implies $y=0$. 
Applied to our case this in turn implies $B=0$. 
Noting that $V_j^{(a)}-V_a^{(a)}=S_{aj}$ we therefore have the equations
\bq
 \sum\limits_{j=1, j \neq a}^b G_{ij}^{(a)} b_j & = & 0,
 \nonumber \\
 \sum\limits_{j=1}^n S_{aj}  b_j & = & 1.
\eq
To solve these equations one first determines $(n-1)$ numbers 
$(b_1,...,b_{a-1},b_{a+1},...,b_n)$, which define a vector in the kernel of $G$.
The second equation is then used to fix the value of $b_a$.
The matrix $G$ has the singular value decomposition
\bq
G_{ij} & = & \sum\limits_{k=1}^4 U_{ik} w_k \left(V^T\right)_{kj},
\eq
where $U$ and $V$ are orthogonal $(n-1)\times(n-1)$ matrices.
The kernel of $G$ is spanned by the vectors
$V_{i 5}$, $V_{i 6}$, ..., $V_{i (n-1)}$.
We can therefore set 
\bq
  b_i & = & \frac{V_{i 5}}{W_5}, \;\;\; i < a,
 \nonumber \\
  b_{i+1} & = & \frac{V_{i 5}}{W_5}, \;\;\; i \ge a,
 \nonumber \\
  b_a & = & - \sum\limits_{j=1, j\neq a}^n b_j.
\eq
The normalisation factor $W_5$ we get from 
$\sum\limits_{j=1}^n S_{aj}  b_j = 1$:
\bq
 W_5 & = & 
 \sum\limits_{j<a} \left( S_{aj} - S_{aa} \right) V_{j5}
 +
 \sum\limits_{j>a} \left( S_{aj} - S_{aa} \right) V_{(j-1)\;5}.
\eq
Let us now consider the four-vector $r=\sum\limits_{i=1}^n b_i q_i$. For $n \ge 6$ we can show that
\bq
 q_a \cdot r 
 & = &
 q_b \cdot r
\eq
for all $a,b \in \{1,...,n\}$. In particular this implies
\bq
 q_a \cdot r
 & = &
 q_n \cdot r = 0.
\eq
Therefore it follow that $r=0$.
\\
\\
Let us summarise the situation: We distinguish three cases.
\begin{enumerate}
\item $n\le 5$. For generic external momenta the kinematical matrix $S$ is invertible 
and the coefficients $b_i$ are unique. The sum of all coefficients $B$ is non-zero and the four-vector $r$
is non-zero.
\item $n=6$. For generic external momenta the kinematical matrix $S$ is invertible 
and the coefficients $b_i$ are unique. The sum of all coefficients $B$ is zero and the four-vector $r$ 
is zero.
\item $n\ge 7$. The kinematical matrix $S$ is not invertible 
and the coefficients $b_i$ are not unique. The sum of all coefficients $B$ is zero and the four-vector $r$
is zero.
\end{enumerate}
At the end of this appendix we consider the special case where a constant quantity $\mu_{\mathrm{IR}}^2$
is added to all masses squared:
\bq
 {m_i'}^2 & = & m_i^2 + \mu_{\mathrm{IR}}^2.
\eq
The quantity $\mu_{\mathrm{IR}}^2$ may be complex, however we assume that the masses $m_i$ are real.
As before we have the kinematical matrix $S_{ij}$ defined by
\bq
 S_{ij} & = & \left( q_i-q_j \right)^2 - m_i^2 - m_j^2,
\eq
and the corresponding matrix $S_{ij}'$ with the primed masses:
\bq
 S_{ij}' & = & S_{ij} - 2 \mu_{\mathrm{IR}}^2.
\eq
The kinematical matrix $S_{ij}$ is real, whereas the matrix $S_{ij}'$ may be complex.
Suppose we are interested in the coefficients $b_i'$, which satisfy
\bq
\label{eq_S_prime_b_prime}
 \sum\limits_{j=1}^n S_{ij}' b_j' & = & 1.
\eq
These can be obtained without complex arithmetic. We first determine the real coefficients $b_i$ defined by
\bq
 \sum\limits_{j=1}^n S_{ij} b_j & = & 1
\eq
and set then
\bq
\label{b_i_prime}
 b_i' & = & \frac{b_i}{1-2\mu_{\mathrm{IR}}^2 B}.
\eq
One easily verifies that the coefficients in eq.~(\ref{b_i_prime}) satisfy eq.~(\ref{eq_S_prime_b_prime}).
In particular we have $b_i'=b_i$ for $n\ge 6$.

\end{appendix}

\bibliography{/home/stefanw/notes/biblio}
\bibliographystyle{/home/stefanw/latex-style/h-physrev3}

\end{document}